\documentclass[a4paper,11pt]{article}
\usepackage{jheppub} 

\usepackage{physics}

\usepackage{xcolor,tikz}
\usetikzlibrary{patterns}
\usetikzlibrary{decorations.markings}
\usepackage{tikz-cd}
\usetikzlibrary{arrows}
\usepackage{cleveref}

\usepackage[breakable, theorems, skins]{tcolorbox}

\renewcommand{\P}{\mathbb{P}}

\newtheorem{theorem}{Theorem}[section]

\newtheorem{corollary}[theorem]{Corollary}
\newtheorem{lemma}[theorem]{Lemma}
\newtheorem{proposition}[theorem]{Proposition}
\newtheorem{definition}[theorem]{Definition}

\usepackage{enumitem}

\usepackage{array}
\newcolumntype{C}[1]{>{\centering\arraybackslash}m{#1}}

\def\bc{\begin{corollary}}
\def\ec{\end{corollary}}
\def\bcl{\begin{claim}}
\def\ecl{\end{claim}}
\def\bd{\begin{definition}}
\def\ed{\end{definition}}
\def\ben{\begin{enumerate}}
\def\een{\end{enumerate}}
\def\be{\begin{equation}}
\def\ee{\end{equation}}
\def\bse{\begin{equation*}}
\def\ese{\end{equation*}}
\def\bex{\begin{example}}
\def\eex{\end{example}}
\def\bit{\begin{itemize}}
\def\eit{\end{itemize}}
\def\bl{\begin{lemma}}
\def\el{\end{lemma}}
\def\bnn{\begin{notation}}
\def\enn{\end{notation}}
\def\bn{\begin{note}}
\def\en{\end{note}}
\def\bp{\begin{proposition}}
\def\ep{\end{proposition}}
\def\bq{\begin{proof}}
\def\eq{\end{proof}}
\def\br{\begin{remark}}
\def\er{\end{remark}}
\def\bs{\begin{solution}}
\def\es{\end{solution}}
\def\btab{\begin{table}}
\def\etab{\end{table}}
\def\btb{\begin{tabular}}
\def\etb{\end{tabular}}
\def\bt{\begin{theorem}}
\def\et{\end{theorem}}
\def\btik{\begin{tikzpicture}}
\def\etik{\end{tikzpicture}}

\def\a{\alpha}
\def\b{\beta}
\def\del{\delta}
\def\g{\gamma}
\def\l{\lambda}

\def\sig{\sigma}
\def\Sig{\Sigma}

\def\b1{\mathbb{1}} 
\def\C{\mathbb{C}} 

\def\bP{\mathbb{P}} 
\def\R{\mathbb{R}} 

\def\Z{\mathbb{Z}} 

\def\cJ{\mathcal{J}}

\def\cL{\mathcal{L}}
\def\cM{\mathcal{M}}
\def\cN{\mathcal{N}}

\def\cT{\mathcal{T}}

\def\so{\mathfrak{so}}

\def\fR{\mathfrak{R}}

\def\p{\partial}

\def\ss{\subset}

\def\la{\langle}

\def\ra{\rangle}
\def\qand{\quad \text{and} \quad}

\def\bbox{\begin{tcolorbox}[colback=white!5,colframe=blue!75!black,title=Exercise]}
\def\ebox{\end{tcolorbox}}

\newcommand*\xbar[1]{%
  \hbox{%
    \vbox{%
      \hrule height 0.5pt 
      \kern0.5ex
      \hbox{%
        \kern-0.1em
        \ensuremath{#1}%
        \kern-0.1em
      }%
    }%
  }%
} 

\title{\boldmath $G_2$ Mirrors from Calabi-Yau Mirrors}

\author{Andreas P. Braun}
\author{and Richie Dadhley}

\affiliation{Department of Mathematical and Computing Sciences,\\ 
Durham University Upper Mountjoy Campus, \\
Stockton Rd, Durham DH1 3LE, UK}

\emailAdd{andreas.braun@durham.ac.uk}
\emailAdd{richie.s.dadhley@durham.ac.uk}

\abstract{We study the worldsheet CFTs of type II strings on compact $G_2$ orbifolds obtained as quotients of a product of a Calabi-Yau threefold and a circle. For such models, we argue that the Calabi-Yau mirror map implies a mirror map for the associated $G_2$ varieties by examining how anti-holomorphic involutions behave under Calabi-Yau mirror symmetry. The mirror geometries identified by the worldsheet CFT are consistent with earlier proposals for twisted connected sum $G_2$ manifolds.}

\begin{document}
\maketitle
\flushbottom

\newpage 
\section{Introduction}
\label{sec:intro}

Since its discovery \cite{Dixon:1988,lerche1989chiral,Candelas:1989hd}, mirror symmetry for type II strings on Calabi-Yau manifolds quickly evolved into a powerful tool \cite{Candelas:1990rm} with intricate mathematical implications such as homological mirror symmetry\cite{Kontsevich:1994dn}. This development has been driven by the wealth of examples that can be readily constructed and analyzed using techniques from toric geometry \cite{Batyrev94dualpolyhedra,Hosono:1993qy}, and a detailed understanding of the worldsheet SCFT in which the equivalence for distinct target spaces could be proven directly \cite{greene1990duality,Aspinwall:1990xe,hori2000mirror}. Key technical advances in this development were Gepner models \cite{Gepner:1987vz,gepner1989space}, which give direct access to the worldsheet SCFT, as well as the detailed study of $\mathcal{N} = (2,2)$ models and in particular the correspondence between Calabi-Yau sigma-models and Landau-Ginzburg models \cite{Vafa:1988uu,Greene:1988ut,Cecotti:1989jc, Cecotti:1990kz,Witten:1993yc}. Extending the equivalence from the worldsheet theory to the full string theories, which includes BPS states associated with wrapped branes, not only vastly extended the scope of this duality, but also led to the geometric idea of mirror symmetry being T-duality along the fibres of a torus fibration \cite{Strominger:1996it}. This picture becomes particularly clear 
for toroidal orbifolds \cite{vafa1995orbifolds}.

For $G_2$ manifolds, a similar phenomenon in which topologically distinct target spaces lead to isomorphic worldsheet SCFTs has been conjectured in \cite{Shatashvili:1994zw}, and has been dubbed `$G_2$ mirror symmetry'. Whereas a necessary condition for a pair of Calabi-Yau threefolds $X$, $X^\vee$ to be mirror is that their Hodge numbers satisfy
\begin{equation}
\left(h^{1,1}(X),h^{2,1}(X)\right) = \left(h^{2,1}(X^\vee),h^{1,1}(X^\vee)\right)  
\end{equation}
the corresponding condition for a pair of $G_2$ manifolds $M$ and $M^\vee$ is weaker and says merely that their Betti numbers obey
\begin{equation}\label{eq:G2mirrorcondition}
 b^2(M) + b^3(M) =  b^2(M^\vee) + b^3(M^\vee)\, .
\end{equation}
Coincident with this work in physics was the first construction of compact $G_2$ manifolds as smoothings of toroidal orbifolds \cite{joyce1996compactI,joyce1996compactII}, which intriguingly also produced the first examples satisfying the above relation. As for Calabi-Yau orbifolds, mirror symmetry can be understood as a consequence of T-duality for these cases \cite{Acharya:1997rh,gaberdiel2004generalised}. 

$G_2$ target spaces only preserve half of the supercharges of their Calabi-Yau cousins, which makes it significantly harder to establish robust results for the worldsheet SCFT, and most of the techniques which proved invaluable for Calabi-Yau targets become unavailable. For models based on Calabi-Yau quotients, detailed studies based on Gepner models were undertaken in \cite{eguchi2002string,blumenhagen2002superconformal,roiban2002rational,Roiban:2002iv}. 

Geometrically, the reduced supersymmetry manifests itself in the difficulty to construct manifolds of $G_2$ holonomy. There is no analogue of Yau's theorem, i.e. no topological condition which guarantees the existence of a Ricci-flat metric with holonomy $G_2$, and almost all examples are ultimately based on Calabi-Yau geometry. Mirror manifolds were proposed for orbifolds based on Calabi-Yau threefolds in \cite{Harvey:1999as,Partouche:2000uq}, for twisted connected sum (TCS) $G_2$ manifolds in \cite{Braun:2017ryx,Braun:2017csz}, and for non-compact $G_2$ manifolds with adiabatic circle fibrations in \cite{Aganagic:2001ug}.

The main purpose of the present work is to strengthen the understanding of $G_2$ mirror symmetry from the perspective of the worldsheet SCFT, and in particular to provide more evidence for some of the geometric mirror constructions that have appeared in the literature. Our approach is based on the work of 
\cite{hori2000mirror}, which showed how mirror symmetry for Calabi-Yau hypersurfaces in toric varieties can be demonstrated by using duality in $\mathcal{N}=(2,2)$ gauged linear sigma models (GLSMs). For a Calabi-Yau threefold $X$ of this type, their analysis together with T-duality shows the equivalence of the worldsheet theory for the following models 
    \begin{center}
        \btik 
            \node[left] at (0,0) {IIA/B on $X \times S^1$};
            \node[right] at (4,0) {IIB/A on $X^{\vee} \times S^1$};
            \node[left] at (0,-3) {IIB/A on $X \times \left(S^1\right)^\vee$};
            \node[right] at (4,-3) {IIA/B on $X^\vee \times \left(S^1\right)^\vee$};
            \draw[thick, ->] (0,0) -- (4,0);
            \node[above] at (2,0.25) {mirror};
            \draw[thick, ->] (0,-3) -- (4,-3);
            \node[below] at (2,-3.25) {mirror};
            \draw[thick, ->] (-0.5,-0.5) -- (-0.5,-2.5);
            \node[left] at (-0.75,-1.5) {T-duality};
            \draw[thick, ->] (4.5,-0.5) -- (4.5,-2.5);
            \node[right] at (4.75,-1.5) {T-duality};
        \etik 
    \end{center}
For every one of those four geometries we may form a $G_2$ variety\footnote{We will loosely use the term $G_2$ variety in analogy to its usage in algebraic geometry to refer to both $G_2$ manifolds, as well as singular geometries which become $G_2$ manifolds after an appropriate smoothing, or after their singularities are excised.} by modding out an isometry $(\sigma,-)$ which acts as a anti-holomorphic involution $\sigma$ on the Calabi-Yau, and an inversion on the circle. Using the explicit dualisation of \cite{hori2000mirror} allows us to show that appropriate anti-holomorphic involutions are identified under the mirror map. This implies that we not only have four isomorphic SCFTs, but four isomorphic instances of an SCFT together with an involution. Taking the quotient hence results in four isomorphic $\mathcal{N}=(1,1)$ theories after including appropriate twisted sectors. There hence exist three $G_2$ mirror maps as a consequence of Calabi-Yau mirror symmetry and T-duality for such models. A similar line of reasoning has already appeared in \cite{Harvey:1999as} for `barely' $G_2$ manifolds resulting from free actions of $\sigma$, and for non-compact $G_2$ manifolds in \cite{Aganagic:2001ug}.

Using the equivalence of a large class of TCS $G_2$ manifolds with (smoothings of) $(X \times S^1)/(\sigma,-)$  found in \cite{Braun:2019wnj} in turn allows us to compare our results with the proposals made for TCS $G_2$ manifolds in \cite{Braun:2017ryx,Braun:2017csz}. TCS $G_2$ manifolds are based on gluing two asymptotically cylindrical Calabi-Yau threefolds $X_\pm$ times a circle, and \cite{Braun:2017ryx,Braun:2017csz} described three mirror constructions which either swap $X_+$, $X_-$, or both with an appropriate mirror. For TCS $G_2$ manifolds which have an equivalent realization as $(X \times S^1)/(\sigma,-)$, we find that these three mirrors are consistent with the three mirrors found above.

This paper is organized as follows. As an extended introduction and to set notation we review the constructions of compact $G_2$ manifolds and the mirror geometries that have been proposed (mostly from a space-time perspective) in \Cref{sect:spacetime_G2}, as well as aspects of worldsheet theories for type II strings on Calabi-Yau threefolds in  \Cref{sect:CY_mirrors}. After reviewing some aspects of type II strings with $G_2$ targets, we then show how anti-holomorphic involutions of the GLSM are identified under the mirror map of \cite{hori2000mirror} in \Cref{sect:G2_mirror_ws}. We end with a discussion of our results and several avenues for further work in \Cref{sect:discussion}. As an elementary exemplification of the lift of Calabi-Yau mirror symmetry to associated $G_2$ manifolds, we treat the case of a Joyce orbifold from this perspective in detail in an \Cref{sec:TorodialOrbifold}. The largest classes of geometric mirror constructions for Calabi-Yau and $G_2$ are based on toric geometry, and we review the necessary tools in \Cref{sec:BatyrevMirrorSym}.

\section{Constructions of $G_2$ Manifolds and Mirror Symmetry}\label{sect:spacetime_G2}

In this section we will review those constructions of compact $G_2$ manifolds needed later in this paper, explain the overlap between them, and state the 
mirror pairs which have been proposed in the literature. For a general introduction to $G_2$ geometry see \cite{joyce2000compact}.

\subsection{Joyce Orbifolds}

The first non-trivial examples of compact $G_2$ manifolds were found by Joyce \cite{joyce1996compactI,joyce1996compactII} as smoothings of torus 
orbifolds $T^7/\Gamma$ with $\Gamma$ a finite subgroup of $G_2$. At the time of writing, there exists no classification of such finite subgroups, and only a few examples 
have been studied from the CFT perspective using a free field realization \cite{Acharya:1997rh,gaberdiel2004generalised,Braun:2019lnn}. A particularly instructive example 
is based on $\Gamma = \Z_2^3$ with generators $\alpha,\beta,\gamma$ acting on $T^7$ with coordinates 
$x_i \sim x_i+1$, $i=1...7$ as 
    \be
        \begin{split}
            \a : (x_1,x_2,x_3,x_4,x_5,x_6,x_7) & \mapsto (x_1,x_2,x_3,-x_4,-x_5,-x_6,-x_7) \\
            \beta : (x_1,x_2,x_3,x_4,x_5,x_6,x_7) & \mapsto (x_1,-x_2,-x_3,x_4,-x_5,\tfrac{1}{2}-x_6,x_7) \\
            \gamma : (x_1,x_2,x_3,x_4,x_5,x_6,x_7) & \mapsto (-x_1,x_2,-x_3,x_4,-x_5,x_6,\tfrac12-x_7),
        \end{split}
    \ee
As shown in \cite{joyce1996compactII} there are $9$ topologically inequivalent smoothings $M_l$ of this orbifold that have a Ricci-flat $G_2$ metric. 
The resulting Betti numbers are 
\begin{equation}
 b^2(M_l) = 8 + l \qquad b^3(M_l) = 47 - l \, ,
\end{equation}
for $l = 0 ... 8$. All of these share the same $b^2 + b^3$ and hence satisfy the condition of \eqref{eq:G2mirrorcondition}. By analysing different assignments of discrete 
torsion, \cite{Acharya:1997rh,gaberdiel2004generalised} reproduced those Betti numbers. There are two non-trivial mirror maps $\mathcal{T}_3$ and 
$\mathcal{T}_4$ based on performing three or four T-dualities along suitable tori, which map 
\begin{equation}
 \begin{aligned}
  \mathcal{T}_3: \mbox{IIA/B on }  M_l \rightarrow  \mbox{IIB/A on }  M_{8-l} \\
  \mathcal{T}_4: \mbox{IIA/B on }  M_l \rightarrow  \mbox{IIA/B on }  M_{8-l} \\
 \end{aligned}\, .
\end{equation}

\subsection{Quotients of Calabi-Yau Manifolds}

Given a Calabi-Yau threefold $X$, an anti-holomorphic involution $\sigma$ is an isometry of $X$ which maps the complex structure of $X$ to minus itself. In suitable local coordinates it can be described as complex conjugation and it acts as
\begin{equation}
\sigma: \qquad\begin{aligned}
J& \rightarrow -J \\
\Omega^{3,0}& \rightarrow  \xbar{\Omega^{3,0}} 
 \end{aligned}
\end{equation}
on the K\"ahler form $J$ and the holomorphic three-form $\Omega^{3,0}$ of $X$. As a consequence of the oddness of the volume form and Poincar\'e duality, 
the dimensions of even/odd subspaces of cohomology are such that \cite{Grimm:2004ua}
\begin{equation}
b^2_\pm = b^4_\mp \qquad
 b^3_+ = b^3_-\, .
\end{equation}
While freely acting anti-holomorphic involutions of complete intersection Calabi-Yau threefolds in products of projective spaces have been classified in \cite{Grigorian:2009nx}, there is no such classification for other classes of Calabi-Yau threefolds. In this work we will focus on Calabi-Yau hypersurfaces in
toric varieties, where there is always at least one anti-holomorphic involution which maps all of the homogeneous coordinates to their complex conjugates. We will call this the `vanilla' involution $\sigma_v$. Of course, its definition depends on the embedding of $X$ into an ambient toric variety and is not intrinsic to the Calabi-Yau, but this is not inappropriate for our formulation.

The product $X \times S^1$ becomes a $G_2$ manifold with the associative and coassociative forms 
\begin{equation}
\begin{aligned}
 \Phi &:= \mbox{Re}\left(\Omega^{3,0}\right) + J \wedge dt \\
 \Psi &:= \mbox{Im}\left(\Omega^{3,0}\right) \wedge dt + \tfrac12 J \wedge J
\end{aligned}
\end{equation}
and the associated $G_2$ structure persists upon taking a quotient by $(\sigma,-)$ where $(-)$ acts on the coordinate $\theta \sim \theta+1$ on $S^1$ as $\theta \rightarrow -\theta$. We will call such involutions $G_2$ involutions. 

The fixed point locus $L_\sigma$ of an anti-holomorphic involution is either empty, or a special Lagrangian submanifold of $X$. In the former case 
\begin{equation}
 M_\sigma = \left( X \times S^1 \right)/(\sigma,-)
\end{equation}
is smooth and is called a barely $G_2$ manifold. Its holonomy is not all of $G_2$, but only $SU(3) \ltimes \Z_2$ and its Betti numbers are
\begin{equation}\label{eq:bettibarelyg2}
b^2 = h^{1,1}_+(X) \hspace{1cm} b^3 = 1 + h^{1,1}_- + h^{2,1} 
\end{equation}

If the fixed locus $L_\sigma$ of $\sigma$ on $X$ is not empty, $M_\sigma$ has two copies of $L_\sigma$ worth of singularities locally modelled on $A_1 \times \R^3$. 
As shown in \cite{2017arXiv170709325J}, $M_\sigma$ can be smoothed to a $G_2$ manifold $M$ if there exists a $\Z_2$ bundle $\mathcal{Z}$ on $L_\sigma$, such that there is 
a nowhere vanishing harmonic\footnote{With respect to the Ricci-flat K\"ahler metric on $X$.} one-form $\lambda$ valued in $\mathcal{Z}$ on $L_\sigma$. The Betti numbers of $M$ are then given by
\begin{equation}\label{eq:joyceKara_betti}
 b^2(M) = b^2(M_\sigma)+ b^0(L_\sigma,\mathcal{Z})\qquad b^3(M)= b^3(M_\sigma)+b^1(L_\sigma,\mathcal{Z})\, .
\end{equation}
Here $b^k(M_\sigma)$ are counting those cohomology classes that are even under $\sigma$ and $b^i(L_\sigma,\mathcal{Z})$ are the $\mathcal{Z}$-twisted Betti numbers. 
Even though it is not difficult to write down Calabi-Yau threefolds with anti-holomorphic involutions, it is in general hard to show the existence of such a form $\lambda$. 

The worldsheet theories associated to $G_2$ manifolds of this type have been investigated from the perspective of Gepner models in \cite{eguchi2002string,blumenhagen2002superconformal,roiban2002rational,Roiban:2002iv}. Here, different choices of discrete torsion also give rise to different smoothings. As argued in \cite{blumenhagen2002superconformal}, the usual mirror map for Gepner models leads to isomorphic SCFTs after the involution as well. 

An argument leading to a similar conclusion was sketched in \cite{Harvey:1999as} for barely $G_2$ manifolds, i.e. Calabi-Yau mirror symmetry of $X$ induces a pair of $G_2$ mirrors in the above construction. In this case $L_\sigma = \emptyset$ and \eqref{eq:bettibarelyg2} shows that \eqref{eq:G2mirrorcondition} is satisfied for Calabi-Yau mirrors with free involutions.

\subsection{Twisted Connected Sums}\label{sect:TCS}

Twisted connected sum $G_2$ manifolds are obtained from a gluing construction \cite{MR2024648,MR3109862,Corti:2012kd} using a pair of asymptotically cylindrical (acyl) Calabi-Yau threefolds. Detailed discussions can also be found in the physics literature \cite{Halverson:2014tya,Halverson:2015vta,Braun:2016igl,Guio:2017zfn}, see \cite{Braun:2017uku} for a derivation of the TCS construction from the duality between M-Theory and heterotic strings. 


A non-compact Calabi-Yau threefold $X$ is called asymptotically cylindrical if it is diffeomorphic to the product of a K3 surface $S_{0}$ and a cylinder 
$S^1_b \times I$ outside a compact submanifold. Asymptotically cylindrical Calabi-Yau threefolds can be obtained as follows: let $Z$ be a K3 fibred 
K\"ahler threefold whose first Chern class, $c_1(Z)$, is given by the Poincar\'{e} dual $[S_0]_{PD}$ of the homology class $[S_0]$, with $S_0$ the K3 fibre over $p_0 \in \bP^1$. Then 
\begin{equation}
 X = Z \setminus S_{0}\, 
\end{equation}
is an asymptotically cylindrical Calabi-Yau threefold if there is no monodromy from a small loop around $p_0$. The auxiliary threefold $Z$ is called a building block.

For a pair of acyl Calabi-Yau threefolds $X_\pm$ with cylinder regions $S_{0,\pm} \times S^1_{b,\pm} \times I$, a topological manifold $M$ is formed by gluing $X_\pm \times S^1_{e, \pm}$ along the cylindrical regions of $X_\pm$ by identifying
\begin{equation}
S^1_{b,\pm} = S^1_{e,\mp}\, ,
\end{equation}
as well as the interval direction and the K3 surfaces $S_{0,\pm}$. The isometry $\varphi$ between the K3 surfaces $S_{0,\pm}$ must be such that it identifies
\begin{equation}
\begin{aligned}\label{eq:gluingconditions}
J(S_{0,\pm}) & = \Re\left(\Omega^{2,0}(S_{0,\mp})\right) \\
\Im\left(\Omega^{2,0}(S_{0,+})\right) & = -\Im\left(\Omega^{2,0}(S_{0,-})\right)
\end{aligned} 
\end{equation} 
where $J(S_{0,\pm})$ is the K\"ahler forms and $\Omega^{2,0}(S_{0,\pm})$ the holomorphic two-form on $S_{0,\pm}$ in the complex structure inherited from $Z_\pm$.

The resulting topological manifold $M$ admits a Ricci-flat metric with holonomy group $G_2$ which asymptotes to the Ricci-flat Calabi-Yau metrics on $X_\pm$ in the limit in which the interval along which the gluing takes place becomes very long (the `Kovalev limit'). We will use the notation
\begin{equation}
 M(Z_-,Z_+) := \left(Z_- \times S^1_{e-} \right) \#_\varphi \left(Z_+ \times S^1_{e+} \right) \, .
\end{equation}
as a short-hand for this construction. 

Let $\rho_\pm: H^{1,1}(Z_\pm) \rightarrow  H^{1,1}(S_{0 \pm})$ be the natural restriction maps. Then 
\begin{equation}\label{eq:NK_buildingblocks}
\begin{aligned}
N_\pm & := \mbox{im}(\rho_\pm) \\
T_\pm & := N_\pm^\perp \in H^2(S_{0\pm},\mathbb{Z})\\
K(Z_\pm) & := \ker(\rho_\pm)/[S_0]_{PD}
\end{aligned}\, .
\end{equation}

The isometry $\varphi$ induces an isomorphism $\varphi: H^2(S_{0+},\mathbb{Z}) \cong H^2(S_{0-},\mathbb{Z})$, which in turn allows us to think of 
$N_\pm$ and $T_\pm$ as both sitting in $\Gamma^{3,19} = U^{\oplus 3} \oplus (-E_8)^{\oplus 2}$. The integral cohomology groups of $M$ can then be computed as 
\begin{equation}\label{eq:bettitcsz}
\begin{aligned}
H^1(M,\mathbb{Z}) & =   0 \\
H^2(M,\mathbb{Z}) & =  N_+ \cap N_- \oplus K(Z_+) \oplus K(Z_-) \\
H^3(M,\mathbb{Z}) & = \mathbb{Z}[S] \oplus \Gamma^{3,19} /(N_+ + N_-) \oplus (N_- \cap T_+) \oplus (N_+ \cap T_-)\\
& \hspace{1cm} \oplus H^3(Z_+)\oplus H^3(Z_-) \oplus K(Z_+) \oplus K(Z_-) \\
H^4(M,\mathbb{Z}) & = H^4(S) \oplus (T_+ \cap T_-) \oplus \ \Gamma^{3,19} /(N_- + T_+) \oplus \Gamma^{3,19} /(N_+ + T_-) \\
& \hspace{1cm} \oplus  H^3(Z_+)\oplus H^3(Z_-) \oplus K(Z_+)^* \oplus K(Z_-)^* \\
H^5(M,\mathbb{Z}) & = \Gamma^{3,19} /(T_+ + T_-) \oplus K(Z_+) \oplus K(Z_-) \,.
\end{aligned}
\end{equation}

In this paper we will only consider isometries $\varphi$ such that 
\begin{equation}
N_\pm \otimes \mathbb{R} = \left(N_\pm\otimes \mathbb{R}  \cap N_\mp\otimes \mathbb{R}  \right) \oplus \left(N_\pm\otimes \mathbb{R}  \cap T_\mp\otimes \mathbb{R} \right)\, ,
\end{equation}
which is called orthogonal gluing. In this case, the simplified relation
\begin{equation}\label{eq:bettitcs}
b^2 +b^3 = 23 + 2 \left[ |K(Z_+)| + |K(Z_-)| + h^{2,1}(Z_+) + h^{2,1}(Z_-) \right]
\end{equation}
for the sum of the Betti numbers holds. 

In the light of the Shatashvili-Vafa relation \eqref{eq:G2mirrorcondition} this formula is very suggestive: swapping one (or both) of the building blocks $Z_\pm$
for another building block $Z_\pm^\vee$ such that 
\begin{equation}\label{eq:mirror_building_block_property}
\begin{aligned}
  h^{2,1}(Z_\pm^\vee) & = |K(Z_\pm)|\\
  h^{2,1}(Z_\pm) & = |K(Z_\pm^\vee)|
\end{aligned}
\end{equation}
preserves $b^2+b^3$ for the resulting TCS $G_2$ manifolds. 

A construction of building blocks in analogy to Batyrev's construction of Calabi-Yau hypersurfaces in toric varieties was given in \cite{Braun:2016igl}, and to be reasonably self-contained we have reviewed this construction in \Cref{app:proj_tops}. The key combinatorial objects are a pair of projecting tops $(\Diamond,\Diamond^\circ)$, and each such pair gives rise to a family of building blocks $Z_{\Diamond,\Diamond^\circ}$. Crucially, exchanging the role played by $\Diamond$ and $\Diamond^\circ$ gives us a pair of building blocks such that \eqref{eq:mirror_building_block_property} is satisfied \cite{Braun:2017ryx}. A consequence of this 
construction is that the K3 fibres of $Z = Z_{\Diamond,\Diamond^\circ}$ are from the (algebraic) mirror family of the K3 fibres of $Z^\vee =Z_{\Diamond^\circ,\Diamond}$.

To define a $G_2$ mirror we not only need to construct appropriate mirror building blocks, but furthermore need to find an isometry $\varphi^\vee$ to glue the asymptotically cylindrical Calabi-Yau threefolds $X_{\pm}^\vee$ to a TCS $G_2$ manifold. That such a `mirror gluing' always exists was shown in \cite{Braun:2017ryx,Braun:2017csz} by employing the following arguments. For type II strings on a $G_2$ variety $M$ we not only need to specify the geometry of the target,
but furthermore the $B$-field. If $M$ is TCS, the $B$-field in general restricts non-trivially to $X_{\pm}$ and the asymptotic K3 fibres $S_{0 \pm}$. Consistency of the gluing then implies that
\begin{equation} \label{eq:Bfieldgluing}
B|_{S_{0-}}=  B|_{S_{0+}} \, .
\end{equation}
In the asymptotically cylindrical regions of $X_{\pm}$, mirror symmetry acting on $X_{\pm}$ implies that the $K3$ fibres $S_{0\pm}$ are mapped to 
their mirrors. Mirror symmetry for a $K3$ surface $S$ can be understood as a linear map acting on $J(S),\Re \Omega^{2,0}(S), \Im \Omega^{2,0}(S), B_S$ that is specified by a choice of special Lagrangian fibration of $S$, and results in a mere reinterpretation of the same point in the CFT moduli space \cite{Aspinwall:1994rg,gross1999special}. 

Replacing both $Z_{\pm}$ by $Z_{\pm}^\vee$ then replaces $S_{0 \pm}$ by  $S_{0\pm}^\vee$, which in turn implies that $J(S_{0\pm}^\vee), \Omega^{2,0}(S_{0\pm}^\vee),B_{S_{0\pm}^\vee}$
satisfy the relations \eqref{eq:gluingconditions} and \eqref{eq:Bfieldgluing}, so that the mirror symmetry canonically identifies a mirror gluing 
$\varphi^\vee$ that can be used to construct 
\begin{equation}
\left[ M(Z_-,S_{0-},Z_+,S_{0+},\varphi)\right]^\vee :=  M(Z_-^\vee,S_{0-}^\vee,Z_+^\vee,S_{0+}^\vee,\varphi^\vee) \, .
\end{equation}
By using a similar logic as in the original SYZ argument, this mirror map is associated with performing $4$ T-dualities along a coassociative $T^4$ 
fibration of $M$. Here, both $S_{e \pm}$ are contained in the coassociative $T^4$. 

Using a similar analysis one can show that there are gluings $\varphi^{\wedge \pm}$ which allow to construct
\begin{equation}
\begin{aligned}
\left[ M(Z_-,S_{0-},Z_+,S_{0+},\varphi)\right]^{\wedge - } :=  M(Z_-^\vee,S_{0-}^\vee,Z_+,S_{0+},\varphi^{\wedge -}) \\
\left[ M(Z_-,S_{0-},Z_+,S_{0+},\varphi)\right]^{\wedge + } :=  M(Z_-,S_{0-},Z_+^\vee,S_{0+}^\vee,\varphi^{\wedge +})  
\end{aligned}
\end{equation}
and that these mirror maps are associated with associative $T^3$ fibrations. For ${}^{\wedge \pm}$, the SYZ picture 
implies that $S^1_{e \mp}$ are contained in the associative $T^3$ fibre, but $S^1_{e \pm}$ are not.

Besides sharing $b^2(M) +b^3(M)$, the total integral cohomology
\begin{equation}
 H^\bullet(M)= \bigoplus_{k} H^k(M,\Z)
\end{equation}
satisfies the stronger condition
\begin{equation}
 H^\bullet(M) =  H^\bullet(M^\vee) = H^\bullet(M^{\wedge \pm}) 
\end{equation}
for any type of gluing, not just orthogonal gluing. Note that this implies that $b^2+b^3+b^4+b^5 = 2(b^2+b^3)$ is the same for all of these geometries. 

Another interesting aspect of these mirror maps is that smooth $G_2$ manifolds can potentially have (geometrically) singular mirrors \cite{Braun:2017csz}. A TCS $G_2$ variety necessarily contains ADE singularities if there is a non-trivial ADE root lattice contained in $N_+ \cap N_-$, which is a possible realization of non-Higgsable clusters \cite{Morrison:2012np} in M-Theory \cite{Braun:2017uku}. Whereas a TCS $G_2$ manifold $M$ might 
be such that $N_+ \cap N_-$ contains no roots, this does not necessarily hold for all of its mirrors. However, the presence of such singularities does not imply a non-abelian gauge group as there is necessarily a non-trivial $B$-field along the corresponding $\mathbb{P}^1$s in such cases. 

\subsection{Calabi-Yau Quotients as Twisted Connected Sums}\label{sect:TCSasquotient}

The sets of $G_2$ manifolds constructed as (resolutions of) quotients of Calabi-Yau threefolds and TCS $G_2$ manifolds are not disjoint, which already follows from the fact that a number of Joyce orbifolds can be cast into both descriptions. The relationship between the two constructions was explored more generally in \cite{Braun:2019wnj} by studying the M-Theory lift of IIA orientifolds. The geometric aspects of this work are applicable more generally to $G_2$ geometries, and the upshot is that whenever a $G_2$ variety is realized as (a resolution of) $\left(X \times S^1\right)/(\sigma, -)$ for an anti-holomorphic involution $\sigma$, the resulting $G_2$ variety can also be described as a TCS if $X$ carries a K3 fibration with base
$\P^1_b$ which is respected by $\sigma$. This can be expressed as 
\begin{equation}
 \sigma|_{\P^1_b}: [z_0:z_1] \mapsto [\bar{z}_0,\bar{z}_1] \hspace{1cm} \sigma|_{S_0} =  \sigma_{S_0}
\end{equation}
where $[z_0:z_1]$ are homogeneous coordinates on $\P^1_b$ and $\sigma_{S_0}$ is an anti-holomorphic involution acting on $S_0$: 
\begin{equation}
 \sigma_{S_0}: J(S_0) \mapsto - J(S_0) \hspace{1cm} \sigma_{S_0}: \Omega^{2,0}(S_0) \mapsto \overline{\Omega^{2,0}(S_0)}\, .
\end{equation}
Let us further assume that the Calabi-Yau threefold $X$ is realized as a hypersurface in a toric variety corresponding to a reflexive pair of polytopes 
$X = X_{\Delta,\Delta^\circ}$, and that the K3 fibration on $X_{\Delta,\Delta^\circ}$ is made manifest by $\Delta^\circ$ containing a reflexive three-dimensional subpolytope $\Delta_F^\circ$. We have collected some technical details of this construction in
\Cref{sec:BatyrevMirrorSym}. In such a situation, anti-holomorphic involutions of the type we are interested in are guaranteed to exist if 
$\Delta^\circ$ is cut into two isomorphic copies $\Diamond^\circ_{1},\Diamond^\circ_{2}$ of a projecting top $\Diamond^\circ$, and the anti-holomorphic involution is inherited from an automorphism of $\Delta^\circ$ that exchanges $\Diamond^\circ_{1} \leftrightarrow \Diamond^\circ_{2}$. This becomes the vanilla anti-holomorphic involution given above upon redefining the coordinates on the base.

To better understand possible $\sigma_{S_0}$, we can perform a hyper-K\"ahler rotation $\psi$ to a new complex structure on $S_0$ such that 
\begin{equation}
J(S_0^\psi) =  \Re \Omega^{2,0}(S_0)  \hspace{1cm} \Omega^{2,0}(S_0^\psi) = J(S_0) - i\Im \Omega^{2,0}(S_0)   \, ,
\end{equation}
where we denote the K3 surface in the hyper-K\"ahler rotated complex structure by $S_0^\psi$.  
On $S_0^\psi$,  $\sigma_{S_0}$ acts as 
\begin{equation}
 \sigma_{S_0}: J(S_0^\psi) \mapsto J(S_0^\psi) \hspace{1cm} \sigma_{S_0}: \Omega^{2,0}(S_0^\psi) \mapsto -\Omega^{2,0}(S_0^\psi)\, .
\end{equation}
and such involutions have been classified \cite{nikulin1976finite,0025-5726-14-1-A06,2004math......6536A} in terms of three 
invariants $(r,a,\delta)$. The fixed point locus of $\sigma_{S_0}$ on $S_0^\psi$ is given as the union of a Riemann surface of genus $g$ together with 
$f-1$ $\P^1$s, where $f$ and $g$ are related to $(r,a,\delta)$ by 
\begin{equation}
\begin{aligned}
f &= (r-a)/2 + 1 \\
g &= (20-r-a)/2 +1 
\end{aligned}
\end{equation}
if $(r,a,\delta) \neq (10,10,0)$ or $(10,8,0)$.  
 
As detailed in \cite{Braun:2019wnj}, one may then construct $(X_{\Delta,\Delta^\circ} \times S^1)/(\sigma,-)$ as a TCS and we have
\begin{equation}
(X_{\Delta,\Delta^\circ} \times S^1)/(\sigma, -) = M(Z_{\Diamond,\Diamond^\circ},S_0,\Upsilon_{r,a,\delta},S_0^\psi,\psi)  \, .
\end{equation}
Here, $Z_{\Diamond,\Diamond^\circ}$ is the building block constructed from $(\Diamond, \Diamond^{\circ})$, $\psi$ is the hyper-K\"{a}hler rotation considered above and 
\begin{equation}
    \Upsilon_{r,a,\delta} = \left(S_0^\psi \times \P^1\right)/\left(\sigma_{S_0} ,-\right) 
\end{equation}
with $(-)$ acting on the homogeneous coordinates of $\P^1$ as $(-): [\zeta_1,\zeta_2] \rightarrow  [-\zeta_1,\zeta_2]$. 

Using this construction, $(X_{\Delta,\Delta^\circ} \times S^1)/(\sigma, -)$ can then be smoothed by resolving or deforming the building block $\Upsilon_{r,a,\delta}$ while keeping its asymptotic region and $Z_{\Diamond,\Diamond^\circ}$ fixed. The orbifold $\Upsilon_{r,a,\delta}$ always has a crepant resolution $\tilde{\Upsilon}_{r,a,\delta}\rightarrow \Upsilon_{r,a,\delta}$ with
\begin{equation}
\begin{aligned}
 |K(\tilde{\Upsilon}_{r,a,\delta})| = 2f \\
 h^{2,1}(\tilde{\Upsilon}_{r,a,\delta}) = 2g  
\end{aligned} 
\end{equation}
which results in a $G_2$ manifold $\tilde{M}$ with Betti numbers
\begin{equation}
\begin{aligned}
 b^2(\tilde{M}) &= h^{1,1}_+(X_{\Delta,\Delta^\circ}) + 2f \\ 
 b^2(\tilde{M}) + b^3(\tilde{M}) & = h^{1,1}(X_{\Delta,\Delta^\circ})+h^{2,1}(X_{\Delta,\Delta^\circ}) + 4(f+g) +1
\end{aligned}
\end{equation}
which agrees with \eqref{eq:joyceKara_betti} for trivial $\mathcal{Z}$.

\section{SCFTs with Calabi-Yau Target}
\label{sect:CY_mirrors}

    In this section we will review the relevant material on Calabi-Yau manifolds, their superconformal field theory (SCFT) realisations, and Calabi-Yau mirror symmetry. Reviews containing a guide to the extensive literature on this topic are \cite{cox1999mirror,hori2003mirror,Aspinwall:2009isa}.
     
    The worldsheet SCFT of type II strings on Calabi-Yau manifolds is a $\cN=(2,2)$ field theory in $(1+1)$-dimensions. We therefore start with a review of some of the basic concepts and set notations of $\cN=(2,2)$ SCFTs. More detailed reviews can be found in \cite{hori2003mirror,blumenhagen2009introduction}.

    \subsection{$\cN=2$ SCFTs}

    Although we have $\cN=2$ SUSY for both the left and right moving sectors, we will just focus on one side here, and assume the other is understood implicitly. We will return to how the left and right quantum numbers are related later.

    The $\cN=2$ Virasoro algebra contains four generators: the energy-momentum tensor, two supersymmetry currents and a $U(1)$ current $(T,G^0, G^3,J)$, respectively. It is common to work with $G^{\pm} = \frac{1}{\sqrt{2}}(G^0\pm i G^3)$ instead of $G^0$ and $G^3$, and we do so here. It can be shown \cite{Howe:1991ic} that the $U(1)$ current and one of the supersymmetry currents stem from the K\"{a}hler form in the geometry.\footnote{The general statement is that the existence of a covariantly constant $p$-form on the target manifold gives rise to conformal dimension $\frac{p}{2}$ and $\frac{p+1}{2}$ currents in the algebra, the latter being the superpartner of the former. Here the K\"{a}hler form is a $2$-form and so gives rise to a dimension $1$ and dimension $\frac{3}{2}$ current, which are the $J$ and $G^3$, respectively.}  The algebra is defined via the mode relations 
    \be 
        \begin{split}
            [L_m,L_n] & = (m-n)L_{m+n} + \frac{c}{12}(m^3-m)\del_{m+n,0} \\
            [L_m,J_n] & = -n J_{m+n} \\
            [J_m,J_n] & = \frac{c}{3} m \del_{m+n,0} \\
            [L_m, G^{\pm}_r] & = \bigg(\frac{m}{2}-r\bigg) G^{\pm}_{m+r} \\
            [J_m,G^{\pm}_r] & = \pm G^{\pm}_{m+r} \\
            \{ G^+_r, G^-_s\} & = 2L_{r+s} + (r-s)J_{r+s} + \frac{c}{3}\bigg(r^2 - \frac{1}{4}\bigg)\del_{r+s,0} \\
            \{G^{\pm}_r,G^{\pm}_s\} & = 0.
        \end{split}
    \ee 
    The indices on the supersymmetry operators dictate whether we are in the Neveu-Schwarz (NS) or Ramond (R) sector, with $r,s\in \Z$ being R and $r,s \in \Z + \frac{1}{2}$ being NS.
    
    An important property of the $\cN=2$ algebra is the existence of spectral flow. Spectral flow provides a map between the R and NS sectors, and so it provides a way to map bosons and fermions into each other: it encapsulates the spacetime SUSY. Spectral flow acts by mapping the conformal weight and $U(1)$ charge via 
    \be 
    \label{eqn:SpectralFlowhq}
        (h,q) \mapsto \bigg(h - \eta q + \frac{\eta^2}{6}c, q - \frac{\eta}{3}c \bigg),
    \ee 
    where $c$ is the central charge and $\eta \in \R$. One flows from NS to R (or vice versa) with $\eta = \pm 1/2$.

    States in the NS Hilbert space with the property $(h,q) = (h,\pm 2h)$ are called chiral and anti-chiral states. We have a notion of chiral and anti-chiral for both the left and right $\cN=2$ algebras. We therefore end up with four types of fields $(c,c)$, $(a,a)$, $(c,a)$ and $(a,c)$, where $c$ and $a$ stand for chiral and anti-chiral, respectively. We map from $c$ to $a$ (and vice versa) simply by charge conjugation, and so only two of these are independent. Therefore, we really only need to consider, say, $(c,c)$ and $(a,c)$ fields. These sets of fields actually obey a ring structure \cite{lerche1989chiral} and so we have two rings, which we denote $\fR_{(c,c)}$ and $\fR_{(a,c)}$. We shall denote the union of these rings as $\fR = \fR_{(c,c)} \cup \fR_{(a,c)}$.

    \subsubsection{Minimal Models}

    The simplest class of $\cN=2$ SCFTs are the $\cN=2$ minimal models. These are rational conformal field theories and so contain a finite number of primary fields. They are uniquely defined by their central charge
    \be 
        c = \frac{3k}{k+2},
    \ee    
    where $k$ is an integer known as the level and  $0 < c < 3$. We will denote the level $k$ minimal model as $MM_k$. The superconformal primaries in $MM_k$ are given by a triple $(l,m,s)$ where 
    \be 
        0 \leq l \leq k, \qquad s \sim s + 4 \qand m \sim m + 2(k+2). 
    \ee 
    We identify $s=0,2$ as the NS states while $s=\pm 1$ are the R states. The conformal weights and $U(1)$ charges are given by 
    \be 
    \label{eqn:MMhqValues}
        h^l_{m,s} = \frac{l(l+2)-m^2}{4(k+2)} + \frac{s^2}{8} \qand q^l_{m,s} = -\frac{m}{k+2} + \frac{s}{2}.
    \ee 
    Note that the chiral and anti-chiral conditions are given by $(l,m,s) = (l,\mp l, 0)$, respectively. Spectral flow maps $(l,m,s) \mapsto (l,m-1,s-1)$, and so our R ground states are given by $(l,m,s) = (l,\mp l-1,-1)$.

    \subsection{N=2 SCFTs with Calabi-Yau target}

    So far we have discussed $\cN=2$ SCFTs in general. We now want to specialise to the worldsheet theories of Type II strings compactified on a Calabi-Yau manifold, which is described by a nonlinear sigma model (NLSM) in $2D$. The bosonic fields in this SCFT are the coordinates on the Calabi-Yau target space.

    \subsubsection{Nonlinear Sigma Model}

    The NLSM on a Calabi-Yau manifold is, in particular, a NLSM on a K\"{a}hler manifold. We have already seen that K\"{a}hler implies $\cN=(2,2)$ on the worldsheet, and so this NLSM is a $\cN=(2,2)$ field theory in $(1+1)$-dimensions. Just as with $\cN=1$ in $(3+1)$-dimensions, there is a notion of chiral and anti-chiral superfields, i.e. fields which obey, respectively,
    \be 
        \bar{D}_{\pm} \Phi = 0 \qand D_{\pm}\bar{\Phi} = 0,
    \ee 
    for supercovariant derivatives $D_\pm$ and $\bar{D}_\pm$. 
    
    The Lagrangian (density) for the NLSM on a K\"{a}hler manifold is given by the $D$-term
    \be 
        \cL_{\text{kin}} = \int d^4\theta K\big(\Phi_i,\bar{\Phi}_{\bar{i}}\big)
    \ee 
    with $K(\Phi_i,\bar{\Phi}_{\bar{i}})$ a real function of chiral and anti-chiral superfields. It defines the K\"{a}hler metric 
    \be 
        g_{i\bar{j}} := \p_i\p_{\bar{j}} K\big(\Phi_i,\bar{\Phi}_{\bar{i}}\big).
    \ee 

    $2D$ field theories with $\cN=(2,2)$ contain two $U(1)$ R-symmetries: $U(1)_V$ and $U(1)_A$, where $V$ stands for vector and $A$ for axial.\footnote{It turns out that a $\cN=(2,2)$ theory in $(1+1)$ dimensions can be obtained by the dimensional reduction of $\cN=1$ SUSY in $(3+1)$-dimensions. The $U(1)_V$ is the R-symmetry of the $4D$ theory and $U(1)_A$ corresponds to rotations in the compactified directions.} In what follows we shall often work in a different charge basis, defined by
    \be
        U(1)_L = \frac{U(1)_V+U(1)_A}{2} \qand  U(1)_R = \frac{U(1)_V-U(1)_A}{2},
    \ee 
    where $L/R$ stands for left/right, respectively.
    
    For these to be symmetries of our theory, we need to show that the action is invariant under their action. It is easy to show that the kinetic term is invariant provided $K(\Phi_i,\bar{\Phi}_{\bar{i}})$ has $(q_V,q_A) = (0,0)$. We note that if $K(\Phi_i,\bar{\Phi}_{\bar{i}}) = K(|\Phi_i|^2)$, then we can assign \textit{any} charges to the individual chiral superfields (the anti-chiral superfields then have opposite charge). 
    
    However this only guarantees classical invariance and we need to check for the existence of anomalies. It can be shown \cite{hori2003mirror} that $U(1)_V$ is not anomalous but that $U(1)_A$ can be, depending on the value of the first Chern class of the target manifold $\cM$. In particular $U(1)_A$ is also anomaly free if, and only if, $c_1(\cM)=0$. This is the condition that makes the target spacetime a Calabi-Yau manifold.

    \subsubsection{Ramond Ground States and Chiral Rings}
    \label{sec:StatesAndForms}

    The Witten index was evaluated for an $\cN=1$ supersymmetric NLSM in \cite{witten1982constraints}, where the famous result that only Ramond ground states contribute was obtained. Identifying the fermionic creation and annihilation operators with adding and removing differential forms on the target manifold, it was also shown that there is a one-to-one correspondence between Ramond ground states and harmonic forms on the target manifold. 
    
    \label{sec:ChiralRings}

    It is important to note that, at this level, we cannot identify individual Betti/Hodge numbers of our supersymmetric sigma model: all we can say is that there is the same number of Ramond ground states as there are harmonic forms. If we are to introduce more structure to our target space, then it is possible to obtain further information and potentially get further topological relationships.

    The case with $\cN=2$, with the constraint that $q_L-q_R\in \Z$, was considered in \cite{lerche1989chiral}, where it was shown that the elements of the chiral ring $\fR$ are related to the Hodge numbers of the target manifold. In particular, for $\cN=(2,2)$ theories there is the same number of states of charge $(p,q)$ in our ring $\fR$ as there are $(\dim \cM-p,q)$-forms on $\cM$.\footnote{We have changed convention compared to \cite{lerche1989chiral}, which identifies $(p,q)$ charge with $(p,q)$-forms. We pick this convention for later convenience.} 
    
    While this is clearly related to the previous result, it strengthens it in that we can now look at individual Hodge numbers. This is related directly to the fact that our algebra has two $U(1)$ charges (i.e. two generators $J_0$ and $\bar{J}_0$). Indeed it can be shown for a theory with only $U(1)_V$ (i.e. a K\"{a}hler target) that we would only be able to compute the Hodge numbers $h^{p,q}$ up to a set value of $p-q$. The existence of the $U(1)_A$ symmetry allows us to compute Hodge numbers up to set value of $p+q-\dim\cM$, and so it is the combination of these two that allowed us to get the above result. This observation leads to the following important result: we can only determine the Hodge numbers of the target space up to the ambiguity
    \be 
    \label{eqn:HodgeMirrorMap}
        h^{p,q} \leftrightarrow h^{\dim \cM - p,q}.
    \ee 
    which is Calabi-Yau mirror symmetry seen as an ambiguity of associating a geometry with a given SCFT.
    

    \subsubsection{Odake Algebra}
    
    The above discussion holds for any $\cN=(2,2)$ NLSM. Here we want to specialise to the case where our target manifold is a Calabi-Yau. Firstly, we note that, by central charge arguments, the CFT for our Calabi-Yau must have $c=9$: i.e. we have a total of $c=15$ but the $4$-dimensional spacetime takes up $c_{ST}=6$ of these. 

    As a Calabi-Yau is, in particular, a K\"{a}hler manifold, the $\cN=(2,2)$ SCFT is a good starting point: the $U(1)$ current gives us the  K\"{a}hler form. However, it is not sufficient: we still need the holomorphic $(3,0)$-form, which we denote $\Omega$. We account for $\Omega$ in the CFT by extending the $\cN=2$ Virasoro by a field with quantum numbers $(h,q)_{NS} = (3/2,3)$, where the subscript indicates that the field lives in the NS sector. The resulting Odake algebra with this central charge was first written down in \cite{odake1989extension}\footnote{A generic Odake algebra corresponds to an extension of the $\cN=2$ Virasoro algebra by a $(n/2,n)$ field, and the central charge is $c=3n$. Here we are just considering $n=3$ as this is the relevant value for $3$-folds.} The associated field $\Omega$ is decomposed as 
    \be 
    \label{eqn:OmegaDecomp}
        \Omega = A + iB
    \ee 
    The complex conjugate of this field (corresponding to the $(0,3)$-form) is $\Omega^* = A -iB$, and it has $(h,q)_{NS} = (3/2,-3)$. The superpartner of this field is $\Upsilon = \frac{1}{\sqrt{2}}(C + iD)$, such that $(A,C)$ and $(B,D)$ are pairs of superpartners. So, in total, the generators of our Odake algebra are $(T,G^0,J,G^3,A,B,C,D)$. The OPEs of these generators can be found in \cite{fiset2018superconformal,figueroa1997extended}.
    
    As detailed in the original paper, these theories only admit a finite number of irreducible, unitary highest weight representations. The key thing for us will be the allowed massless representations, of which there are three for NS and three for R. As mentioned before, these representations are linked by spectral flow so that we only need to consider one set. The allowed values in the NS sector are\footnote{The $(3/2,\pm3)$ states are actually related to a single state, $(0,0)$, by spectral flow with $\eta = \pm 1$. However, for future simplicity we treat them as their own fields here.}
    \be 
        (h,q)_{NS} = (3/2,-3), \quad  (1/2,1), \quad (1/2,-1) \qand (3/2,3)
    \ee 
    which have corresponding R values
    \be 
        (h,q)_R = (3/8,3/2), \quad (3/8,-1/2), \quad (3/8, 1/2) \qand (3/8,-3/2),
    \ee 
    respectively. We note at this point that every R ground state has $h=3/8$, and so the R ground states are specified simply by their $q$ values. 

    Here we have only written down the quantum numbers for one side (say the left side) of our SCFT. The discussion is completely identical for the right hand side, and a general state is given by a product of two of the above states.
    
    As we will see, all the models we consider actually have $q_L = \pm q_R$. We now see that this is important: the $(c,c)$ ring corresponds to $3$-forms, i.e. if $q_L=q_R =q \in \{0,1,2,3\}$\footnote{The $0,2$ cases are obtained by spectral flow of the $-3,-1$ cases, respectively.} then $(q,q) \cong \a \in h^{3 -q,q}$; the $(c,a)$ then give us our diagonal forms $(q,-q) \overset{\eta = -1}{\longrightarrow} (q,3-q) \cong \beta \in h^{3-q,3-q}$, where we have made use of spectral flow in order to ensure the degree of our form is positive. It is therefore important that \textit{both} $\fR_{(c,c)}$ and $\fR_{(a,c)}$ are non-trivial in our theory to reflect the whole cohomology of a Calabi-Yau threefold. 
 
    We can also relate the Hodge numbers to the charges in the R sector by spectral flow. We simply use \eqref{eqn:SpectralFlowhq} with $\eta=\pm1/2$ and $c=9$ so that $q_{NS} \mapsto q_R = q_{NS} \mp \frac{3}{2}$. From here we can say that to every state with charges $(q_L,q_R)_R$ there is a $(m,n)$-form with
    \be
    \label{eqn:qlqrRmn}
        (q_L,q_R)_R = \bigg(\frac{3}{2}-m,n-\frac{3}{2}\bigg).
    \ee 

    We emphasise here that we can only equate the \textit{number} of these things. That is, if $V_{q_L,q_R}$ denotes the vector space of states with charges $(q_L,q_R)_R$, then 
    \be 
        \dim \big(V_{\frac{3}{2}-m,n-\frac{3}{2}}\big) = h^{m,n}.
    \ee 
    It is generally true that a differential form can be represented by some state in the CFT, however we are not guaranteed that such a state will have definite charge.

    \subsubsection{Mirror Symmetry}

    An $\cN=2$ sigma-model with Calabi-Yau target has the following automorphism
    \be 
    \label{eqn:MirrorCYGenerators}
        \cM_{\text{CY}} : (T,G^0,J,G^3,A,B,C,D) \mapsto (T,G^0,-J,-G^3,A,-B,C,-D).
    \ee 
    Note, in particular, that it flips the sign of any state, $q \mapsto -q$. We are dealing with two copies of the algebra, and we have seen that the charges of the states are related to the degrees of the forms on the target manifold. In this context, mirror symmetry is understood as applying \eqref{eqn:MirrorCYGenerators} to one side, say the right side: $(q_L,q_R) \mapsto (q_L,-q_R)$. Note that this maps an element in $(c,c)$ to an element of $(c,a)$, and vice versa. From our relation to differential forms, this recovers the well known replacement $h^{p,q} \mapsto h^{p,3-q}$.

    \subsection{Gauged Linear Sigma Models}

    NLSMs have the key disadvantage of being confined to working in coordinate patches of the target geometry. In this section we briefly review gauged linear sigma models (GLSM), which can be seen as the worldsheet analogue of the elegant geometric construction of Calabi-Yau manifolds as hypersurfaces in simpler spaces with a global description in terms of homogeneous coordinates. 

    A GLSM is a $\cN=(2,2)$ field theory, written in superspace, with a collection of $n$ chiral superfields $\{\Phi_i\}$ along with a $U(1)$ gauge group.\footnote{The generalisation to $U(1)^s$ is straight forward.} The gauge field associated to the $U(1)$ is $V$, and it enjoys the gauge symmetry $V \mapsto V + i(\Lambda-\bar{\Lambda})$, where $\Lambda$ is a chiral superfield that labels the $U(1)$ action. 
    
    The Lagrangian of the GLSM contains four pieces: 
    \be \label{eq:GLSM_action}
        \cL = \cL_{\text{kin}} + \cL_W + \cL_{\text{gauge}} + \cL_{FI,\theta}.
    \ee 
    which are given by
    \be 
        \begin{split}
            \cL_{\text{kin}} & = \int d^4\theta \sum_i \bar{\Phi}_i e^{2Q_iV} \Phi_i \\
            \cL_W & = \int d^2\theta W(\Phi_i) + c.c. \\
            \cL_{\text{gauge}} & = -\frac{1}{2e^2} \int d^4 \theta \bar{\Sigma}\Sigma  \\
            \cL_{FI,\theta} & = \frac{1}{2}\bigg( - \int d\bar{\theta}^- d\theta^+ t \Sigma  + c.c.\bigg).
        \end{split}
    \ee 
    where $W(\Phi_i)$ is the superpotential, $e$ is the gauge coupling constant and $t = r-i\theta$. Here $r$ is the FI parameter and $\theta$ the theta angle. The field $\Sigma = \bar{D}_+ D_- V$ is the field strength of $V$ which is a twisted chiral superfield which obeys $\bar{D}_+\Sigma = D_-\Sigma = 0$. Viewing $\cL_{\text{gauge}}$ as the twisted equivalent of $\cL_{\text{kin}}$, we can then use $\cL_{FI,\theta}$ to define a linear twisted superpotential, $\widetilde{W}(\Sigma) = -t\Sigma$. Explicit expressions for the component expansions of these Lagrangians can be found in \cite{Witten:1993yc}. 
    
    We again need to ask about the $U(1)_V\times U(1)_A$ symmetries and anomaly conditions. The invariance of $\cL_{\text{kin}}$ is of course the same as the NLSM discussion and gives the same result. The F-term, $\cL_W$, tells us that the superpotential is required to have $(q_V,q_A) = (2,0)$. This constrains the form it can take, namely we require it to be quasi-homogeneous:
    \be 
    \label{eqn:QuasihomogeneousW}
        W\big(\l^{q_V^i} \Phi_i\big) = \l^2 W\big(\Phi_i\big).
    \ee 
    The twisted F-term, $\cL_{FI,\theta}$, tells us that $\Sigma$ must have $(q_V,q_A) = (0,2)$. The anomaly conditions again carry over, along with the requirement that the $U(1)$ gauge group charges cancel \cite{hori2000mirror,hori2003mirror}, i.e. 
    \be 
    \label{eqn:AnomalyChargesVanish}
        \sum_i Q_i = 0.
    \ee 
    This is required to ensure $U(1)_A$ is non-anomalous. In particular if $\sum_i Q_i = p$ then $U(1)_A$ is broken to $\Z_{2p}$.

    \subsubsection*{Connection to NLSMs and The Landau-Ginzburg/Calabi-Yau Correspondence}

    In order to see the connection between GLSMs and NLSMs, we solve the equations of motion for the auxiliary fields:
    \be 
        \begin{split}
            D & = -e^2 \bigg(\sum_i Q_i |\phi_i|^2 - r\bigg) \qand F_i = \frac{\p W}{\p\phi_i},
        \end{split}
    \ee 
    where the lower case indicates the lowest component of the superfield.
    Doing this leaves us with a dynamical theory for the fields $(\phi_i,\sig)$, which has potential energy
    \be 
        U(\phi_i,\sig) = \frac{1}{2e^2} D^2 + \sum_i |F_i|^2 + 2|\sig|^2 \sum_i Q_i^2|\phi_i|^2
    \ee  

    Let $\cM_{\text{Vac}}$ denote the vacuum manifold of the GLSM.  That is, in the GLSM we identify the chiral superfields $\{\Phi_1,...,\Phi_n\}$ with coordinates on $\C^n$ and then consider the surface defined via minimising the potential energy. It can in fact be shown that the IR limit of a GLSM is the NLSM on $\cM_{\text{Vac}}$.
    
    For example, consider a theory of $n$ chiral superfields all with charge $Q_i = 1$\footnote{Note that this doesn't obey \eqref{eqn:AnomalyChargesVanish}, and so the NLSM is anomalous and therefore cannot correspond to  a Calabi-Yau.} and vanishing superpotential, $W=0$. Then we have 
    \be 
        U(\phi_i,\sig) = \sum_i |\sig|^2|\phi_i|^2 + \frac{e^2}{2}\bigg(\sum_i |\phi_i|^2 - r\bigg)^2.
    \ee 
    If $r >0$ then $U=0$ is given by $\sig=0$ and 
    \be 
        \sum_i |\phi_i|^2 = r.
    \ee 
    This defines a sphere $S^{n-1}$. However we now need to account for the $U(1)$ action, so that in total the vacuum manifold is 
    \be 
        \C\bP^{n-1} = \frac{\big\{ (\phi_1,...,\phi_n) \big| \sum_i |\phi_i|^2 = r \big\}}{U(1)}.
    \ee   
    By an identical calculation, assigning different charges to the fields will produce a weighted projective space. 

    We now need to account for $F$-terms, i.e. non-vanishing superpotential. We will now show that appropriately chosen superpotentials lead to hypersurfaces in the ambient toric manifolds. For the sake of simplicity, we focus on the simplest case of a hypersurface in $\bP^{n-1}$, but more general results can be found in \cite{Witten:1993yc}. 
    
    Consider a GLSM with $n+1$ chiral superfields $\{P,\Phi_1,...,\Phi_n\}$ with gauge group charges $q_i = 1$ and $q_P = -n$, and superpotential 
    \be 
        W = P \cdot G\big(\Phi_1,...,\Phi_n\big)
    \ee 
    where $G(\Phi)$ is a homogeneous polynomial of degree $n$. We assume $G(\Phi)$ is generic, in the sense that
    \be
    \label{eqn:GenericG}
        G = \frac{\p G}{\p\Phi_1} = \frac{\p G}{\p\Phi_2} = ... = \frac{\p G}{\p\Phi_n} = 0 \qquad \implies \qquad \Phi_1 = \Phi_2 = ... = \Phi_n = 0.
    \ee 
    The potential energy for this system is given by
    \be 
        U = \big|G(\phi_i)\big|^2 + |p|^2 \sum_i \bigg|\frac{\p G}{\p \phi_i}\bigg|^2 + \frac{1}{2e^2}D^2 + 2|\sig|^2 \bigg(\sum_i |\phi_i|^2 + n^2|p|^2\bigg)
    \ee 
    where 
    \be 
        D = -e^2 \bigg(\sum_i |\phi_i|^2 -n|p|^2 - r\bigg).
    \ee 
    The vacuum manifold of this theory is defined by $U=0$ and is $r$ dependent. The case $r >> 0$ requires at least one of the $\phi_i$ to be non-zero. From here, the $|\sig|^2 \sum |\phi_i|^2$ term gives $\sig=0$, while the $|p|^2 \sum |\p_i G|^2$ term (along with \eqref{eqn:GenericG}) tells us that $p=0$. Finally we require $G=0$. We are thus left in exactly the case as before, but now with the constraint $G=0$. In other words, the GLSM flows to the NLSM on $X \subset \C\bP^{n-1}$, defined by degree $n$ homogenous polynomial. These are precisely the conditions for a Calabi-Yau manifold.
    
    The case $r << 0$ can similarly be shown to require $p\neq 0$ and so the field $P$ picks up a vev, and breaks the $U(1)$ gauge group to a $\Z_n$ subgroup: $\phi_i \mapsto e^{\frac{2\pi i}{n}}\phi_i$. This leads to a theory with superpotential $W^{\prime} = \sqrt{-r} \cdot G(\phi_i)$ subject to this $\Z_n$ action. This defines a Landau-Ginzburg (LG) orbifold. This recovers the well known Calabi-Yau/Landau-Ginzburg correspondence: we can view them as two different phases of the same GLSM.
    
     It is important that we have a LG \textit{orbifold}, as it is known that a LG theory only has non-trivial $\fR_{(c,c)}$, while $\fR_{(a,c)}$ contains just the identity, but we need both to be non-trivial for strings on Calabi-Yaus. However, as demonstrated in \cite{vafa1989string}, the twisted states in the orbifold theory give rise to elements in $\fR_{(a,c)}$.

    We can alter the action of the gauge group on this system to account for hypersurfaces in weighted projective spaces. We can pick $Q_P = -H$ and $Q_i = w_i$, where
    \be
    \label{eqn:wiH}
        w_i = \frac{H}{k_i+2} \qand H = \text{lcm}(k_i+2),
    \ee 
    and then the anomaly condition enforces 
    \be 
    \label{eqn:LGAnomalyCondition}
        \sum_i\frac{1}{k_i+2} = 1.
    \ee 
    Then we set 
    \be 
        G(\Phi_i) = \Phi_1^{k_1+2} + ... + \Phi_n^{k_n+2}
    \ee 
    in the superpotential. Our Calabi-Yau is then defined by this degree $H$ Fermat hypersurface in $\bP^{n-1}_{w_1,...,w_n}$. The LG orbifold is then given by $W^{\prime} = \Phi_1^{k_1+2} + ... + \Phi_n^{k_n+2}$ with $\Z_H$ quotient 
    \be 
        \Phi_i \mapsto e^{\frac{2\pi i \g }{k_i+2}}\Phi_i.
    \ee

    \subsubsection{Gepner Models}

    It is known \cite{lerche1989chiral} that the IR limit of a LG model with $W = \Phi^{k+2}$ is a $(2,2)$ SCFT with central charge
    \be 
        c = \frac{3k}{k+2}.
    \ee 
    which is the level $k$ $\cN=2$ minimal model, $MM_k$. The idea is then that the worldsheet SCFT (i.e. the nonlinear sigma model) is isomorphic to the SCFT obtained by the IR limit of the LG orbifold. We can therefore use the minimal models to construct and study the worldsheet SCFT. 

    Gepner \cite{gepner1989space} proposed a method for constructing the CFT of a Calabi-Yau as a direct product of $\cN=2$ minimal models. At the GLSM level each term in the product corresponds to a different $\Phi_i^{k_i+2}$ in $G(\Phi_i)$.
    
    The key observation is that the central charge adds under products, and so we could form a $c=9$, $\cN=2$ theory out of a collection of minimal models. That is 
    \be 
        \big(\cN=2\big)_{c=9} = \bigotimes_{i=1}^r \big(\cN=2\big)^{MM}_{c_i} \qquad \text{with} \qquad \sum_{i=1}^r c_i = \sum_{i=1}^r \frac{3k_i}{k_i+2} = 9.
    \ee 
    The remaining part of our CFT corresponds to the $4D$ spacetime. Working in lightcone gauge, this is a CFT with central charge $c=3$ and consists of two bosons and their accompanying fermions. The fermions are described by an $\so(2)_1$ affine Lie algebra, which has four representations $(O_2)_{h=0,q=0}$, $(V_2)_{h=1/2,q=1}$, $(S_2)_{h=1/8,q=1/2}$ and $(C_2)_{h=1/8,q=-1/2}$. The NS sectors are $O_2$ and $V_2$ while $S_2$ and $C_2$ are the R sectors. As we are focusing on the Gepner model part here, we drop the fermions for now but shall return to them later.

    Recall that the superconformal primaries in $MM_k$ are defined by the triple $(l,m,s)$ where the conformal dimension and $U(1)$ charge is given by \eqref{eqn:MMhqValues}. The conformal weights and charges add under products of different $MM_k$. Therefore all we need to do is account for the orbifold action. As detailed in \cite{vafa1989string}, at the level of the CFT the orbifold acts as a projection on the charges via
    \be 
    \label{eqn:gAction}
        g = e^{2\pi i J_0}.
    \ee 
    We therefore require our states to have integer charge, and our Gepner model is defined via 
    \be 
    \label{eqn:Gepner}
        (Gep) = \big[MM_{k_1}, MM_{k_2}, ..., MM_{k_r}]\big|_{U(1)\text{-projection}},
    \ee 
    where the $U(1)$-projection enforces 
    \be 
    \label{eqn:GepnerLiCondition}
        \sum_{i=1}^r \bigg[\frac{l_i}{k_i+2}\bigg] = 0,1,2,3,
    \ee 
    The restriction on the right-hand side follows from \eqref{eqn:LGAnomalyCondition} along with $l_i \leq k_i$. The integrality of the charges is also required to ensure spacetime SUSY (see \cite{greene1990duality} and references therein). This result is actually not surprising: we have already seen that our Odake algebra limits the NS charges to be $q= \pm 3, \pm1$ which are equivalent, via spectral flow, to $q=0,1,2,3$. The above equation is nothing other than the NS charges of our states. 

    As we are considering an orbifold, we obtain both untwisted and twisted sectors. Let's start with the untwisted sector. Here we have $q_L=q_R$ and the charges of a state are simply given by the sum of the individual MM charges. In the R sector (which is where we will predominantly work), we therefore have the untwisted charges 
    \be 
        \sum_{i=1}^5 \bigg(\frac{l_i+1}{k_i+2} - \frac{1}{2}\bigg) = \sum_{i=1}^5 \bigg(\frac{l_i}{k_i+2}\bigg) - \frac{3}{2},
    \ee 
    where we made use of our anomaly condition, \eqref{eqn:LGAnomalyCondition}. If we then impose the Gepner condition,  \eqref{eqn:GepnerLiCondition}, we see that the R charges are restricted to $q = \pm \frac{3}{2}, \pm \frac{1}{2}$. This same result is, of course, obtained by applying spectral flow to the allowed NS charges.

    A chiral field $\Phi$ is identified with $(l,m,s) = (1,-1,0)$, and so the $l_i$ value determines the power of $\Phi_i$. Therefore, states with 
    \be 
        \sum_{i=1}^5 \frac{l_i}{k_i+2} = 0,1,2,3
    \ee 
    correspond to degree $0,H,2H$ and $3H$ polynomials, where $H=\text{lcm}(k_i+2)$. In particular the state $\ket{l_i}$ is identified geometrically with $\Phi_i^{l_i}$. Note that $l_i \leq k_i$ and so we must set $\Phi_i^{k_i+1} = 0$.\footnote{This is related to the fact that the $(c,c)$ ring of a Landau Ginzburg orbifold is obtained by a Jacobian ring (see, e.g. \cite{greene1997string}): the quotient in the Jacobian sets these monomials to zero.} This result is directly related to the Griffiths residue giving us the primitive cohomology of the Calabi-Yau manifold \cite{GriffithsI}. 

    Let's now discuss the twisted sectors of our Gepner model. As detailed in \cite{vafa1989string} these states have $q_L=-q_R$, and the charge of a state depends on which twisted sector we are in:
    \be 
        q^{\nu}_L = \sum_{i | \nu \notin (k_i+2)\Z} \bigg( \frac{\nu}{k_i+2} - \bigg[\frac{\nu}{k_i+2}\bigg] - \frac{1}{2}\bigg),
    \ee 
    where $[...]$ stands for the integer part of the argument, and $\nu = 1,...,H-1$ labels the twisted sector.

    States in the twisted sector can become untwisted when $\nu \in (k_i+2)\Z$, in which case their charge is computed simply using 
    \be 
        q_i = \frac{l_i+1}{k_i+2} - \frac{1}{2}.
    \ee   
    and $(q_i)_L = (q_i)_R$ for these factors. 

    So, in total, a charge of a generic state is given by 
    \be 
        \begin{split}
            q^{\nu}_L = \sum_{i | \nu \in (k_i+2)\Z} \bigg( \frac{l_i+1}{k_i+2} - \frac{1}{2}\bigg) + \sum_{i | \nu \notin (k_i+2)\Z} \bigg( \frac{\nu}{k_i+2} - \bigg[\frac{\nu}{k_i+2}\bigg] - \frac{1}{2}\bigg) \\
            q^{\nu}_R = \sum_{i | \nu \in (k_i+2)\Z} \bigg( \frac{l_i+1}{k_i+2} - \frac{1}{2}\bigg) - \sum_{i | \nu \notin (k_i+2)\Z} \bigg( \frac{\nu}{k_i+2} - \bigg[\frac{\nu}{k_i+2}\bigg] - \frac{1}{2}\bigg)
        \end{split}
    \ee 
    where the fully untwisted sector is identified with $\nu=0$. We can write this in a more symmetric manner by defining $l_i^{(\nu)} + 1 := \nu \mod(k_i+2)$, then the two sums above take the same form. An overall state is considered untwisted if $q_L=q_R$ and twisted if $q_L = - q_R$, despite what the individual $(q_i)_L$ and $(q_i)_R$ obey.

    Before moving on, we note an interesting point. Suppose that $H$ is even, then we can set $\nu = H/2 \mod(k_i+2)$. We now claim that if $w_i = \frac{H}{k_i+2}$ is even, then the twist is trivial, i.e. $\nu = 0$. Let's see this: for some $n\in \Z$, we can write $\nu = H/2$ as
    \be 
        \frac{H}{2} + n(k_i+2) = \bigg( \frac{H}{2(k_i+2)} + n \bigg) (k_i+2) = \bigg(\frac{w_i}{2} + n\bigg) (k_i+2),
    \ee 
    but if $w_i$ is even, then we can always pick $n = - \frac{w_i}{2}$, and so $\nu = 0$. When $w_i$ is odd, the above calculation shows us that 
    \be 
        l^{(\frac{H}{2})}_i = \frac{k_i}{2},
    \ee 
    in which case the $U(1)$ charges vanish, $q_L=q_R = 0$. We are then left with the untwisted states $\ket{l_i}_R$, which have $(q_L)_i=(q_R)_i = \frac{l_i +1}{k_i+2} -\frac{1}{2}$, and correspond to $(2,1)$ or $(1,2)$ forms. 

    \subsubsection{Mirror Symmetry for Gepner Models}

    In \cite{greene1990duality}, Greene and Plesser took the observation made in  \cite{gepner1987modular}, that the quotient of a Gepner model by its full symmetry group yields an isomorphic theory, and extended it to more general quotients. In particular they looked at the geometrical phase of such a duality. 
    
    Consider the Gepner model obtained by the minimal model product $(k_1,...,k_r)$. Let $d$ denote the order of $g = e^{2\pi i J_0}$. Then, this model has discrete symmetry group 
    \be 
        G = \bigg(\prod_{i=1}^r \Z_{k_i+2}\bigg)\bigg{/} \Z_n,
    \ee 
    where $n=d$ for $d$ odd and $n=d/2$ for $d$ even. The mirror model is given by quotienting by $H \subset G$ defined such that 
    \be 
    \label{eqn:GPgammaCondition}
        \sum_{i=1}^r \frac{\g_i}{k_i+2} \in \Z,
    \ee 
    where $\g_i \in \Z_{k_i+2}$ represents an element of $G$. The mirror theory is isomorphic to the original Gepner model with one of the $U(1)$ charges reversed.

    In terms of the corresponding LG orbifold, the statement is that the LG orbifold with superpotential 
    \be 
        W(\Phi_i) = \sum_{i=1}^r \Phi_i^{k_i+2} 
    \ee 
    with orbifold action $\Z_H$ has a mirror LG orbifold with the same form of the superpotential\footnote{The subscript $F$ is to indicate ``Fermat type".}
    \be 
        \widetilde{W}_F(\Phi_i^{\vee}) = \sum_{i=1}^r (\Phi_i^{\vee})^{k_i+2},
    \ee 
    but now the quotient is by $\Gamma^{\vee} \subset \prod_{i=1}^r \Z_{k_i+2}$ acting on the fields as
    \be 
        \Phi_i^{\vee} \mapsto e^{\frac{2\pi i \g_i}{k_i+2}}\Phi_i^{\vee}
    \ee 
    subject to \eqref{eqn:GPgammaCondition}. The $(2,1)$-forms of this dual theory are then related to the deformations of this equation, where in particular the product $\Phi_1^{\vee}\Phi_2^{\vee}...\Phi_r^{\vee}$ is always present. In fact, the mirror theory for a non-zero value of the $(FI,\theta)$ parameter $t$
    is the LG orbifold with superpotential
        \be 
            \widetilde{W}(\Phi_i^{\vee}) = \sum_{i=1}^r (\Phi_i^{\vee})^{k_i+2} + e^{t/H} \prod_{i=1}^r \Phi_i^{\vee}\, .
        \ee 

   The geometric phase is given by the Calabi-Yau defined by a hypersurface in a toric manifold, with defining polynomial given by $W(z_i) = 0$. For example, for $(k_i+2) = 5$ for all $i$ (and $r=5$), we recover the quintic and mirror quintic Calabi-Yaus.

    We can actually see the generation of this dual superpotential by looking at the states in the Gepner model. We go to the case of interest, namely $r=5$. States with $q_L=q_R$ (i.e. elements of $\fR_{(c,c)}$) are the untwisted states, while states with $q_L=-q_R$ (elements of $\fR_{(a,c)}$) are the twisted states. Therefore, mirror symmetry acts on the Gepner model by mapping the twisted and untwisted states to untwisted and twisted states, respectively. The original twisted states should now be interpreted as the untwisted states in the mirror model and so, as per the previous discussion, should be interpreted as monomials of degrees $0,H, 2H$ and $3H$. We can indeed see that this is the case as follows: we are now essentially mapping $\nu \mapsto -\nu$. This follows from the fact that the twisted states come from quotienting by $g =e^{2\pi i \nu J_0}$, but if we send $J_0 \mapsto -J_0$ this is the same as sending $\nu\mapsto-\nu$ in $g$. From here we simply interpret the $l_i^{(-\nu)}$ as the powers of the corresponding mirror homogeneous coordinates, i.e.  
    \be 
        \ket{l_i^{(-\nu)}} \cong (\Phi_i^{\vee})^{l_i^{-\nu}}.
    \ee 
    
    Indeed this ties in nicely with the mirror description in terms of LG models. Let's look at the allowed deformations of $W^{\vee}_F$. The monomial $\Phi_1^{\vee} ... \Phi_5^{\vee}$, which is always present (by construction), would correspond to a state with $l^{(-\nu)}_i = 1$ for all $i$, and it is indeed true that this state always appears. This is seen simply from 
    \be 
        l_i^{(-\nu)} +1 = -\nu \mod(k_i+2) \qquad \implies \qquad l_i^{(-H+2)} = 1 \qquad \forall i.
    \ee 
    Also note that $\nu=H-2$ always gives a $(1,1)$-form in the original theory. This follows simply from 
    \be 
        l^{(H-2)}_i +1 = H-2 \mod(k_i+2) \qquad \implies \qquad l^{(H-2)}_i = k_i-1,
    \ee 
    which together with 
    \be 
        \sum_{i=1}^5 \frac{k_i}{k_i+2} = 3 \qand \sum_{i=1}^5 \frac{1}{k_i+2} = 1
    \ee
    gives $\sum_i\frac{l^{(H-2)}_i}{k_i+2} = 2$, which is $q_L = -q_R = \frac{1}{2}$ and is the criteria for a $(1,1)$-form.

    For the original untwisted states, we simply take the $l_i$ values and plug them into $l_i+1 = - \nu \mod(k_i+2)$, and use this. For example, $l_i=0$ for all $i$ is the unique state that always gives the $(3,0)$-form, which should be mirrored to the $(0,0)$-form. Under this mirror map, this would give $\nu = H-1$ for all $i$, but we know that this is the unique twist that gives the $(0,0)$-form, as required.

    We defined a twisted contribution to a state as one in which $\nu \notin (k_i+2)\Z$, but that simply mapping $\nu \mapsto -\nu$ won't change this condition. However, the mirror of a twisted state is meant to be untwisted. The key thing is that it is untwisted w.r.t. the mirror Gepner model, i.e. we have a $\nu^{\vee}$ and a twisted contribution to a state in the mirror Gepner model obeys $\nu^{\vee} \notin (k_i+2)\Z$. This $\nu^{\vee}$ must account for the orbifold group of the mirror Gepner model, and so it is not easy to write down a general relationship between $\nu$ and $\nu^{\vee}$. However, it is in principal not too difficult to obtain the relationship for specific cases.

    \subsubsection{Mirror Symmetry for Gauged Linear Sigma Models}

    As we have seen, Gepner models are a particular phase of the more general theory of GLSMs. The concept of mirror symmetry can be elevated to the general construction following \cite{hori2000mirror}. 

    Given a GLSM with a set of chiral superfields $\{\Phi_i\}$ with charges $Q_i$ and vanishing superpotential, the mirror theory is given by the LG theory with twisted superpotential 
    \be 
        \widetilde{W} = \bigg(\sum_i Q_i Y_i - t\bigg) \Sigma + \sum_i e^{-Y_i}.
    \ee 
    The fields $Y_i$ are anti-chiral superfields and their imaginary parts, $\vartheta_i := \frac{1}{2}(Y_i-\bar{Y}_i)$, are periodic in $2\pi$. These fields are dual to the chiral superfields $\Phi_i$, and the duality relation is given by 
    \be 
    \label{eqn:RealPartYPhi}   
        Y_i + \bar{Y}_i = 2 \bar{\Phi}_i e^{2Q_iV}\Phi_i.
    \ee 
    We can also relate the imaginary part of $Y_i$ to the phase of $\Phi_i$. This is not easily done in terms of the superfields, but can be seen if we consider a component expansion of the fields. If the lowest component of $\Phi_i$ is $\phi_i = \rho_i e^{i\psi_i}$, then the result is
    \be
    \label{eqn:ImPartYPhi}
        d\vartheta_i = \star d \psi_i.
    \ee
    Solving the equations of motion for the dynamical field $\Sigma$ results in the $D$-term constraint 
    \be 
    \label{eqn:DTermConstraint}
        \p_{\Sigma} \widetilde{W} = 0 \qquad \implies \qquad \sum_iQ_iY_i = t.
    \ee 
    Finally, defining 
    \be 
        X_i = e^{-Y_i},
    \ee 
    we see that the twisted superpotential takes the form 
    \be 
        \widetilde{W}(X_i) = \sum_i X_i \qquad \text{subject to} \qquad \prod X_i^{Q_i} = e^{-t}.
    \ee    
    A theory of a superfield with (twisted) superpotential is a LG theory. So here we have a LG theory for the twisted chiral superfields $Y_i$ (expressed in terms of $X_i$) with twisted superpotential $\widetilde{W}(X_i)$ as above.

    Let's now modify this construction slightly by introducing another chiral superfield $P$ to our set $\{\Phi_1,...,\Phi_n\}$, and we set the charge of $P$ to be negative, $Q_P = -H$. Let $\widetilde{P} = e^{-Y_P}$ be the dual field to $P$, then it follows from the constraint above that 
    \be 
        \widetilde{P}^{-H}X_1^{Q_1} ... X_n^{Q_n} = e^{-t}.
    \ee    
    Defining
    \be 
        \Phi_i^{\vee} = X_i^{1/H},
    \ee 
    then results in the condition 
    \be 
        \widetilde{P} = e^{t/H} (\Phi_1^{\vee})^{Q_1}... (\Phi_n^{\vee})^{Q_n}.
    \ee 
    The twisted superpotential then takes the form 
    \be
    \label{eqn:TwistedSuperpotential}
        \widetilde{W}(\Phi_i^{\vee}) = (\Phi_1^{\vee})^{Q_1} + ... + (\Phi_n^{\vee})^{Q_n} + e^{t/H}(\Phi_1^{\vee})^{Q_1}... (\Phi_n^{\vee})^{Q_n}.
    \ee
    Note that this superpotential is invariant under the action $\Phi_i^{\vee} \mapsto \exp(2\pi i \g_i/Q_i) \Phi_i^{\vee}$. We still need to account for the periodicity $Y_i \sim Y_i + 2\pi i$. We have 
    \be 
        \Phi_i^{\vee} = X_i^{1/H} = e^{-Y_i/H},
    \ee 
    and so $\Phi_i^{\vee} \sim e^{2\pi i/H}\Phi_i^{\vee}$. Our twisted superpotential is then subject to an orbifold action $\Gamma^{\vee} \subset \prod_i \Z_{Q_i/H}$. Specifically it acts on the fields as 
    \be 
    \label{eqn:GammaPhiVeeAction}
        \Phi_i^{\vee} \mapsto \exp(\frac{2\pi i \g_i H}{Q_i}) \Phi_i^{\vee}, \qquad \text{subject to} \qquad \sum_i \frac{\g_i H}{Q_i} \in \Z.
    \ee 
    The constraint condition comes from the fact that $e^{t/H}\prod_i (\Phi_i^{\vee})^{Q_i} \in \widetilde{W}(\Phi_i^{\vee})$.

    We therefore arrive at the result that the mirror of the NLSM on a weighted projective space is a LG model with the above superpotential. If we allow for negative charges then we obtain a LG orbifold. This is not quite what we want yet: we want to show that the mirror of a Calabi-Yau is again Calabi-Yau. 
    
    On the LG orbifold side, we can simply apply the Calabi-Yau/Landau-Ginzburg correspondence to find another Calabi-Yau. On the NLSM side, we are so far only working on a weighted projective space and we need a superpotential to restrict the theory to a NLSM on a Calabi-Yau hypersurface. Happily, it turns out that the resulting mirror LG orbifold is unchanged by introducing such a superpotential. The difference between the two cases is actually encapsulated in what are considered to be the fundamental fields on the mirror side: for the case with a superpotential the fundamental fields are the $X_i$ while in the absence of the superpotential the fundamental fields are the $Y_i$. However we have a very simple relation between the two, and the LG orbifold remains unchanged.

    We therefore arrive at mirror symmetry as a map between two Calabi-Yaus. Note that if we pick the charges as $Q_i = w_i = H/(k_i+2)$ and $Q_P = -H$, then we arrive at the result of Greene and Plesser. Namely, the mirror of a LG orbifold with Fermat type superpotential is again a (deformation of) a LG orbifold with the same Fermat type superpotential but now with a different quotient group. Indeed \eqref{eqn:GammaPhiVeeAction} becomes exactly the result of \cite{greene1990duality}.

    \subsubsection{Mirror Symmetry for Toric Hypersurfaces}
    \label{sec:CYMirrorGeneralisation}

    The above geometrical mirror map can be generalised to hypersurfaces in a generic toric variety, akin to the construction of Batyrev \cite{Batyrev94dualpolyhedra}, details of which can be found in \Cref{sec:BatyrevMirrorSym}.

  The starting point is a GLSM with $(h+1)$ chiral superfields $(\Phi_1,...\Phi_h,P)$ and gauge group $U(1)^k$. For each $U(1)$ we have a field strength $\Sigma_a$ and associated FI parameter $t_a$. Let $Q_{i,a}$ denote the charge of $\Phi_i$ under the $a^{\text{th}}$ $U(1)$ factor. Set 
    \be
        d_a := \sum_{i=1}^h Q_{i,a},
    \ee
    and define $t_i$ via\footnote{The $t_i$ are defined up to redefinition of the $Q_{i,a}$.}
    \be
    \label{eqn:tati}
        t_a = \sum_{i=1}^h Q_{i,a} t_i.
    \ee   
    In order to have a Calabi-Yau we must obey the anomaly condition, \eqref{eqn:AnomalyChargesVanish}. This implies that the charges of $P$ under the $a^{\text{th}}$ $U(1)$ is $-d_a$.

    We can again dualise this theory in order to obtain a theory with $(h+1)$ twisted chiral superfields $(Y_1,...,Y_h,Y_P)$ and then define 
    \be 
        \widetilde{P} := e^{-Y_P} \qand X_i := e^{-Y_i}.
    \ee 
    The $D$-term constraint, \eqref{eqn:DTermConstraint}, gives $k$ relations:
    \be 
        \sum_{i=1}^h Q_{i,a} Y_i -d_aY_P = t_a.
    \ee 
    In terms of the new variables this is 
    \be 
    \label{eqn:XiTidlePta}
        \bigg(\prod_{i=1}^h X^{Q_{i,a}}\bigg) \widetilde{P}^{-d_a} = e^{-t_a}
    \ee 
    It the follows from \cite{hori2000mirror}, that in our case the twisted superpotential is actually empty and the defining hypersurface of the mirror Calabi-Yau is given by the above constraint along with 
    \be
    \label{eqn:SumXiTildeP}
        \sum_{i=1}^h X_i + \widetilde{P} = 0.
    \ee 

    To write the superpotential above in terms of chiral fields of the dual theory we now proceed as follows. The fan of the toric variety underlying the GLSM has $h$ ray generators $n_i$ sitting in the N lattice which obey the $k$ relations 
    \be
    \label{eqn:nuiQia}
        \sum_{i=1}^h n_i Q_{i,a} = 0\, .
    \ee 
    Let the superpotential take the form $W(\Phi_i,P) = P \cdot G(\Phi_i)$, where $G(\Phi_i)$ is a homogeneous polynomial of degrees $\{d_1,...,d_k\}$ with respect to the $U(1)^k$. Next, let $M$ denote the dual lattice to $N$, and define $m_\ell$ such that we can write
    \be
        G(\Phi_i) = \sum_{\ell=1}^{h^{\vee}} \prod_{i=1}^h \Phi_i^{\langle m_{\ell},n_i\rangle + 1}\, ,
    \ee 
    where necessarily $\langle m_{\ell},n_i\rangle \geq -1$ for all $\ell$ and $i$. 
    
    The Calabi-Yau/Landau-Ginzburg story carries over and again we get two phases of the GLSM: $p = 0$ gives the nonlinear sigma model on a Calabi-Yau defined by the vacuum manifold; $p\neq 0$ gives a LG orbifold with superpotential $W= G(\Phi_i)$.

    We can now solve \eqref{eqn:XiTidlePta} by introducing $\{\Phi_1^{\vee},...,\Phi_{h^{\vee}}^{\vee}\}$, and defining
    \be
    \label{eqn:XiPhiVeeGeneral}
        \widetilde{P} = \prod_{\ell=1}^{h^{\vee}} \Phi_{\ell}^{\vee} \qand X_i  = e^{-t_i} \prod_{\ell=1}^{h^{\vee}} \big(\Phi_{\ell}^{\vee}\big)^{\langle m_{\ell},n_i\rangle + 1}\, .
    \ee 
    Plugging this into \eqref{eqn:SumXiTildeP} then gives the hypersurface equation 
    \be 
    \label{eqn:DualHypersurfaceEqn}
        \sum_{i=1}^h   e^{-t_i} \prod_{\ell=1}^{h^{\vee}} \big(\Phi_{\ell}^{\vee}\big)^{\langle m_{\ell}, n_i\rangle +1} + \prod_{\ell=1}^{h^{\vee}} \Phi_{\ell}^{\vee}= 0\, .
    \ee 
    This is the family of Calabi-Yau hypersurfaces identified by Batryrev's construction.

\section{$G_2$ Mirror Symmetry From the SCFT Perspective}\label{sect:G2_mirror_ws}

    We now move on to the main part of this paper: we want to demonstrate $G_2$ mirror symmetry at the CFT level by showing how the involution carries over to the mirror side. The logic is straight forward: consider the theory of a product of a Calabi-Yau and circle. Acting on this theory with mirror symmetry generates an isomorphic theory. If we can define an involution in the original theory, there must exist an isomorphic involution on the mirror. The crucial question is if we can find this involution and if it acts in a way that can be understood geometrically. 

    \subsection{$G_2$ Algebra}

    We can construct the superconformal algebra of a $G_2$ manifold in a similar manner to that of the Calabi-Yau. We start with the $\cN=1$ super Virasoro generators $(T,G)$ and then, again by the results of \cite{Howe:1991ic}, add in additional generators corresponding to the $3$-form and $4$-form of the $G_2$. The $3$-form gives rise to a pair of fields $(\Phi,K)$ with conformal weights $(\frac{3}{2},2)$ which are related by $G$. The $4$-form gives a further pair of fields $(X,M)$ which have conformal weights $(2,\frac{5}{2})$. The OPEs between these fields can be found in \cite{Shatashvili:1994zw,figueroa1997extended}.\footnote{Note the typo in \cite{Shatashvili:1994zw} in the $K(z)M(w)$ OPE, as pointed out in \cite{figueroa1997extended}.}
    
    It is generally true that a Ramond ground state in a superconformal field theory has  conformal weight $h = \frac{d}{16}$, where $d$ is the number of Ramond fermions, i.e. the dimension of the target space in NLSMs. For a $G_2$-manifold we therefore require $h=7/16$. The reason this is important to us is that we have seen that a Ramond ground state in the Odake algebra has $h = \frac{3}{8}$. Putting this together with the fact that the R ground state of the SCFT of a circle (a boson-fermion pair) has $h=\frac{1}{16}$, provides support that we can form the $G_2$ algebra from the product of an  Odake algebra and the boson-fermion pair, reflecting the geometrical construction.

    Indeed \cite{figueroa1997extended} explicitly demonstrates how one combines the two algebras to obtain the $G_2$ algebra. In terms of the Odake generators $(T_{\text{CY}}, G^0, J, G^3, A,B,C,D)$ and the boson-fermion generators $(j,\psi)$, we obtain the $G_2$ generators as
    \be 
    \label{eqn:G2GeneratorsCYS1}
        \begin{split}
            T & = T_{\text{CY}} + T_{S^1} \\
            G & = G^0 + G_{S^1} \\
            \Phi & = A + (J\psi) \\
            X & = (B\psi) + \frac{1}{2}(JJ) - \frac{1}{2}(\p\psi\psi) \\
            K & = C + (Jj) + (G^3\psi) \\
            M & = (D\psi) - (Bj) + (j\p \psi) + (JG^3) - \frac{1}{2}\p G,
        \end{split}
    \ee 
    where $(...)$ stands for normal ordering and we defined
    \be 
        T_{S^1} = \frac{1}{2}(jj) + \frac{1}{2}(\p\psi\psi) \qand G_{S^1} = (j\psi).
    \ee 
    We are dealing with an $\cN=(1,1)$ algebra, and so we have two copies of this: the left and right copies. 

    \subsubsection*{Automorphisms}

   We now want to ask how a $G_2$ involution acts on the generators of the SCFT. Geometrically, an anti-holomorphic involution is defined via the action on the K\"{a}hler form, $J_K \mapsto -J_K$, and on the holomorphic $(3,0)$ form, $\Omega^{3,0} \mapsto \bar{\Omega}^{3,0}$. Recalling that the results of \cite{Howe:1991ic} tell us that the K\"{a}hler form gives rise to $(J,G^3)$ generators and that the imaginary part of $\Omega^{3,0}$ gives rise to $(B,D)$, we conclude that an anti-holomorphic involution inverts the sign of these four generators. We can make similar arguments for the $S^1$ factor, where we see that the signs of both generators $(j,\psi)$ are changed. So, in total, we see that a $G_2$ involution acts on the generators as
    \be 
    \label{eqn:ahiGenerators}
        (\sig,-) : (T_{\text{CY}}, G^0,J, G^3, A,B,C,D, j,\psi) \mapsto (T_{\text{CY}}, G^0, -J, -G^3, A,-B,C,-D, -j,-\psi)
    \ee    
    on both the left and right algebras simultaneously. This is clearly an automorphism of the $G_2$ algebra as the generators $(T,G,\Phi,X,K,M)$ are all invariant. Indeed this is one way to obtain the decomposition in \eqref{eqn:G2GeneratorsCYS1}: they generate the subalgebra of $\text{Od}^3\times S^1$ fixed by $\sig$.

    In the description at the level of the algebra we have given, the anti-holomorphic involution is not unique in its action on all states of the theory. This is not surprising as the same can be said geometrically, i.e. a given Calabi-Yau threefold can have many inequivalent anti-holomorphic involutions.  
   
    Besides the anti-holomorphic involution automorphism, the $G_2$ algebra contains three other interesting automorphisms $(\cM_{\text{CY}},T_{S^1},\cM_{G_2})$, who's actions on the generators are given in \Cref{tab:G2Mirrors}. Contrary to the automorphism associated with anti-holomorphic involutions, we want to let these automorphisms only act on \textit{one side} of the $\cN=(1,1)$ algebra, say the right side. 

    \begin{table}[h!]
        \begin{center}
        	\begin{tabular}{@{}  C{1cm} | C{1cm} | C{1cm} | C{1cm} | C{1cm} | C{1cm} | C{1cm} | C{1cm} | C{1cm} | C{1cm} | C{1cm} @{}}
                    \, & $T_{\text{CY}}$ & $G^0$ & $J$ & $G^3$ & $A$ & $B$ & $C$ & $D$ & $j$ & $\psi$  \\
                    \hline
                    $\cM_{\text{CY}}$ & $+$ & $+$ & $-$ & $-$ & $+$ & $-$ & $+$ & $-$ & $+$ & $+$  \\
                    $T_{S^1}$ & $+$ & $+$ & $+$ & $+$ & $+$ & $+$ & $+$ & $+$ & $-$ & $-$  \\
                    $\cM_{G_2}$ & $+$ & $+$ & $-$ & $-$ & $+$ & $-$ & $+$ & $-$ & $-$ & $-$  \\
        	\end{tabular}
        \end{center} 
        \caption{\label{tab:G2Mirrors} Three automorphisms of the $G_2$ algebra formed via the product of the Calabi-Yau and circle algebras. The action is written via its action on the generators, with $(T_{\text{CY}},G^0,J,G^3,A,B,C,D)$ corresponding to the Calabi-Yau and $(j,\psi)$ the circle.}
    \end{table}
    The first automorphism is nothing other than the Calabi-Yau mirror map (\eqref{eqn:MirrorCYGenerators}), and the second is simply $T$-duality on the boson-fermion pair. The final map is just the composition of these two\footnote{\label{footnote:PhaseAutomorphism} The mirror automorphism given in \cite{gaberdiel2004generalised} takes the form 
    \be 
        \cM_{GK} : (T,G,\Phi,X,K,M) \mapsto (T,G,-\Phi,X,-K,M).
    \ee 
    As pointed out in \cite{fiset2018superconformal}, this automorphism is related to $\cM_{\text{CY}}$ and $T_{S^1}$ via a phase rotation on the Calabi-Yau generators:
    \be 
        Ph^{\pi} : (T_{\text{CY}},G^0,J,G^3,A,B,C,D) \mapsto (T_{\text{CY}},G^0,J,G^3,-A,-B,-C,-D).
    \ee }. The latter acts on the $G_2$ generators trivially, i.e. 
    \be 
        \cM_{G_2} : (T,G,\Phi,X,K,M) \mapsto (T,G,\Phi,X,K,M),
    \ee 
    whereas the other two do not have easily defined action on these generators, e.g. 
    \be
        \cM_{\text{CY}} : \Phi = A + (J\psi) \mapsto \Phi^{\prime} = A - (J\psi).
    \ee 
    Nevertheless, it is straightforward to show that both $\cM_{\text{CY}}$ and $T_{S^1}$ are automorphisms of the algebra. This follows simply from the fact that the Calabi-Yau subalgebra does not speak to the boson-fermion subalgebra, i.e. the OPEs between $(T_{\text{CY}},G^0,J,G^3,A,B,C,D)$ and $(j,\psi)$ all vanish. Therefore if our map is an automorphism of the subalgebras, it must be an automorphism of the full algebra.

    \subsubsection*{Mirror Symmetry}

    We are now in a position to make the observation that all three mirror automorphisms commute with the anti-holomorphic involution automorphism. It follows from this that for every anti-holomorphic involution on the original theory, there is a corresponding action in the mirror theory which also acts on the algebra in the same way that an anti-holomorphic involution does. This statement has been made simply at the level of the generators of the algebra, and so we are blind to details such as which anti-holomorphic involution we are doing. In the following, we will give a more detailed description in terms of Gepner models and 
    GLSMs.

    \subsection{$G_2$ Gepner Models}
    \label{sec:G2GepnerModels}

    The construction of $G_2$ Gepner models have been studied in  \cite{blumenhagen2002superconformal,eguchi2002string,roiban2002rational}. 
    Under Gepner's construction the full SCFT (in light-cone gauge) is given by a Gepner model and an $\so(2)_1$ affine Lie algebra, which gives the two fermions in the non-compact directions. It can be shown using simple current arguments (see \cite{blumenhagen2009introduction} for a review), that the NS vs. R sectors of the two parts must agree, i.e. if we have a NS state in our Gepner model, we must take a NS state from our $\so(2)_1$ model. Similarly we can show that the overall $U(1)$ charge of a state must be an odd integer. 

    When adding the $\so(2)_1$ factor for a NS state in our Gepner model, we have two options: $O_2$ and $V_2$. These have $(h,q)_O = (0,0)$ and $(h,q)_V = (1/2,1)$. Now, we know that the NS states should have total charge being an odd integer, however we chose our spectral flow such that our Gepner models always had odd integer NS charge, and so we can only couple to the $O_2$ rep. This is all consistent: if we had taken a state of the form $q = q_{Gep} + q_{\mathfrak{so}(2)_1} = 2+1$, so that we were using the $V_2$ rep, we could use spectral flow to go to the state with $q_{Gep}=-1$. The spectral flow from NS to R in the $\mathfrak{so}(2)_1$ theory is given by either $C_2$ or $S_2$ (depending on which direction you flow). Either way, we are doing this same spectral flow twice (to go NS to R to NS) and so we are using $C_2^2 = S_2^2 = V_2$, which follows from the fusion rules of $\so(2)_1$. Putting this together with $V_2\times V_2 = O_2$, we see that our $V_2$ rep flows to an $O_2$ rep, as needed. All together, that is 
    \be 
        (q_{Gep}=2)\times V_2 \mapsto (q_{Gep}=-1)\times O_2.
    \ee 
    Therefore we can always represent a state in the NS sector as a state in the Gepner model with odd integer charge along with the $O_2$ rep.
    
    In order to construct our $G_2$ Gepner model, we need to split the $\so(2)_1$ factor into two copies of $\so(1)_1$. In other words, we want to treat the two fermions separately, as one will remain a flat direction, whereas the other will be compactified on $S^1$. There are three representations of $\so(1)_1$: $(O_1,V_1,S_1)$, which have conformal weights $(0,1/2,1/16)$, respectively. The R representation $S_1$ has $h=1/16$, which is \textit{exactly} the conformal weight required in order to take the R ground states of a Calabi-Yau CFT and produce R ground states of a $G_2$ CFT, i.e. $h_{G_2} = 7/16 = 3/8 + 1/16 = h_{Gep} + h_{S_1}$.

    One can form the four reps of $\so(2)_1$ out of the three reps of $\so(1)_1$ as follows:
    \be 
        \begin{split}
            O_2 & = O_1O_1 + V_1V_1 \\
            V_2 & = O_1V_1 + V_1O_1 \\
            S_2 & = S_1S_1 \\
            C_2 & = S_1S_1 \\
        \end{split}
    \ee 
    The at-face-value equality of $S_2$ and $C_2$ is dealt with via arguments related to fixed points of simple current orbits in the $\so(1)_1\times \so(1)_1$ theory (see, \cite{blumenhagen2002superconformal} for details). We can use this to write our generic NS $(Gep)\times \so(2)_1$ state in terms of $\so(1)_1$ reps, namely 
    \be 
        \bigg(h=\frac{|q|}{2}, q \in \{\pm 3, \pm 1\} \bigg) \otimes \big(O_1O_1 + V_1V_1\big),
    \ee 
    and similarly for the right states (i.e. tildes everywhere).
    
    \subsubsection*{Anti-holomorphic Involution}
    
    The $G_2$ involution maps the $\so(1)_1$ NS reps via $(O_1,V_1) \mapsto (O_1,-V_1)$.\footnote{It's action to $S_1$ is less easily written, but it acts via $S_2 \leftrightarrow C_2$.} As we have seen, it also maps states in our Gepner models by changing the sign of the $U(1)$ charges. In terms of the tuples $(l_i,m_i,s_i)$ of the minimal model factors, the involution acts as 
    \be
    \label{eqn:GepnerInvolutionlms}
        (l_i,m_i,s_i) \mapsto (l_i,-m_i,-s_i)
    \ee 
    on the states in the highest weight representation. This actually gives the vanilla involution (i.e. simply complex conjugation), but we can easily generalise this to involutions that swap homogeneous coordinates that have the same weight. At the Gepner level, this would be a map that swaps two minimal model factors that have the same levels. 

    Using the general change of sign argument, the states that exist in the Calabi-Yau Gepner model split into even and odd under the involution. Namely, working in a basis of states with definite charge, our states are paired in their charge conjugates. We form the even and odd combinations in these pairs: the even ones couple with  $O_1O_1$ and survive while the odd ones couple with $V_1V_1$ and survive. 
    
    Using the equivalence between the charges of the states in a Gepner model and the number of differential forms, along with identifying the presence of $V_1$ as wedging with $d\theta$ (the differential form on the $S^1$), the above reproduces the geometrical argument that the differential forms that survive give Betti numbers
    \be 
        b^0 = b^7 = 1 \qquad b^2 = h^{1,1}_+ \qand b^3 = h^{1,1}_-+h^{2,1}+1.
    \ee 
    These are only the Betti numbers corresponding to the untwisted states under the involution $\sigma$ as we have ignored the twisted sectors. 

    \subsubsection*{Mirror Map}

    Next we want to look at the action of the mirror map on this construction and demonstrate that it gives rise to a mirror anti-holomorphic involution. Given the above arguments, this is straightforward; the key thing is that both maps act as a reversal of charges and commute. If we denote the mirror minimal model tuples as 
    \be 
        (l_i,m_i,s_i) \mapsto (l_i^{\vee}, m_i^{\vee}, s_i^{\vee}),
    \ee 
    then the charges of the mirror states are given in terms of $(m_i^{\vee},s_i^{\vee})$. As the mirror involution acts as a change of sign, it must act as
    \be 
        (l_i^{\vee},m_i^{\vee},s_i^{\vee}) \mapsto (l_i^{\vee},-m_i^{\vee},-s_i^{\vee}),
    \ee 
    which is exactly equivalent to \eqref{eqn:GepnerInvolutionlms}. This tells us that the mirror involution has the same geometrical interpretation, namely it is an anti-holomorphic involution. 
    
   At this point one might be tempted to identify states of definite charge with differential forms of a specific Hodge type using \eqref{eqn:qlqrRmn}. Consider a state corresponding to a $(2,1)$-form in the original Gepner model: under the involution, this state is mapped to a state who's corresponding form is of Hodge type $(1,2)$. Acting with the mirror map on both of these states we find a $(1,1)$-form and a $(2,2)$-form. This now seems to imply that the involution $\sigma$ on the mirror side has to map a $(1,1)$-form to a $(2,2)$-form, which cannot be achieved by an anti-holomorphic involution. 
    However, eigenstates of the charge operators do not need to correspond to forms of fixed degree. This can be made very explicit in orbifold models and we have treated one example in detail in \Cref{sec:TorodialOrbifold}.

    As mentioned before, we expect our $G_2$ model to have three different mirrors. In the quotient construction, geometrically these three mirrors correspond to: (i) mirroring the Calabi-Yau but leaving the circle factor alone, (ii) leaving the Calabi-Yau alone and doing T-duality on the circle, and (iii) doing both Calabi-Yau mirror and T-duality on the circle. For our Gepner model here we have only obtained one mirror map, corresponding to case (ii). By studying the interplay of T-duality and the action of the involution on the $\so(1)_1$ factor, one should be able to obtain similar results for the other two mirror maps. We do not do this calculation here, but claim that this construction exists, and provide evidence of this below.

    \subsection{$G_2$ Sigma Models}

    Given a Calabi-Yau threefold $X$, a sigma model on the metric product $X \times S^1$ splits into the sum of the Calabi-Yau and the circle models. The latter is simply the theory of a boson fermion pair, and the former has been discussed in detail above. As $G_2$ manifolds are not K\"ahler this is a $(1,1)$ theory, however this is realized here as the tensor product of a $(2,2)$ theory (the Calabi-Yau part) and a $(1,1)$ theory (the circle). As before, we are interested in how $G_2$ involutions act on mirror pairs.

    \subsubsection*{Anti-holomorphic Involutions and GLSMs}

    The key observation which allows us to immediately write down anti-holomorphic involutions for GLSMs is that the chiral superfields $\Phi_i$ are identified with the homogeneous coordinates of the toric Calabi-Yau ambient space. Therefore the anti-holomorphic involution acts on these chiral superfields in the same way that it acts on the coordinates. As the anti-holomorphic involution furthermore needs to map $G_\pm \rightarrow G_\mp$ it follows that $\theta_\pm \rightarrow \theta_\mp$ which implies that also the twisted chiral superfields $\Sigma_a$ are sent to their complex conjugates. 
    The vanilla anti-holomorphic involution hence acts as
    \be 
        \sig_v :\hspace{1cm}
        \begin{aligned}
        \Phi_i \mapsto \bar{\Phi}_i\\
        \Sigma_a \mapsto \bar{\Sigma}_a
        \end{aligned}  
    \ee 
    We can now trace this through the dualisation procedure of \cite{hori2000mirror} to find the action of $\sig_v$ on the mirror. We find that 
    \begin{equation}
    \sig_v: \hspace{1cm}\begin{aligned}
     \Re(Y_i) &=& \bar{\Phi}_i e^{2Q_{i,a}V_a}\Phi_i &\rightarrow&  &\bar{\Phi}_i e^{2Q_iV}\Phi_i& & = & \Re(Y_i)\\
    \Im(Y_i) &=& \vartheta_i &\rightarrow &- &\vartheta_i&  & =& - \Im(Y_i)
    \end{aligned}
    \end{equation}
    by using \eqref{eqn:RealPartYPhi} and \eqref{eqn:ImPartYPhi}. 
    
    For the case of weighted projective spaces we can directly track this action through to an action on the fields $\Phi_i^{\vee}$: We have $\Phi_i^{\vee} = X_i^{1/H} = e^{-Y_i/H}$, and so 
    \be 
        \sig_v:\Phi_i^{\vee} \mapsto \overline{\Phi}_i^{\vee},
    \ee 
    which is simply the `vanilla' anti-holomorphic involution on the mirror side again. 
    
    For more general toric varieties we have that the fields in the dual theory are
    \be
        \widetilde{P} = \prod_{\ell=1}^{h^{\vee}} \Phi_{\ell}^{\vee} \qand X_i  = e^{-t_i} \prod_{\ell=1}^{h^{\vee}} \big(\Phi_{\ell}^{\vee}\big)^{\langle m_{\ell},n_i \rangle + 1}\, .
    \ee 
    where the $X_i$ are dual variables for the $\Phi_i$ and $\widetilde{P}$ is the dual variable of $P$. To trace the action of $\sigma_v$ through the duality, we can now simply think of the vanilla involution $\sigma_v$ as being defined on $X^\vee$ instead of $X$, where it acts as $\sigma^\vee_v:\Phi^\vee \rightarrow \bar{\Phi}^\vee$. This then implies immediately that 
    \be 
        \sigma_v:  \hspace{1cm}
        \begin{aligned}
        X_i & \mapsto \bar{X}_i \\ 
         \widetilde{P} &\mapsto \overline{\widetilde{P}}
        \end{aligned}
    \ee    
    and hence
    \be 
        \sigma_v:  \hspace{1cm}
        \begin{aligned}
        \Phi_i & \mapsto \bar{\Phi}_i \\ 
         P &\mapsto \overline{P}
        \end{aligned}
    \ee   

To have a symmetry of the GLSM, and not just the fields, we also need to make sure the anti-holomorphic involution is a symmetry of the action \eqref{eq:GLSM_action}. This means that we need to restrict the complex parameters in the superpotential such that 
\begin{equation}\label{eq:superpotcond}
     W(\overline{\Phi}_i) = \overline{W(\Phi_i)}\, .
\end{equation}
and furthermore have to take the $t_a$ to be real. In the LG theory after dualisation, the resulting superpotential then again satisfies \eqref{eq:superpotcond}. We hence recover the result that the vanilla anti-holomorphic involution of a GLSM is mapped to an involution of the same type for its mirror.

    As the number $h^\vee$ of dual fields $\Phi^\vee_\ell$ can be larger than the number $h$ of fields $\Phi_i$, we cannot in general solve the above equations for $\Phi^\vee_\ell$ to directly show that complex conjugation of the $\Phi_i$ implies complex conjugation (and nothing else) of the $\Phi_\ell^\vee$. This does not prevent us from associating $\sigma_v$ with $\sigma_v^\vee$. The action of an involution on an isomorphic theory must be unique up to automorphism, so that any freedom to associate $\sigma_v$ with a different involution implies that this simply gives the vanilla involution in disguise. 
    
    A similar argument holds for the action of the involution of the circle part of the sigma model, where the action of T-duality identifies it with another involution inverting the coordinate on the circle. In summary, we hence have four isomorphic tuples 
    \begin{equation}
    \begin{aligned}
     & \big( X \times S^1, (\sig,-)\big) \\
      \cong &\big( X^{\vee} \times (S^1)^{\vee}, (\sig^{\vee},-)\big) \\
      \cong &\big( X^{\vee} \times (S^1), (\sig^{\vee},-)\big) \\
       \cong& \big( X \times (S^1)^{\vee}, (\sig,-)\big)    
    \end{aligned}
    \end{equation}

    As a Gepner model is a particular limit of the GLSM, our proof of the existence of the three mirror GLSMs also provides a proof of our claim above that there are three mirror maps for the $G_2$ Gepner model.

     Even though we have focused the discussion on the vanilla involution $\sigma_v$, which always exists, it is clear that analogous results can be obtained for any other anti-holomorphic involution $\sigma$. By following through the same analysis investigating the dualisation procedure in the GLSM, such an involution $\sigma$ will also have a mirror $\sigma^\vee$ which acts geometrically on $X^\vee$. An upshot of this realization is that the set of anti-holomorphic involutions on $X$ is isomorphic to the set of anti-holomorphic involutions of $X^\vee$. We expect that this can be made precise for toric hypersurfaces by relating $\sigma$ to automorphisms of $\Delta^\circ$.

    \subsection{$G_2$ mirror maps}

    Above, we have shown the equivalence of the vanilla anti-holomorphic involutions in the dual descriptions found after mirror symmetry and/or T-duality for both Gepner models and GLSMs. Of course, merely specifying the tuple $\left(X\times S^1,(\sigma,-)\right)$ does not yet define a $G_2$ model as we need to include an appropriate twisted sector, which is in general not unique. 

    For a given choice of superpotential obeying \eqref{eq:superpotcond}, and a choice of real parameters $t_a$, the mirror map identifies a corresponding superpotential and FI parameters on the mirror. This in particular means that the fixed loci $L_\sigma$ and $L_{\sigma^\vee}$ are completely determined. However, there will in general be several inequivalent (partial) smoothings of the orbifold singularities of $\left(X \times S^1\right)/ (\sigma_v,-)$ by choosing different bundles $\mathcal{Z}$ in the construction of \cite{2017arXiv170709325J}. In a TCS description, this freedom will appear as the freedom to resolve or deform the building block $\Upsilon_{r,a,\delta}$.
    
    Given a point in the moduli space of the worldsheet SCFT of type II strings on $X\times S^1$,  our analysis hence implies that pairwise isomorphic worldsheet CFTs must exists among the four isomorphic sets 
       \begin{equation}
    \begin{aligned}
            & \left\{ (X \times S^1,(\sigma_v,-),X_\sigma^i) \right\} \\
      \cong &\left\{ (X^\vee \times (S^1)^\vee,(\sigma_v^\vee,-),X^{\vee}_\sigma{}^i) \right\}  \\
      \cong &\left\{ (X^\vee \times S^1,(\sigma_v^\vee,-),X^{\wedge-}_\sigma{}^i) \right\}  \\
       \cong& \left\{ (X \times (S^1)^\vee,(\sigma_v,-),X^{\wedge+}_\sigma{}^i) \right\}   
    \end{aligned}
    \end{equation}
    where we have denoted different twisted sectors of $X_\sigma^i, X^{\vee}_\sigma{}^i, X^{\wedge-}_\sigma{}^i, X^{\wedge+}_\sigma{}^i$. 
    
    It is beyond the scope of the present work to investigate these sets and the precise identification between their elements. For specific models, some results can be found in \cite{blumenhagen2002superconformal,roiban2002rational,Roiban:2002iv}. 
    
    It is intriguing to compare what we have found here with the mirror maps that were proposed for twisted connected sum $G_2$ manifolds, reviewed in \Cref{sect:TCS} and \Cref{sect:TCSasquotient}. For models of this type that can equally be realized as $(X \times S^1)/(\sigma,-)$, $Z_-$ captures the geometry of $X$, and $Z_+ = \Upsilon_{r,a,\delta}$ that of the twisted sector. For the three mirror maps found using the GLSM description, there are obvious candidates for a corresponding geometrical construction as a TCS, as indicated by the notation used. Making this precise requires an in-depth analysis of twisted sectors, and an identification of the twisted sectors in the GLSM with different TCS realizations.

\section{Discussion}\label{sect:discussion}

In this work we studied $G_2$ mirror symmetry from the worldsheet perspective for models which can be realized as a quotient of a Calabi-Yau threefold $X$ times a circle. The worldsheet CFT of Type II strings propagating on the covering space $X \times S^1$ then enjoys Calabi-Yau mirror symmetry as well as T-duality, giving rise to three distinct duality maps when combined. 

It hence becomes possible to lift these duality maps to duality maps for the CFT describing the $G_2$ quotient as well. Given a pair of isomorphic CFTs and a pair of involutions that are identified under the isomorphism, one must find isomorphic theories after performing the quotient. What is not immediately obvious however, is if one can have a geometrical description in terms of a $G_2$ mirror pair on both sides. The involution which is used to form a $G_2$ variety from $X \times S^1$ must act geometrically as an anti-holomorphic involution on $X$. In the context of Calabi-Yau threefolds realized as toric hypersurfaces, the explicit description of mirror symmetry in terms of a gauged linear sigma model made it possible for us to show that the mirror map precisely identifies pairs of anti-holomorphic involutions, so that Calabi-Yau mirror symmetry identifies pairs of dual quotients realized geometrically as $G_2$ varieties. This identification agrees with equivalences found using other techniques, such as a free-field description of toroidal orbifolds and Gepner models, where these are available. 

Specifying an involution of a CFT is not enough to uniquely capture the quotient theory due to the presence of twisted sectors. Given a CFT and an involution, the set of possible twisted sectors must be uniquely determined however, so that our argument really identifies sets of models. For each possible twisted sector of the quotient CFT, there must be a possible twisted sector of its mirror. In the context of $G_2$ Gepner models, twisted sectors have received attention in \cite{blumenhagen2002superconformal,eguchi2002string,roiban2002rational,Roiban:2002iv}, but it remains an interesting question for future study how to understand the general picture. 

The results obtained in this work are consistent with earlier proposals of $G_2$ mirrors for twisted connected sum $G_2$ manifolds. Whenever quotients $\left(X \times S^1\right)/(\sigma,-)$ can be realized as twisted connected sums, this in particular gives us access to various smoothings by resolving and/or deforming the building blocks. It would be very interesting to work out in detail how this approach relates to twisted sectors both before and after mirror symmetry. 

It would also be very interesting to start investigating enumerative problems for $G_2$ mirrors that are based on the equality of effective superpotentials, e.g. along the lines of \cite{Harvey:1999as, Braun:2018fdp,Acharya:2018nbo}. This is bound to be a hard problem with interesting mathematical ramifications \cite{Joyce:2016fij}.

For $Spin(7)$ mirror symmetry \cite{Shatashvili:1994zw}, we expect that results analogous to the ones presented here can be found. $Spin(7)$ manifolds 
can be realized as quotients of Calabi-Yau fourfolds by anti-holomorphic involutions \cite{Joyce:1999nk}, a construction which can be recast as 
a gluing of two simpler pieces analogous to TCS \cite{Braun:2018joh}. This was in turn used to propose a mirror construction in \cite{Braun:2019lnn}
which can be studied using similar techniques as used in the present work. $Spin(7)$ mirror symmetry has been studied from the perspective of Gepner models in 
\cite{Blumenhagen:2001qx}.

\acknowledgments

We wish to thank Ralph Blumenhagen for helpful correspondence and Madalena Lemos for dicussion. 
The work of R.D. has been supported by the STFC grants ST/T000708/1 and ST/X000591/1.

\appendix 

    \section{Toroidal Orbifold}
    \label{sec:TorodialOrbifold}

     Here we present a detailed analysis of the content of this paper for the case of a toroidal orbifold. Here we have a lot of control over the content of the theory and so it really helps to highlight how everything works together.

    \subsection{Calabi-Yau}
    
    Consider the Calabi-Yau formed as the orbifold $T^6/\Z_2^2$, where the $\Z_2^2$ acts via 
    \be 
    \label{eqn:Z2Action}
        \begin{split}
            \a : (x_1,x_2,x_3,x_4,x_5,x_6) & \mapsto (+x_1,+x_2,-x_3,a_4-x_4,-x_5,a_6-x_6) \\
            \beta : (x_1,x_2,x_3,x_4,x_5,x_6) & \mapsto (-x_1,b_2-x_2,+x_3,+x_4,b_5-x_5,b_6-x_6)
        \end{split}
    \ee 
    where $a_4,a_6,b_2,b_5$ and $b_6$ are either $0$ or $\frac{1}{2}$. As different values change the orbifold action, they have an effect on the twisted sector, i.e. the fixed points. Here we will focus solely on the untwisted sector as the notions we need are evident in the simpler untwisted sector. We include them here as it will allow a good connection with the Joyce orbifolds $T^7/\Z_2^3$ \cite{joyce1996compactI,joyce1996compactII}.
    
    As the values of the $a_i$ and $b_i$ won't affect our discussion, we set them all to zero. This orbifold was studied in \cite{vafa1995orbifolds}, and then later generalised in \cite{gaberdiel2004generalised}, in the context of discrete torsion and its role in mirror symmetry. 

    Before discussing the states in the CFT, we can use our orbifold action to compute the expected untwisted cohomology. We note that the values of $a_i$ and $b_i$ don't affect this, as they are simple shifts. Working with complex structure 
    \be 
        z^j = x^{2j-1} + i x^{2j} \qquad \text{where} \qquad j = 1,2,3,
    \ee 
    it is straight forward to check that the invariant forms are the $(0,0)$-form along with
    \be 
    \label{eqn:GKInvariantForms}
        \begin{gathered}
            dz^i_+d\bar{z}^i_+\\
            dz^1_+dz^2_+dz^3_+, \quad dz^i_+dz^j_+d\bar{z}^k_+,  \quad dz^i_+d\bar{z}^j_+d\bar{z}^k_+, \quad d\bar{z}^1_+d\bar{z}^2_+d\bar{z}^3_+ \\
            dz^i_+d\bar{z}^i_+dz^j_+d\bar{z}^j_+, \\            dz^1_+d\bar{z}^1_+dz^2_+d\bar{z}^2_+dz^3_+d\bar{z}^3_+,
        \end{gathered}
    \ee 
    where $i,j,k \in \{1,2,3\}$ but $i\neq j \neq k$. From here we see that the non-zero, even Hodge numbers are $h^{0,0}_+ = h^{3,0}_+ = h^{0,3}_+ = h^{3,3}_+ = 1$ and $h^{1,1}_+ = h^{2,2}_+ = h^{2,1}_+ = h^{1,2}_+ = 3$. 

    Let's now turn to the CFT and ask ``what are the Ramond-Ramond (RR) ground states in our CFT?" For each coordinate $x_j$ we have a left- and right-moving Majorana-Weyl spinor $\psi^i$ and $\widetilde{\psi}^j$ respectively. Given we are working with the flat metric on $T^6$, the zero modes obey the anticommutation relations
    \be 
        \{\psi^i_0,\psi^j_0\} = \{\widetilde{\psi}^i_0, \widetilde{\psi}^j_0\} = 2\del^{ij} \qand \{\psi^i_0,\widetilde{\psi}^j_0\} = 0.
    \ee   
    We then define the complexified
    \be 
    \label{eqn:GKPsipm}
        \psi^j_{\pm} = \frac{1}{2}\big( \psi^j_0 \pm i \widetilde{\psi}^j_0\big),
    \ee 
    which can easily be checked to obey the standard creation and annihilation anticommutators
    \be 
        \{\psi^i_{\pm}, \psi^j_{\mp}\} = \del^{ij} \qand \{\psi^i_{\pm},\psi^j_{\pm}\} = 0.
    \ee    
    We then adopt the convention that $\psi^i_+$ are creation and $\psi^i_-$ are annihilation operators. We note that these operators are left-right symmetric and so create left-right symmetric states.

    Let's now look at the untwisted sector states, i.e. those states that are invariant under \eqref{eqn:Z2Action}. This action was defined on the coordinates $x_i$, but it acts on the fermions in the same way, as required by SUSY. The invariant states, i.e. the untwisted sector, are then easiest expressed using the $\psi^i_{\pm}$ algebra: 
    \be 
    \label{eqn:GKUntwistedStates}
        \begin{gathered}
            \ket{0} \\ 
            \ket{12}, \quad \ket{34}, \quad \ket{56} \\
            \ket{135}, \quad \ket{136}, \quad \ket{145}, \quad \ket{146}, \quad \ket{235}, \quad \ket{236}, \quad \ket{245}, \quad \ket{246} \\
            \ket{1234}, \quad \ket{1256}, \quad \ket{3456} \\
            \ket{123456},
        \end{gathered}
    \ee    
    where we have introduced the notation $\ket{i...j} := \psi_+^i...\psi_+^j\ket{0}$. The twisted sector is straight forward to compute, but will not play a role here, instead the interested reader is directed to \cite{gaberdiel2004generalised}. 

    \subsubsection{Link To Cohomology}

    We now want to find a relationship between the above RR states and the cohomology of the target Calabi-Yau manifold. The identification is straight forward: 
    \be 
        \ket{ij...k} \cong dx^i \wedge dx^j \wedge ... \wedge dx^k\, ,
    \ee 
    but really we want \textit{complex} differential forms (e.g. the $(3,0)$-form $\Omega$). For this reason we work in a different basis for our creation and annihilation operators. We define
    \be
        \phi^i_{\pm} = \frac{1}{2}\big(\psi^{2i-1}_{\pm} + i \psi^{2i}_{\pm}\big) \qand \bar{\phi}^i_{\pm} = \frac{1}{2}\big(\psi^{2i-1}_{\pm} - i \psi^{2i}_{\pm}\big),
    \ee 
    which obey 
    \be 
        \{\phi^i_{\pm}, \bar{\phi}^j_{\mp}\} = \del^{ij}
    \ee 
    and all others vanishing. We identify the creation operators via the $+$ subscript: i.e. $\phi^i_+$ and $\bar{\phi}^i_+$. This set of operators will create states that are left-right symmetric and also form complex pairs. We then have
    \be 
        \phi^i_+\ket{0} \cong dz^i \qand \bar{\phi}^i_+ \ket{0} \cong d\bar{z}^i.
    \ee 
    We can now form the cohomology easily: we simply take products of the $\phi^i_+$ and $\bar{\phi}^i_+$s and use the anticommutation properties. Of course we can only keep those that can be formed using the untwisted states listed above. It is not hard to verify that the only allowed combinations are 
    \be 
        \begin{gathered}
            \ket{0} \\ 
            \phi^i_+\bar{\phi}^i_+\ket{0} \\
            \phi^1_+\phi^2_+\phi^3_+\ket{0}, \quad \phi^i_+\phi^j_+\bar{\phi}^k_+\ket{0},  \quad \phi^i_+\bar{\phi}^j_+\bar{\phi}^k_+\ket{0}, \quad \bar{\phi}^1_+\bar{\phi}^2_+\bar{\phi}^3_+\ket{0} \\
            \phi^i_+\bar{\phi}^i_+\phi^j_+\bar{\phi}^j_+\ket{0} \\            \phi^1_+\bar{\phi}^1_+\phi^2_+\bar{\phi}^2_+\phi^3_+\bar{\phi}^3_+\ket{0},
        \end{gathered}
    \ee 
    where $i,j,k \in \{1,2,3\}$ but $i\neq j \neq k$. These are, of course, the same results we arrived at in \eqref{eqn:GKInvariantForms}.
    
    The question we want to ask is: how do we write these forms in terms of our states $\ket{ij...k}$? The answer is to simply expand the $\phi^i_+$ and $\bar{\phi}^i_+$ in terms of the $\psi^i_+$s. Let's first look at the $0,2,4$ and $6$-forms (i.e. the diagonal forms) and consider 
    \be 
        \begin{split}
            \phi^1_+\bar{\phi}^1_+ & = \frac{1}{4}(\psi_+^1 + i\psi_+^2)(\psi_+^1 - i\psi_+^2) \\
            & = \frac{1}{2i}\psi^1_+\psi^2_+,
        \end{split}
    \ee 
    which up to a rescaling is simply $\ket{12}$. The same argument applies for all the other forms, and we get that the states with an even number of $\psi^i_+$s can simply be replaced with $\phi^i_+\bar{\phi}^i_+$. We shall call such states the \textit{diagonal} states.
    
    All that is left are the $3$-forms. We compute these in the same manner and obtain: 

    \be 
    \label{eqn:GKNon-DiagonalStates}
        \begin{split}
            \Omega = \phi_+^1\phi_+^2\phi_+^3 & = \ket{135} - \ket{245} - \ket{146} - \ket{236} + i \big[\ket{136} - \ket{246} + \ket{145} + \ket{235}] \\
            \bar{\Omega} = \bar{\phi}_+^1\bar{\phi}_+^2
            \bar{\phi}_+^3 & = \ket{135} - \ket{245} - \ket{146} - \ket{236} - i \big[\ket{136} - \ket{246} + \ket{145} + \ket{235}] \\
            \omega_1 = \bar{\phi}_+^1\phi_+^2\phi_+^3 & = \ket{135} + \ket{245} - \ket{146} + \ket{236} + i \big[\ket{136} + \ket{246} + \ket{145} - \ket{235}] \\ 
            \bar{\omega}_1 = \phi_+^1\bar{\phi}_+^2\bar{\phi}_+^3 & = \ket{135} + \ket{245} - \ket{146} + \ket{236} - i \big[\ket{136} + \ket{246} + \ket{145} - \ket{235}] \\
            \omega_2 = \phi_+^1\bar{\phi}_+^2\phi_+^3 & = \ket{135} + \ket{245} + \ket{146} - \ket{236} + i \big[\ket{136} + \ket{246} - \ket{145} + \ket{235}] \\ 
            \bar{\omega}_2 = \bar{\phi}_+^1\phi_+^2\bar{\phi}_+^3 & = \ket{135} + \ket{245} + \ket{146} - \ket{236} - i \big[\ket{136} + \ket{246} - \ket{145} + \ket{235}] \\ 
            \omega_3 = \phi_+^1\phi_+^2\bar{\phi}_+^3 & = \ket{135} - \ket{245} + \ket{146} + \ket{236} + i \big[ -\ket{136} + \ket{246} + \ket{145} + \ket{235}] \\ 
            \bar{\omega}_3 = \bar{\phi}_+^1\bar{\phi}_+^2\phi_+^3 & = \ket{135} - \ket{245} + \ket{146} + \ket{236} - i \big[ -\ket{136} + \ket{246} + \ket{145} + \ket{235}] \\
        \end{split}
    \ee 
    We shall refer to this collection of states as the \textit{non-diagonal} states from now on. The $(3,0)$-form has a decomposition $\Omega = A + i B$ with 
    \be 
        A = \ket{135} - \ket{245} - \ket{146} - \ket{236} \qand B = \ket{136} - \ket{246} + \ket{145} + \ket{235}.
    \ee 
    One can use the OPE between two fermions to then check that these expressions do indeed obey the OPEs required of $A$ and $B$.
    
    We immediately notice the difference between the diagonal and non-diagonal states: the former are given by single RR states, whereas the latter are given by a complex linear combination of all the RR states with three $\psi^i_+$s. This gives a first hint at a subtly: we know mirror symmetry is meant to map middle cohomology states to non-middle cohomology states, however we have just seen that these two classes of states take distinctly different forms. As we will see in \Cref{sec:GKMirrorSymmetry}, the fix to this problem is that our $3$-forms don't simply map to a single diagonal form, but to a complex linear combination of all the diagonal states. However, first it is instructive to compute the charges of our states under our $U(1)$ current. 

    \subsubsection{Charges}

    In order to compute the charges of our states, we of course need to know the form the $U(1)$ current takes. For our theory of complex fermions, the left-moving $U(1)$ current takes the simple form
    \be 
        J = -\sum_{i=1}^3 N\big(\phi^i \bar{\phi}^i\big) = \sum_{i=1}^3 N\big(\psi^{2i-1}\psi^{2i}\big)
    \ee 
    where the $N(...)$ stand for normal ordering. We have an analogous result for the right-moving current, but with tildes everywhere. Note that the current takes the form of a sum over the K\"{a}hler forms for the three $T^2$s that make up our $T^6$, i.e. $\omega_i \sim \psi^{2i-1}\psi^{2i}$. To compute the charges of our states, we need to find the zero mode in the expansion of $J$:
    \be 
        j_0 =  -i \sum_{r\in \Z} \sum_{j=1}^3 \psi^{2j-1}_{-r} \psi^{2j}_r. 
    \ee 
    
    We now make use of the following fact: the modes $\psi^j_r$ with $r>0$ will annihilate the vacuum, and because $j_0$ contains products of $\psi^{2j-1}_{-r}\psi^j_r$, along with the fact that we can anticommute the different $\psi^j$s and the fact that all our states are simply actions of $\psi_0^j$s on the vacuum, means that we can effectively drop all the terms in $j_0$ that don't have $r=0$. That is, we can instead simply consider the terms 
    \be 
    \label{eqn:ModifiedCurrent}
        \cJ = -i \big( \psi^1_0\psi^2_0 + \psi^3_0\psi^4_0 + \psi^5_0\psi^6_0\big) \in j_0.
    \ee 
    
    The idea now is to express $\cJ$ in terms of the $\psi^i_{\pm}$s, as this will allow us to easily compute the charges of our states. Using \eqref{eqn:GKPsipm}, we decompose $\cJ$ into two pieces $\cJ = \cJ_d + \cJ_{n-d}$ given by 
    \be 
    \label{eqn:DiagonalCurrent}
        \cJ_d = -i\big( \psi^1_+\psi^2_+ + \psi^1_-\psi^2_- + \psi^3_+\psi^4_+ + \psi^3_-\psi^4_- + \psi^5_+\psi^6_+ + \psi^5_-\psi^6_-\big).
    \ee 
    and 
    \be 
    \label{eqn:Non-DiagonalCurrent}
        \cJ_{n-d} = -i\big(\psi^1_+\psi^2_- + \psi^1_-\psi^2_+ + \psi^3_+\psi^4_- + \psi^3_-\psi^4_+ + \psi^5_+\psi^6_- + \psi^5_-\psi^6_+\big).
    \ee 
    The subscripts come from the fact that $\cJ_d$ is the only part of $\cJ$ that has any effect on the diagonal states and similarly $\cJ_{n-d}$ for the non-diagonal states.
    
    \subsubsection*{Diagonal States}
    
    Let's start by looking at the diagonal states and using $\cJ_{d}$. It is then clear that, up to signs, this current is going to take our diagonal states and either add two $\psi^i_+$s or take away two in the pairs $\ket{12},\ket{34}$ or $\ket{56}$. Noting that when we remove the creation operators, we must first anticommute the $\psi^{2j-1}_-\psi^{2j}_-$ in $\cJ_d$, we see that these states come with a minus sign. For example 
    \be 
        \begin{split}
            \psi^1_-\psi^2_- (\psi^1_+\psi^2_+\ket{0}) & = - \psi^2_-\psi^1_-\psi^1_+\psi^2_+\ket{0} \\
            &= - \psi^2_-\psi^2_+\ket{0} + \psi^2_-\psi^1_+\psi^1_-\psi^2_+\ket{0} \\
            & = -\ket{0} + \psi^2_+\psi^2_-\ket{0} - \psi^2_-\psi^1_+\psi^2_+\psi^1_-\ket{0} \\
            & = -\ket{0}.
        \end{split}
    \ee 
    where we have used $\{\psi^i_{\pm},\psi^j_{\mp}\} = \del^{ij}$ and $\{\psi^i_{\pm},\psi^j_{\pm}\} = 0$. The same calculation holds for all other states -- note that we can move bilinears in fermions freely, i.e. we can ``jump" a $\psi^3_-\psi^4_-$ over the $\psi^1_+\psi^2_+$ in $\ket{1234}$ without the cost of any signs.
    
    So we see our diagonal forms are mapped under $\cJ_d$ in the following way
    \be 
    \label{eqn:DiagonalStatesJMapping}
        \begin{array}{lr}
        {\Large \text{$\cJ_d : \quad $}}
            \begin{split}
                \ket{0} & \mapsto -i \big[ \ket{12} + \ket{34} + \ket{56}\big] \\
                \ket{12} & \mapsto -i \big[ -\ket{0} + \ket{1234} + \ket{1256}\big] \\
                \ket{34} & \mapsto -i \big[ -\ket{0} + \ket{1234} + \ket{3456}\big] \\
                \ket{56} & \mapsto -i \big[ -\ket{0} + \ket{3456} + \ket{1256}\big] \\
                \ket{1234} & \mapsto -i \big[ -\ket{12} - \ket{34} + \ket{123456}\big]\ \\
                \ket{1256} & \mapsto -i \big[ -\ket{12} - \ket{56} + \ket{123456}\big] \\
                \ket{3456} & \mapsto -i \big[ -\ket{34} - \ket{56} + \ket{123456}\big] \\
                \ket{123456} & \mapsto -i \big[ -\ket{1234} - \ket{1256} - \ket{3456}\big] \\
            \end{split}
        \end{array}
    \ee 
    The important thing to note is that \textit{none} of these states are eigenstates of our current. We need to take a linear combination of states in order to get an eigenstate. By considering the $(8\times 8)$ matrix defining the action of $\cJ_d$ on our diagonal states, we can compute the eigenvalues and eigenvectors. The results are presented in \Cref{tab:DiagonalEigenstates}.
    \begin{table}[h!]
        \begin{center}
        	\begin{tabular}{@{}  C{2.5cm} | C{12cm} @{}}
                    Left-Charge & Eigenstate \\
                    \hline
                    $+3$ & $ \Sigma = -\big[\ket{0} - \ket{1234} - \ket{3456} - \ket{1256}\big] + i \big[\ket{56} - \ket{123456} + \ket{34} + \ket{12} \big]$ \\
                    $-3$ & $ \bar{\Sigma} = -\big[\ket{0} - \ket{1234} - \ket{3456} - \ket{1256}\big] - i \big[\ket{56} - \ket{123456} + \ket{34} + \ket{12} \big]$ \\
                    $+1$ & $ \sig_1 = -\big[\ket{0} + \ket{1234} - \ket{3456} + \ket{1256}\big] + i \big[\ket{56} + \ket{123456} + \ket{34} - \ket{12} \big]$ \\
                    $-1$ & $\bar{\sig}_1 = -\big[\ket{0} + \ket{1234} - \ket{3456} + \ket{1256}\big] - i \big[\ket{56} + \ket{123456} + \ket{34} - \ket{12} \big] $ \\
                    $+1$ & $\sig_2 = -\big[\ket{0} + \ket{1234} + \ket{3456} - \ket{1256}\big] + i \big[\ket{56} + \ket{123456} - \ket{34} + \ket{12} \big]$ \\
                    $-1$ & $\bar{\sig}_2 =  -\big[\ket{0} + \ket{1234} + \ket{3456} - \ket{1256}\big] - i \big[\ket{56} + \ket{123456} - \ket{34} + \ket{12} \big]$ \\
                    $+1$ & $\sig_3 = -\big[\ket{0} - \ket{1234} + \ket{3456} + \ket{1256}\big] + i \big[-\ket{56} + \ket{123456} + \ket{34} + \ket{12} \big]$ \\
                    $-1$ & $\bar{\sig}_3 = -\big[\ket{0} - \ket{1234} + \ket{3456} + \ket{1256}\big] - i \big[-\ket{56} + \ket{123456} + \ket{34} + \ket{12} \big] $ \\
        	\end{tabular}
        \end{center} 
        \caption{\label{tab:DiagonalEigenstates} Eigenstates of the diagonal left $U(1)$ current $\cJ_d$, and their corresponding charges.}
    \end{table}
    
    The relative coefficients of this matches that of our non-diagonal states, \eqref{eqn:GKNon-DiagonalStates}, i.e. $\Sigma$ and $\Omega$ etc have the same coefficients. We shall return to this in \Cref{sec:GKMirrorSymmetry}.
    
    \subsubsection*{Non-Diagonal States}
    
    We can proceed to compute how $\cJ_{n-d}$ affects the non-diagonal states in a similar fashion. Here we have the rule: if the state contains $\psi^{1,3,5}_+$ then it is replaced with $\psi^{2,4,6}_+$ and vice versa. In order to get the minus signs correct, we first write $\cJ_{n-d}$ with all annihilation operators to the right
    \be
        \cJ_{n-d} = -i\big(\psi^1_+\psi^2_- - \psi^2_+\psi^1_- + \psi^3_+\psi^4_- -\psi^4_+\psi^3_- + \psi^5_+\psi^6_- - \psi^6_+\psi^5_-\big),
    \ee  
    so we see replacing $\psi^{2,4,6}_+ \mapsto \psi^{1,3,5}_+$ but $\psi^{1,3,5}_+ \mapsto -\psi^{2,4,6}_+$. As before, we see how each of the individual $\ket{ijk}$ states are mapped under $\cJ_{n-d}$, and from there check that our non-diagonal states are eigenstates and compute the eigenvalues. We have:
    \be
        {\Large \text{$\cJ_{n-d} : \quad $}} \begin{array}{lr}
        \begin{split}
                \ket{135} & \mapsto -i \big[ -\ket{145} - \ket{145} - \ket{136}\big] \\
                \ket{245} & \mapsto -i \big[ \ket{145} + \ket{145} -\ket{246} \big] \\
                \ket{146} & \mapsto -i \big[ -\ket{246} + \ket{136} + (156)\big]\\
                \ket{236} & \mapsto -i \big[ \ket{136} - \ket{246} + \ket{145}\big] \\
                \ket{136} & \mapsto -i \big[ -\ket{236} - \ket{146} + \ket{135}\big] \\
                \ket{246} & \mapsto -i \big[ \ket{146} + (256) + \ket{245}\big] \\
                \ket{145} & \mapsto -i \big[ -\ket{245} + \ket{135} - \ket{146}\big] \\
                \ket{235} & \mapsto -i \big[ \ket{135} - \ket{245} - \ket{236}\big] \\
        \end{split}
        \end{array}
    \ee
    Recalling \eqref{eqn:GKNon-DiagonalStates} we therefore see that our non-diagonal states are our eigenstates with charges $\pm3, \pm1$, specifically:
    \be 
    \label{eqn:Non-DiagonalStatesCharge}
        q(\phi^1_+ \phi^2_+ \phi^3_+) = 3, \quad q(\bar{\phi}^1_+ \bar{\phi}^2_+ \bar{\phi}^3_+) = -3, \quad q(\bar{\phi}^i_+ \phi^j_+ \phi^k_+) = 1, \qand q(\phi^i_+ \bar{\phi}^j_+ \bar{\phi}^k_+) = -1.
    \ee 

    In fact we have been a little careless here: the above charges are what we expect for the NS states, but here we are dealing with the R states. We go between these via spectral flow and this maps $j_0 \mapsto j_0 \pm \frac{3}{2}$, and so these charges should be shifted. 
    
    \subsubsection*{Right-Charge}
    
    We should also compute the right charge $q_R$. This comes from a similar derivation but now with tildes everywhere 
    \be 
        \widetilde{\cJ} = -i \big( \widetilde{\psi}^1_0\widetilde{\psi}^2_0 + \widetilde{\psi}^3_0\widetilde{\psi}^4_0 + \widetilde{\psi}^5_0\widetilde{\psi}^6_0\big) \in \widetilde{j}_0.
    \ee 
    However now we have 
    \be 
        \widetilde{\psi}^i_0 = -i\big( \psi^i_+ - \psi^i_-\big),
    \ee 
    and so we need to carry this factor of $-i$ through along with the relative sign between $\psi^i_{\pm}$. Everything in $\widetilde{\cJ}$ is bilinear, so we are really dealing with $(-i)^2 = -1$. We therefore get 
    \be 
        \widetilde{\psi}^1_0\widetilde{\psi}^2_0 = -\big[ \psi^1_+ \psi^2_+ +\psi^1_-\psi^2_- - \psi^1_+\psi^2_- - \psi^1_-\psi^2_+ \big]
    \ee 
    etc. We therefore see that the signs on the non-diagonal forms cancel, i.e. we simply have 
    \be 
        \widetilde{\cJ}_{n-d} = \cJ_{n-d}
    \ee 
    however on the diagonal forms we get a relative sign
    \be 
        \widetilde{\cJ}_d = -\cJ_d
    \ee 
    In other words, our non-diagonal states will have $q_L = q_R$ while the diagonal eigenstates have $q_L = -q_R$. This agrees with the result of \Cref{sec:ChiralRings}: the non-diagonal states are elements of the $(c,c)$ ring while the diagonal eigenstates are elements of the $(a,c)$ ring.
    
    \subsubsection{Mirror Symmetry}
    \label{sec:GKMirrorSymmetry}


    We now want to look at how mirror symmetry acts on our Calabi-Yau. As explained in \cite{gaberdiel2004generalised}, in this context mirror symmetry is generated by three $T$-dualities along the coordinates 
    \be
    \label{eqn:CYTdualCoords}
        (j_1,j_2,j_3) \in \big\{(1,3,6),(1,4,5),(2,3,5),(2,4,6) \big\},
    \ee 
    which are exactly the combinations that appear in the imaginary parts of our $3$-forms above.\footnote{We note that in \cite{gaberdiel2004generalised} they also allow for T-dualising along $(1,3,5)$, $(1,4,6)$, $(2,3,6)$ and $(2,4,5)$, which correspond to the real parts of our $3$-forms. Here we do not include these as they define the mirror map as one that maps the right-moving part of $\Omega$ to its complex conjugate, but these additional maps map $\Omega_R \mapsto - \Omega^*_R$. This additional minus sign can be accounted for by an additional automorphism of the algebra that introduces a phase: $\Omega_R \mapsto e^{i\phi}\Omega_R$, which is the one mentioned in \Cref{footnote:PhaseAutomorphism}. In this work we will ignore this additional phase automorphism, and so we ignore these additional maps.}

    We now want to ask how $T$-duality affects our Clifford algebra: it changes the sign of the right-moving fermion zero mode, and so it replaces the creation operator $\psi_+^j$ with the annihilation operator $\psi_-^j$. This modifies the definition of the ground state to be in terms of the mapped operators, namely 
    \be 
        \big(\psi_-^i\big)^{\prime}\ket{0}^{\prime} = 0,
    \ee 
    which is equivalent to 
    \be 
        \psi_-^i\ket{0}^{\prime} = \psi_+^j\ket{0}^{\prime} = 0
    \ee   
    where $j$ labels the coordinates that are $T$-dualised and $i$ labels all others. Using that $(\psi_+^j)^2=0$, we can then express our dual vacuum state in terms of the original one as 
    \be  
    \label{eqn:TDualVacuum}
        \ket{0}^{\prime} = \psi_+^{j_1}\psi_+^{j_2}\psi_+^{j_3}\ket{0},
    \ee 
    with $(j_1,j_2,j_3)$ the dualised indices.
    
    We can now ask how states/forms in the $T$-dual picture relate to states/forms in the original picture. The idea is simple: we work with primes everywhere and then simply substitute in the relations at the end. The untwisted sector is then exactly the same as before, \eqref{eqn:GKUntwistedStates}, but with primes everywhere. For concreteness, let's work with $T$-dualising along $(1,3,6)$. In fact we are going to include a factor of $i$ into our map, for a reason that will be explained shortly. Then we have 
    \be 
        \big(\psi_{\pm}^{2,4,5}\big)^{\prime} = \psi_{\pm}^{2,4,5}, \quad \big(\psi_{\pm}^{1,3,6}\big)^{\prime} = \psi_{\mp}^{1,3,6} \qand \ket{0}^{\prime} = i\ket{136}.
    \ee 
    We now put this together with the anticommutators for the creation/annihilation operators to obtain our relation. The result is the following 
    \be 
    {\Large {\text{$\mathfrak{M} : \quad$}}} \begin{array}{lr}
        \begin{split}
            \ket{135} & \mapsto -i\ket{56} \\
            \ket{245} & \mapsto -i\ket{123456} \\
            \ket{146} & \mapsto i\ket{34} \\
            \ket{236} & \mapsto i\ket{12} \\
            \ket{136} & \mapsto -i\ket{0} \\
            \ket{246} & \mapsto -i\ket{1234} \\
            \ket{145} & \mapsto i\ket{3456} \\
            \ket{235} & \mapsto i\ket{1256} \\
        \end{split}
        \end{array}
    \ee 
    We can easily show from here that the states on the right hand side (i.e. the diagonal states) are mapped with the opposite sign behaviour. As we introduced the $i$ factor we  then get that $T^2=\text{id}$, which is required for it to be an involution.

    The above allows us to ask the question of how our initial states are mapped. For example, the $(3,0)$-form $\Omega$ is mapped as 
    \be 
        T(\Omega = \phi^1\phi^2\phi^3) = -\big[\ket{0} - \ket{1234} - \ket{3456} - \ket{1256}\big] + i \big[ \ket{56} - \ket{123456} + \ket{34} + \ket{12} \big] = \Sigma,
    \ee 
    where $\Sigma$ is as defined in \Cref{tab:DiagonalEigenstates}. Again note that the factor of $i$ we included is needed here, i.e. the real part of $\Omega$ is mapped to the imaginary part of $\Sigma$. A similar calculation will verify that the non-diagonal states in \eqref{eqn:GKNon-DiagonalStates} correspond, respectively, to the diagonal eigenstates in the table, i.e. 
    \be 
        T(\bar{\Omega}) = T(\bar{\Sigma}), \qquad T(\omega_i) = \sig_i \qand T(\bar{\omega}_i) = \bar{\sig}_i
    \ee 
    
    At the level of the charges, this implies that mirror symmetry maps
    \be 
    \label{eqn:GKMirrorMapCharges}
        \mathfrak{M} : (q_L,q_R) \mapsto (q_L,-q_R).
    \ee     
    So we see that mirror symmetry maps charge eigenstates to charge eigenstates. This is the result we expected: mirror symmetry maps $\fR_{(c,c)}$ to $\fR_{(a,c)}$ and vice versa. As we see, such a map takes a $3$-form and maps it to a linear combination of all the diagonal forms. This is not a new result, and is simply related to how BPS states are mapped under mirror symmetry.

    \subsection{$G_2$}

    Let's now look at the corresponding $G_2$ orbifold. Joyce showed \cite{joyce1996compactI,joyce1996compactII} that one can construct a $G_2$ manifold via $T^7/\Z_2^3$, where the $\Z_2^3$ acts via
    \be 
    \label{eqn:z23Action}
        \begin{split}
            \a : (x_1,x_2,x_3,x_4,x_5,x_6,x_7) & \mapsto (+x_1,+x_2,-x_3,a_4-x_4,-x_5,a_6-x_6,x_7) \\
            \beta : (x_1,x_2,x_3,x_4,x_5,x_6,x_7) & \mapsto (-x_1,b_2-x_2,+x_3,+x_4,b_5-x_5,b_6-x_6,x_7) \\
            \sig : (x_1,x_2,x_3,x_4,x_5,x_6,x_7) & \mapsto (x_1,-x_2,x_3,-x_4,x_5,-x_6,-x_7),
        \end{split}
    \ee
    where $a_i,b_i = 0,1/2$. The $\a$ and $\beta$ action here are simply the extension of  \eqref{eqn:Z2Action} to include the $x_7$ coordinate. We note that $\sig$ acts with a minus on $x_{2,4,6}$, which in the complex structure of our $T^6/\Z_2^2$ discussion, is nothing but complex conjugation. From here we see that we can identity\footnote{Strictly speaking we need to resolve the $(\a,\beta)$ action in $T^6$ to obtain the Calabi-Yau. We return to this shortly.} 
    \be 
        \frac{T^7}{\Z_2^3} = \frac{\Big(\frac{T^6}{(\a,\beta)}\Big)\times S^1}{\sig} = \frac{X_{T^6}\times S^1}{\sig},
    \ee   
    Hence we can apply our logic in order to check the involution carries through as we would like.

    \subsubsection{Anti-holomorphic Involution}

    Before discussing the R ground states of our $G_2$ theory, we first want to ask how the anti-holomorphic involution acts on the states in our Calabi-Yau theory. This is particularly easy to do here: our complex structure is given by $z^i = x^{2i-1} + i x^{2i}$, and so complex conjugation simply acts via 
    \be 
        \sig : (x_1,x_2,x_3,x_4,x_5,x_6) \mapsto (x_1,-x_2,x_3,-x_4,x_5,-x_6).
    \ee 
    This mapping is translated directly to the fermions, i.e. we map $\psi^{2,4,6}_+ \mapsto -\psi^{2,4,6}_+$ and the others are left alone. From here we see that our diagonal and non-diagonal states have the desired behaviour, when compared to their differential forms: the $(0,0)$ and $(2,2)$ forms are invariant, the $(1,1)$ and $(3,3)$ forms are odd, and the $3$-forms are mapped in pairs $(m,3-m) \mapsto (3-m,m)$. In particular complex conjugation acts simply on the $i$ appearing in our non-diagonal states, \eqref{eqn:GKNon-DiagonalStates}.

    However we note that the diagonal eigenstates in \Cref{tab:DiagonalEigenstates} are not invariant but are also mapped via complex conjugation on the $i$ factors. Note that this tells us that the real parts of diagonal states are the even forms while the imaginary parts are the odd forms. Putting this together with the non-diagonal states, we see that our charges are mapped via 
    \be
    \label{eqn:GKComplexConjCharges}
        \sig: (q_L,q_R) \mapsto (-q_L,-q_R).
    \ee 
    This is exactly the result we were expecting. 

    \subsubsection{Untwisted Sector}

    We start by looking at the untwisted sector of this theory. Here the values of the $a_i$ and $b_i$ don't matter (as they only affect the fixed points), and so we can ignore them. It is straightforward to check that the invariant states are given by 
    \be 
        \ket{0}, \qquad \ket{i j k}, \qquad \ket{i j k \ell}, \qquad \ket{1234567}
    \ee 
    where 
    \be 
        \begin{split}
            (ijk) & \in \big\{ (127), (347), (567), (135), (146), (236), (245)\big\}, \\
            (ijk\ell) & \in \big\{ (1234), (3456), (1256), (1367), (1457),  (2357), (2467)\big\}.
        \end{split}
    \ee 
    We now note these take the exact form needed to be the extension of our states from our $T^6/\Z_2^2$ calculation: everything that was odd under the anti-holomorphic involution is paired with a $\psi^7_+$ here. Geometrically, this is the statement that forms that are odd under the involution need to be wedged with $d\theta$, the one-form on the $S^1$. In particular, notice that for our $3$-forms, \eqref{eqn:GKNon-DiagonalStates}, we must now work with the real and imaginary parts. For example,
    \be 
        (\omega_1 + \bar{\omega}_1) \qand (\omega_1 - \bar{\omega}_1)\ket{7}
    \ee 
    are the invariant states. This corresponds to taking the real and imaginary linear combinations of a $(2,1)$ and $(1,2)$ form and wedging the imaginary part with $d\theta$. Equally for the diagonal states, those corresponding to $(1,1)$ and $(3,3)$ forms come with a $\ket{7}$, while the $(0,0)$ and $(2,2)$ forms are invariant by themselves.

    Additionally we note that the $G_2$ 3-form and dual 4-form are expressed in the CFT as 
    \be 
        \begin{split}
                \Phi & = \ket{135} - \ket{245} - \ket{146} - \ket{236} + \ket{127} + \ket{347} + \ket{567} \\
                X & = \ket{1457} + \ket{1367} + \ket{2357} - \ket{2467} + \ket{1234} + \ket{3456} + \ket{1256}
        \end{split}
    \ee 
    which matches the geometrical decomposition $\Phi = \Re(\Omega) + J \wedge d\theta$ and $X = \Im(\Omega)\wedge d\theta + \frac{1}{2}J \wedge J$.

    \subsubsection{Twisted Sector}

    We now want to investigate the twisted sector of our action, and ask how this changes the cohomology of our $G_2$ orbifold. This problem has been studied from the perspective of the Joyce orbifold $T^7/\Z_2^3$ in \cite{joyce1996compactII} and then explained at the level of discrete torsion in \cite{gaberdiel2004generalised}. Here we take a slighly different approach, and instead consider 
    \be 
        M_{\sig} = \frac{\widetilde{\Big(\frac{T^6}{(\a,\beta)}\Big)}\times S^1}{\sig},
    \ee 
    where the tilde means we resolve the singularities of the $(\a,\beta)$ action. This will, of course, give the same result as the references above.

\begin{figure}
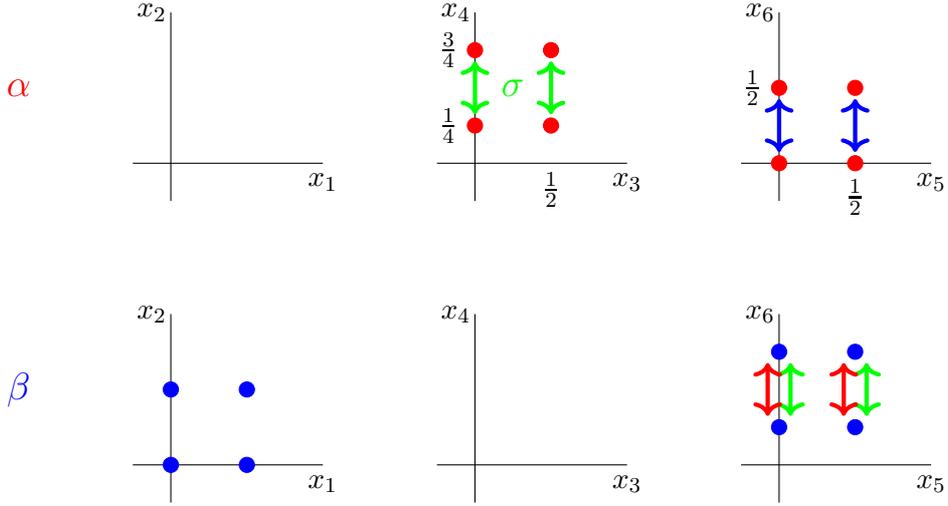

        \begin{center}
            \btik 
                \begin{scope}[xshift = -4cm]
                    \node at (-2,1) {\textcolor{red}{\Large{$\a$}}};
                    \draw[] (-0.5,0) -- (2,0);
                    \node at (2,-0.25) {$x_1$};
                    \draw[] (0,-0.5) -- (0,2);
                    \node at (-0.25,2) {$x_2$};
                \end{scope}
                \begin{scope}
                    \draw[] (-0.5,0) -- (2,0);
                    \node at (2,-0.25) {$x_3$};
                    \draw[] (0,-0.5) -- (0,2);
                    \node at (-0.25,2) {$x_4$};
                    \draw[red, fill=red] (0,0.5) circle [radius=0.1cm];
                    \draw[red,fill=red] (0,1.5) circle [radius=0.1cm];
                    \draw[red,fill=red] (1,0.5) circle [radius=0.1cm];
                    \draw[red,fill=red] (1,1.5) circle [radius=0.1cm];
                    \node at (-0.35,0.5) {$\frac{1}{4}$};
                    \node at (-0.35,1.5) {$\frac{3}{4}$};
                    \node at (1,-0.35) {$\frac{1}{2}$};
                    \draw[green, ultra thick, <->] (0,0.65) -- (0,1.35);
                    \draw[green, ultra thick, <->] (1,0.65) -- (1,1.35);
                    \node at (0.5,1) {\textcolor{green}{\Large{$\sig$}}};
                \end{scope}
                \begin{scope}[xshift = +4cm]
                    \draw[] (-0.5,0) -- (2,0);
                    \node at (2,-0.25) {$x_5$};
                    \draw[] (0,-0.5) -- (0,2);
                    \node at (-0.25,2) {$x_6$};
                    \draw[red,fill=red] (0,0) circle [radius=0.1cm];
                    \draw[red,fill=red] (0,1) circle [radius=0.1cm];
                    \draw[red,fill=red] (1,0) circle [radius=0.1cm];
                    \draw[red,fill=red] (1,1) circle [radius=0.1cm];
                    \node at (1,-0.45) {$\frac{1}{2}$};
                    \node at (-0.35,1) {$\frac{1}{2}$};
                    \draw[blue, ultra thick, <->] (0,0.15) -- (0,0.85);
                    \draw[blue, ultra thick, <->] (1,0.15) -- (1,0.85);
                \end{scope}
                \begin{scope}[xshift = -4cm, yshift = -4cm]
                    \node at (-2,1) {\textcolor{blue}{\Large{$\beta$}}};
                    \draw[] (-0.5,0) -- (2,0);
                    \node at (2,-0.25) {$x_1$};
                    \draw[] (0,-0.5) -- (0,2);
                    \node at (-0.25,2) {$x_2$};
                    \draw[blue,fill=blue] (0,0) circle [radius=0.1cm];
                    \draw[blue,fill=blue] (0,1) circle [radius=0.1cm];
                    \draw[blue,fill=blue] (1,0) circle [radius=0.1cm];
                    \draw[blue,fill=blue] (1,1) circle [radius=0.1cm];
                \end{scope}
                \begin{scope}[yshift = -4cm]
                    \draw[] (-0.5,0) -- (2,0);
                    \node at (2,-0.25) {$x_3$};
                    \draw[] (0,-0.5) -- (0,2);
                    \node at (-0.25,2) {$x_4$};
                \end{scope}
                \begin{scope}[xshift = +4cm, yshift = -4cm]
                    \draw[] (-0.5,0) -- (2,0);
                    \node at (2,-0.25) {$x_5$};
                    \draw[] (0,-0.5) -- (0,2);
                    \node at (-0.25,2) {$x_6$};
                    \draw[blue,fill=blue] (0,0.5) circle [radius=0.1cm];
                    \draw[blue,fill=blue] (0,1.5) circle [radius=0.1cm];
                    \draw[blue,fill=blue] (1,0.5) circle [radius=0.1cm];
                    \draw[blue,fill=blue] (1,1.5) circle [radius=0.1cm];
                    \draw[green, ultra thick, <->] (0.15,0.65) -- (0.15,1.35);
                    \draw[green, ultra thick, <->] (1.15,0.65) -- (1.15,1.35);
                    \draw[red, ultra thick, <->] (-0.15,0.65) -- (-0.15,1.35);
                    \draw[red, ultra thick, <->] (0.85,0.65) -- (0.85,1.35);
                \end{scope}
            \etik 
        \end{center}
        \caption{Fixed points in $T^6/(\a,\beta)$ and how they are mapped under $(\a,\beta,\sig)$. All red objects correspond to $\a$, blue to $\beta$ and green to $\sig$. The graphs indicate a decomposition of $T^6 = T^2\times T^2 \times T^2$, and the number of fixed points understood multiplicitivly. The 16 fixed points of $\a$ are reduced to 4 under the action of $(\beta,\sig)$. However $\a$ and $\sig$ act on the fixed points of $\beta$ in the same manner and so $\beta\sig$ acts trivially. This leads to the introduction of discrete torsion in the $G_2$ manifold.}
        \label{fig:AlphaBetaFixedPoints}
    \end{figure}

    Here the choice of the $a_i,b_i$ in \eqref{eqn:z23Action} matters. We will work with $a_4=b_6=1/2$ and all others vanishing, i.e.
    \be
        \begin{split}
            \a : (x_1,x_2,x_3,x_4,x_5,x_6,x_7) & \mapsto (+x_1,+x_2,-x_3,\frac{1}{2}-x_4,-x_5,-x_6,x_7) \\
            \beta : (x_1,x_2,x_3,x_4,x_5,x_6,x_7) & \mapsto (-x_1,-x_2,+x_3,+x_4,-x_5,\frac{1}{2}-x_6,x_7) \\
            \sig : (x_1,x_2,x_3,x_4,x_5,x_6,x_7) & \mapsto (x_1,-x_2,x_3,-x_4,x_5,-x_6,-x_7),
        \end{split}
    \ee
    First we want to consider the Calabi-Yau given by the resolution of $T^6/(\a,\beta)$. We know from our previous discussion that the untwisted sector gives contributions to the Hodge numbers 
    \be 
        (h^{0,0}, h^{1,1}, h^{2,1}, h^{3,0}) = (1,3,3,1),
    \ee 
    along with their matching Hodge duals, $h^{m,n} = h^{3-m,3-n}$. The contribution from the twisted sector comes from considering the fixed points. Both $\a$ and $\beta$ have $16$ fixed points, however the action of the other on these fixed points leaves $8$ in each case. We show this graphically in \Cref{fig:AlphaBetaFixedPoints}. Locally these fixed points are modelled by $T^2 \times \C^2/\{\pm1\}$, and as standard we can choose to either blow up or deform the $\C^2/\{\pm1\}$. In either case the Hodge numbers are the same, and each fixed point contributes one to both $h^{1,1}$ and $h^{2,1}$ (and their Hodge duals). So in total our Calabi-Yau has $(h^{1,1},h^{2,1}) = (19,19)$, and so is self-mirror. This geometry is known as the Schoen Calabi-Yau $X_S$. In terms of Betti numbers, we have
    \be 
        b^2(X_S) = \underset{T^2}{3} + \underset{\a}{8} + \underset{\beta}{8} = 19 \qand b^3(X_S) = \underset{T^2}{8} + \underset{\a}{16} + \underset{\beta}{16} = 40.
    \ee   

    We now want to consider the action of $\sig$ in $M_{\sig} = (X_S\times S^1)/\sig$, i.e. we want to compute $b^2_{\pm}(X_S)$ and $b^3_{\pm}(X_S)$ and use them to compute the Betti numbers for the smoothing $M = \widetilde{M_{\sig}}$ via
    \be 
        b^2(M) = b^2_+(X_S) + e_2 \qand b^3(M) = b^3_+(X_S) + b^2_-(X_S) + e_3,
    \ee 
    where $e_{2,3}$ denotes the contributions from the fixed points of $\sig$. 
    
    Let's start with $b^2(X_S)$. It is clear that the $3$ contribution from $T^2$ are all odd. The $8$ that arises from the fixed points of $\a$ gives $4$ even and $4$ odd: the fixed points are identified in pairs (see \Cref{fig:AlphaBetaFixedPoints}) and so we can form one even and one odd combination. The $8$ from the $\beta$ fixed points are more subtle: here $\sig$ acts trivially (as its action is equivalent to $\beta$s action, but we have modded out by $\beta$), and so introduces discrete torsion. As detailed in \cite{gaberdiel2004generalised}, if we resolved the singularity in $\beta$ via a blow up then $\sig$ will preserve the orientation of the exceptional divisor and so the ground state $(1,1)$-form is invariant. However, if we had deformed the $\beta$-singularity, then $\sig$ reverses the orientation and so our $(1,1)$-form is odd. Each of the $8$ fixed points can be blown up or deformed independently, and so we have the above choice for each one. Therefore, if $\ell \in \{0,...,8\}$ denotes the number of blow ups, then we get a contribution of $\ell$ to $b^2_+(X_S)$ and $(8-\ell)$ to $b^2_-(X_S)$. 

    Now let's discuss $b^3_{\pm}(X_S)$. The story is very similar to above: the $8+16=24$ that comes from $T^2$ and $\a$ are identified in pairs and so we have $12$ even and $12$ odd contributions. The $16$ from $\beta$ is dependent on the discrete torsion, and contributes $(8-\ell)$ to $b^3_+(X_S)$ and $\ell$ to $b^3_-(X_S)$. 

    Finally we need to add in the contributions from the $\sig$ fixed points, of which there are $4$ independent ones. This is easily seen from the fact that in $T^7/\sig$ we have $16$ fixed points, but these are reduced to $4$ under $(\a,\beta)$, which are modded out to define $X_S$. These four fixed points are $T^3$s and give contributions of $4$ to $b^2(\cM)$ and $12$ to $b^3(\cM)$. In total the smoothed $G_2$ orbifold then has Betti numbers 
    \be
    \label{eqn:b2b3Joyce}
        b^2(M) = 8+\ell \qand b^3(M) = 47-\ell\, ,
    \ee 
    in agreement with \cite{joyce1996compactII,gaberdiel2004generalised}.

    \subsubsection{Mirror Symmetry}
    
    As detailed in \cite{gaberdiel2004generalised}, here we have 4 notions of mirror symmetry. Just as with the Calabi-Yau torodial orbifold considered previously, these take the form of T-dualities: 
    \be 
        \begin{split}
            \cT^+_3 & = \{ (3,4,7), (2,4,5), (1,4,6)\} \\
            \cT^-_3 & = \{ (2,3,6),(5,6,7),(1,2,7),(1,3,5)\} \\
            \cT^+_4 & = \{ (1,2,5,6), (1,3,6,7), (2,3,5,7)\} \\
            \cT^-_4 & = \{ (1,4,5,7), (1,2,3,4), (3,4,5,6), (2,4,6,7)\}.
        \end{split}
    \ee 
    The combinations appearing in here are exactly the terms that appear in $\Phi$ and $X$ above, i.e. these are calibrated submanifolds. 
        The subscripts indicate how many T-dualities we do, and from chirality arguments we can see that $\cT^{\pm}_3$  map compactifications on Type IIA/B to those on Type IIB/A, while $\cT^{\pm}_4$ map Type IIA/B to Type IIA/B. The $\pm$ superscript indicates whether the discrete torsion signs are reversed or not, i.e. whether we blow up or deform the fixed points of $\beta$. This changes the topology of the resulting $G_2$ manifold, i.e. $\ell \mapsto (8-\ell)$ in \eqref{eqn:b2b3Joyce}.

    The key thing we want to notice is that within the $\cT_4$ actions we have $(1,3,6,7)$, $(2,3,5,7)$, $(1,4,5,7)$ and $(2,4,6,7)$ which have the effect of mirroring the Calabi-Yau plus a T-duality in the additional $S^1$ direction. That is, we can take our Calabi-Yau mirror maps in \eqref{eqn:CYTdualCoords} and add on a T-dual along the $S^1$ direction and generate a $G_2$ mirror map.\footnote{We note that, just as in the Calabi-Yau case, mirror symmetry does not simply map a $4$-form to a $3$-form, and vice versa. This we can see from the fact that our $4$-form $X$ contains exactly the terms that appear in $\cT_4$, and so these terms will be mapped to the vacuum. For example, if we did the $(1,3,6,7)$ map, then $X \ni \ket{1367} \mapsto \ket{0}$, which geometrically is the $0$-form. Similarly $\Phi \ni -\ket{245} \mapsto \ket{1234567}$. It is then easy to see that under any of the maps we actually exchange $\Phi + \ket{0}$ and $X + \ket{1234567}$. This is just the equivalent of the fact that a $3$-form in the Calabi-Yau is mapped to a linear combination of the diagonal forms.}  
    
    Here we want to ask how the mirror maps affect the involution action $\sig$. Namely we want to ask how $\sig^{\vee}$ is related to $\sig$. It is clear from the above calculation that 
    \be 
        \bigg(\frac{X_S\times S^1}{\sig}\bigg)^{\vee} = \frac{X_S^{\vee}\times (S^1)^\vee}{\sig},
    \ee 
    and so we can set $\sig^{\vee} = \sig$. There are 9 independent $G_2$ manifolds we can form via the resolution of these spaces, labelled by the number of blow ups, $\ell$. These blow ups appear in $X_S$, and so we can label the 9 $G_2$s via 
    \be 
        G_2^{\ell} = \widetilde{\bigg(\frac{X_S^{\ell}\times S^1}{\sig}\bigg)}.
    \ee 
    Mirror symmetry maps $G_2^{\ell}$ either back to itself or to $G_2^{8-\ell}$, depending on whether we use $\cT_4^+$ or $\cT_4^-$, respectively.
    
    The interesting thing in this case is that $X_S$ is self mirror, and so even though the $X_S^{\ell}$ look different, they are all diffeomorphic. This diffeomorphism alters the action of $\sig$ in the required way, namely we give $\sig$ an $\ell$ index and obtain 
    \be 
        G_2^{\ell} = \widetilde{\bigg(\frac{X_S\times S^1}{\sig^{\ell}}\bigg)}
    \ee 
    We can therefore take two viewpoints on the situation: we either have a collection of different Calabi-Yaus, or we have a collection of different anti-holomorphic involutions.

     \section{Mirror Symmetry and Toric Hypersurfaces}
    \label{sec:BatyrevMirrorSym}
    
    In this appendix we review the geometrical constructions of mirrors for Calabi-Yau manifolds and building blocks using toric geometry.

    \subsection{Polytopes and Hypersurfaces in Toric Varieties}\label{app:polytopes}

    Given a toric variety $\bP_\Sigma$ defined in terms of a fan $\Sigma$ (see, e.g., \cite{cox2011toric} for a detailed review of toric varieties), we can define a hypersurface as the vanishing locus of a section of a line bundle $\cL$. Given $\Sigma$, the divisor class of a line bundle $\mathcal{L}$ is specified as
     \be
    \label{eqn:DivisorClassL}
        D = c_1(\cL) = \sum_i a_i D_i,
    \ee 
    in terms of the toric divisors $D_i$ corresponding to the ray generators of the fan $\Sigma$.

    The relevance of polytopes is that they allow us to encode both a fan and a line bundle in terms of a single object. A generic polytope is defined as follows. 

    \bd[Polytope]
        Let $M_{\R}$ be some real vector space of dimension $d$. Consider some set of points $S \ss M_{\R}$. Then we can define a \textit{polytope} by the convex hull of the set $S$, i.e. 
        \be 
            \Delta = \text{Conv}(S) := \Big\{ \sum_i \l_i m_i \, \Big| \, \sum_i \l_i = 1, \, \forall m_i \in S, \text{ and }  \l_i \in \R^+_0 \Big\} \subseteq M_{\R}. 
        \ee 
        The dimension of the polytope is equal to the dimensional of the smallest affine subspace in $M_{\R}$ that contains $\Delta$, and we will take this dimensional to be equal to $d$. Polytopes can always be made top-dimensional, i.e. $\dim\Delta = \dim M_{\R}$, simply by appropriately reducing the dimension of $M$. We will furthermore focus on the cases where $M_{\R} = M\otimes \R$ for some lattice $M$. We then call a polytope $\Delta \subseteq M_{\R}$ a \textit{lattice polytope} if the vertices of $\Delta$ are lattice points in $M$. 
    \ed

    \bd[Polytope Face]
        Let $(N_{\R}, M_{\R})$ be a set of dual vector spaces and let $\Delta \subseteq M_{\R}$ be a polytope. Then given a non-zero vector $v \in N_{\R}$ and an $a \in \R$, we can define 
        \be 
            H_{v,a} := \{ m \in M_{\R} \, | \, \la m , v \ra = a\} \qand H_{v,a}^+ := \{ m \in M_{\R} \, | \, \la m , v \ra \geq a\}.
        \ee
        $H_{v,a}$ is clearly a hypersurface in $M_{\R}$, and $H_{v,a}^+$ is the upper half plane associated to this hypersurface. We call a subset $\Theta \subseteq \Delta$ a \textit{face} of $\Delta$ if there exists a $H_{v,b}$ and $H_{v,b}^+$ such that 
        \be 
            \Theta = H_{v,a} \cap \Delta, \qand \Delta \subseteq H_{v,a}^+. 
        \ee 
        We will denote a dimension $k$ face by $\Theta^{[k]}$. 
    \ed 

    This definition is intuitively clear: consider some hypersurface in $M_{\R}$ that ``touches" $\Delta$, and then the intersection of this hypersurface with $\Delta$ is a face of $\Delta$. We call a face of codimension-$1$ a \textit{facet}, a face of dimension $1$ an edge and a face of dimension $0$ a vertex. Note we can think of a polytope as the convex hull of its vertices. We give a pictorial example of this for a $2$D polytope corresponding to a triangle below.

    \begin{center}
        \btik 
            \draw[thick, fill=blue!30,opacity=0.8] (0,0) -- (2,0) -- (1,1) -- (0,0);
            \draw[thick, blue] (-0.5,-0.5) -- (1.5,1.5) node [right] {$H_{v,a}$};
            \draw[ultra thick, red] (0,0) -- (1,1) node [midway, above] {$\Theta^{[1]} \quad$}; 
            \node at (1,0.5) {$\Delta$};
        \etik 
    \end{center}
    
    Note that a polytope $\Delta$ is given precisely by the intersection of the a finite number of half planes $H_{v_i,a_i}^+$, i.e. 
    \be 
        \Delta = \bigcap_{i=1}^{\ell} H_{v_i,a_i}^+
    \ee 
    is a polytope. It then follows from the definition of $H_{v_i,a_i}$ that the vectors $v_i$ are perpendicular to surfaces $H_{v_i,a_i}$ and point into the intersection, as this is exactly what we need to ensure that $\la m , v_i \ra \geq a_i$ for all $m \in \Delta$. We give a pictorial example for a $2$D polytope with $\ell=4$ below. 
    
    \begin{center}
        \btik
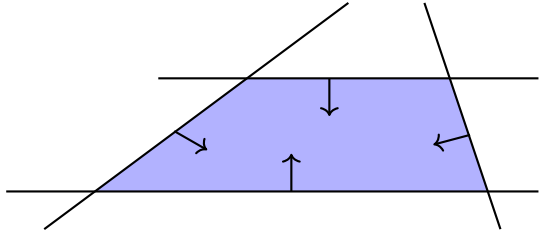
            \begin{scope}
                \clip (-2.5,-1.5) -- (4.5,-1.5) -- (4.5,0) -- (-0.5,0) -- (-2.5,-1.5);
                \fill[blue!30,opacity=0.8] (-2,-2) -- (4,-2) -- (3,1) -- (2,1) -- (-2,-2);
            \end{scope}
            \draw[thick] (-2,-2) -- (2,1);
            \draw[thick] (-2.5,-1.5) -- (4.5,-1.5);
            \draw[thick] (-0.5,0) -- (4.5,0);
            \draw[thick] (4,-2) -- (3,1);
            \draw[thick, ->] (1.25,-1.5) -- (1.25,-1);
            \draw[thick, ->] (1.75,0) -- (1.75,-0.5);
            \draw[thick, ->, rotate around={-30:(-0.29,-0.7)}] (-0.29,-0.7) -- (0.21,-0.7);
            \draw[thick, ->, rotate around={15:(3.6,-0.75)}] (3.6,-0.75) -- (3.1,-0.75);
        \etik 
    \end{center}

    We now note that for a lattice polytope, in which case $\dim\Delta = \dim M_{\R} = d$, each facet $\Theta^{[d-1]}$ has a unique supporting hyperplane. It follows from this that there is a unique choice such that $v_{\Theta}^{[d-1]}$ is a primitive lattice point in $N$, which implies that $a_{\Theta}^{[d-1]}$ is an integer. 
    \be 
        H_{v_{\Theta}^{[d-1]},a_{\Theta}^{[d-1]}} = \{ m \in M_{\R} \, | \, \la m , v_{\Theta}^{[d-1]} \ra = - a_{\Theta}^{[d-1]} \} \qand H_{\Theta^{[d-1]}}^+ = \{ m \in M_{\R} \, | \, \la m , v_{\Theta}^{[d-1]} \ra \geq  - a_{\Theta}^{[d-1]} \}\, .
    \ee 
    Using these, our polytope is given by 
    \be 
    \label{eqn:DeltaFacets} 
        \Delta = \bigcap_{\{\Theta^{[d-1]}\}} H_{\Theta^{[d-1]}}^+ = \{ m \in M_{\R} \, | \, \la m, v_{\Theta}^{[d-1]}\ra \geq -a_{\Theta}^{[d-1]}, \, \forall \text{ facets } \Theta^{[d-1]} \subset \Delta\}, 
    \ee
    The minus sign appearing in the above expressions is included for later convenience. 

    The polytope defines a fan, which in turn can be used to define a toric variety. 
    \bd[Normal Fan]
        To any face $\Theta$ of a polytope $\Delta$, we can associate a cone
        \be 
            \check{\sig}_{\Theta} := \bigcup_{r \geq 0, p_{\Delta} \in \Delta, p_{\Theta} \in \Theta} r \cdot (p_{\Delta} - p_{\Theta}),
        \ee 
     and its dual 
       \be 
       \langle\check{\sig}_{\Theta}, \sig_{\Theta} \rangle \geq 0 \, .
       \ee
    The collection of the cones $\sigma_\Theta$ for all faces $\Theta$ defines a complete fan\footnote{Here, it is customary to consider the whole of 
    $\Delta$ as a face of itself with $\sigma_\Delta$ being the zero-dimensional cone. } called the normal fan $\Sigma_n(\Delta)$ of $\Delta$. 
    \ed 
    Here, a $k$-dimensional face $\Theta^{[k]}$ of $\Delta$ is associated to a $(d-k)$-dimensional cone in $\Sigma_n(\Delta)$. In particular, the facets in $\Delta$ correspond to the rays in $\Sig_{n}(\Delta)$. By an appropriate translation of $\Delta$, the ray generators of this fan become equal to the lattice points $v_{\Theta}^{[d-1]}$. 
    
    We can then use our polytope to define not only a fan $\Sigma_n(\Delta)$, but also the divisor class of a line bundle $\mathcal{L}$ on the associated toric 
    variety $\mathbb{P}_{\Sigma_n(\Delta)}$. This construction relies on the piecewise linear support function on $\Sigma_n(\Delta)$ that is defined by $\Delta$, see \cite{fulton} for details, 
    and implies that we can identify the $a_{\Theta}^{[d-1]}$ in \eqref{eqn:DeltaFacets} with the $a_i$ in \eqref{eqn:DivisorClassL}. 
    
    The Newton polytope describing of global holomorphic sections of $\mathcal{L}$ is then equal to $\Delta$, and the group of holomorphic sections in $\cL$ has monomial basis the elements of which are in one-to-one correspondence with lattice points on $\Delta$. They can be given explicitly as 
    \be 
        p(m) = \prod_i z_i^{\langle m,\nu_i\rangle  + a_i}.
    \ee 
   where $z_i$ are the homogeneous coordinates associated with ray generators $\nu_i$ of $\Sigma_n(\Delta)$. Note that holomorphicity of these sections is guaranteed by the relations $\la m, v_{\Theta}^{[d-1]}\ra \geq -a_{\Theta}^{[d-1]}$. 
    
    The normal fan $\Sigma_n(\Delta)$ does not define a smooth or even orbifold toric variety $\mathbb{P}_{\Sigma_n(\Delta)}$ in general, and we may need to refine it to fan 
    \begin{equation}
     \widetilde{\Sigma}_n(\Delta) \rightarrow \Sigma_n(\Delta)
    \end{equation}
    to resolve singularities. 
    
    \subsection{Calabi-Yau hypersurfaces and Reflexive Polytopes}

    To get a Calabi-Yau hypersurface, adjunction tells us that we need to choose $\cL$ such that its divisor class equals the first Chern class of $\bP_\Sigma$:
    \be
    \label{eqn:DivisorClassL_CY}
        D = c_1(\cL) = \sum_i D_i,
    \ee 
    i.e. $a_i=1$ for all $i$ in \eqref{eqn:DivisorClassL}. For the above construction to work we hence need the polar dual to $\Delta^\circ$ defined by 
    \begin{equation}\label{eq:polar_duals}
     \la \Delta,\Delta^\circ \ra \geq -1
    \end{equation}
    to be a lattice polytope. 
    \bd[Reflexive Polytope]
    A lattice polytope $\Delta$ is called reflexive if its polar dual is also a lattice polytope.
    \ed
    This definition implies that $\Delta^\circ$ is reflexive as well, and that $\Delta^\circ$ and $\Delta$ both have the origin as their unique interior point. Furthermore, one can show that $\Sigma_n(\Delta)$ is equal to the face fan\footnote{The face fan of a polytope $\Delta^\circ$ is the fan found by taking all cones over the faces of $\Delta^\circ$.} $\Sigma_f(\Delta^\circ)$ of $\Delta^\circ$ for a pair of reflexive polytopes. The faces of $\Delta$ and $\Delta^{\circ}$ are related to each other by 
    \be 
        \langle \Theta^{[k]}, \Theta^{\circ [n-k-1]}\rangle = -1,
    \ee 
    where $n$ is the dimension of the polytopes. The important thing is that a $k$-dimensional face in $\Delta$ is related to a $(n-k-1)$-dimensional face in $\Delta^{\circ}$. 
    
    Refinements of the fan $\Sigma_n(\Delta)$ which introduce no new ray generators or ray generators which are lattice points on $\Delta^\circ$ correspond to crepant (partial) resolutions of the hypersurface in the class \eqref{eqn:DivisorClassL_CY}. For a maximal regular\footnote{Regularity of a triangulation is implies projectivity of the toric variety, see \cite{fulton,de2010triangulations} for details and definitions.} triangulation of the polytope we hence find what is called a maximal projective crepant partial (MPCP) desingularisation.   

    As we can use both $\Delta$ and $\Delta^\circ$ as the starting point of this construction, we obtain a pair of Calabi-Yau hypersurfaces in toric varieties with fans $\widetilde{\Sigma}_n(\Delta)$ and $\widetilde{\Sigma}_n(\Delta^\circ)$. The two hypersurface equations are explicitly given as 
    \be 
        \begin{split}
            G(z) & = \sum_{m \in \Delta } \prod_{n \in \Delta^\circ} \a_m z_n^{\langle m, n\rangle +1} = 0 \\
            G^\circ(z^\circ) & = \sum_{n \in \Delta^\circ } \prod_{m \in \Delta} \a_n^\circ z_m^{\circ\,\,\langle m, n\rangle +1}    = 0     \end{split}
    \ee    
    Note that the ray generators $\widetilde{\Sigma}_n(\Delta)(1)$ of $\widetilde{\Sigma}_n(\Delta)$ are equal to the lattice points on $\Delta^\circ$, and the ray generators $\widetilde{\Sigma}_n(\Delta^\circ)(1)$ of $\widetilde{\Sigma}_n(\Delta^\circ)$ are equal to the lattice points on $\Delta$ using MPCP triangulations. We have labelled the homogeneous coordinates associated with a lattice point $n$ ($m$) by $z_n$ ($z_{m}^\circ$). 
    
    In case the MPCP triangulation results in a smooth hypersurface, we will denote a generic member of the resulting family by $X_{\Delta,\Delta^\circ}$.   
    The non-trivial Hodge numbers of $X_{\Delta,\Delta^\circ}$ are \cite{Batyrev94dualpolyhedra}
    \be 
        \begin{split}
            h^{1,1}(X_{\Delta,\Delta^\circ}) & = \ell(\Delta^{\circ}) - (d+1) - \sum_{\Theta^{\circ [n-1]}} \ell^*\left( \Theta^{\circ [d-1]}\right) + \sum_{(\Theta^{\circ [d-2]},\Theta^{[1]})} \ell^*\left( \Theta^{\circ [d-2]}\right) \ell^* \left(\Theta^{[1]}\right) \\
            h^{d-2,1}(X_{\Delta,\Delta^\circ}) & = \ell(\Delta) - (n+1) - \sum_{\Theta^{[d-1]}} \ell^*\left( \Theta^{[d-1]}\right) + \sum_{(\Theta^{\circ [d-2]},\Theta^{[1]})} \ell^*\left( \Theta^{[d-2]}\right) \ell^* \left(\Theta^{\circ [1]}\right) \\
            h^{m,1}(X_{\Delta,\Delta^\circ}) & =  \sum_{(\Theta^{\circ [d-m-1]},\Theta^{[m]})} \ell^*\left( \Theta^{\circ [d-m-1]}\right) \ell^* \left(\Theta^{[m]}\right) \qquad \text{for} \qquad d-2 > m > 1.
        \end{split}
    \ee 
    Here $\ell(...)$ denotes the number of lattice points of its argument, $\ell^*(...)$ only counts the lattice points in the relative interior of its argument,
    and $d-1$ is the complex dimension of the Calabi-Yau hypersurface. 
    
    Exchanging $\Delta \leftrightarrow \Delta^{\circ}$ maps the Hodge numbers as expected for mirror symmetry, i.e. $h^{1,1} \leftrightarrow h^{d-2,1}$. In our main case of interest, which are Calabi-Yau threefolds, we have $d=4$, so that mirror symmetry swaps $h^{1,1} \leftrightarrow h^{2,1}$. In this case, there always exists a MPCP that gives rise to a smooth Calabi-Yau hypersurface \cite{Batyrev94dualpolyhedra}, so we do not have to explicitly check the existence of such a triangulation. 

\subsection{K3 fibrations, Projecting Tops and Building Blocks}\label{app:proj_tops}
    
    Whenever the $N$-lattice reflexive polytope $\Delta^\circ$ contains a reflexive subpolytope, i.e. there is a sublattice $N_F \in N$ such 
    that $\Delta^\circ \cap N_F= \Delta_F^\circ$ with $\Delta_F^\circ$ reflexive, the associated Calabi-Yau hypersurface $X_{\Delta,\Delta^\circ}$
    may admit a fibration with fibres $X_{\Delta_F,\Delta^\circ_F}$ \cite{Klemm:1995tj,Avram:1996pj,Candelas:1996su}. The projection of $N$ to 
    $N/N_f$ gives rise to a projection of the hypersurface if there is an appropriate triangulation of $\Delta^\circ$ that turns this into a toric morphism. 
    
    Our case of interest are K3 fibrations of Calabi-Yau threefolds, where $\Delta^\circ$ is four-dimensional and $\Delta^\circ_F$ is three-dimensional. In this case $\Delta^\circ_F$ separates $\Delta^\circ$ into two halves $\Diamond_1^\circ$ and $\Diamond_2^\circ$ called tops (or top and bottom). Letting $m_0$ be the primitive normal vector to $N_F$, $\langle m_0, N_F \rangle = 0$, we can set 
    \begin{equation}
    \begin{aligned}
     \Diamond_1^\circ &:= \text{Conv}\left(\left\{ n \in \Delta^\circ | \langle m_0,n \rangle \geq 0 \right\}\right) \\
     \Diamond_2^\circ &:= \text{Conv}\left(\left\{ n \in \Delta^\circ | \langle m_0,n \rangle \leq 0 \right\}\right) \, 
         \end{aligned}
    \end{equation}
    so that
    \begin{equation}
     \Delta^\circ = \Diamond_1^\circ \cup \Diamond_2^\circ \hspace{1cm} \Delta^\circ_F = \Diamond_1^\circ \cap \Diamond_2^\circ \, . 
    \end{equation}
    Motivated by the above, we will adopt the following
    \bd[Top]
    A top is a lattice polytope $\Diamond^\circ$ such that 
    \begin{equation}
    \Diamond^\circ =   \text{Conv}\left(\left\{ n \in \Delta^\circ | \langle m_0,n \rangle \geq 0 \right\}\right) 
    \end{equation}
    for a reflexive polytope $\Delta^\circ$ and primitive lattice point $m_0$, and 
    \begin{equation}
     \Delta^\circ_F := \text{Conv}\left(\left\{ n \in \Delta^\circ | \langle m_0,n \rangle = 0 \right\}\right) 
    \end{equation}
    is reflexive. We will make the simple choice $m_0 = (0,0,0,1)$ by exploiting the $SL(4,\Z)$ acting on $N$ in the following. 
    \ed 
    
    \bd[Projecting Top]
    A top $\Diamond^\circ$ is called projecting if the projection of $\Diamond^\circ$ to $N_{F}\otimes\R$ is contained in $\Delta_F^\circ$. 
    \ed
    
    Projecting tops can be used to construct building blocks for TCS $G_2$ manifolds in analogy to Batyrev's construction of Calabi-Yau threefolds \cite{Braun:2016igl}. 
    Given a projecting top, we can define its dual as 
    \begin{equation}
     \langle \Diamond, \Diamond^\circ \rangle \geq -1 \hspace{1cm} \langle \Diamond, n_0 \rangle \geq 0 \, .
    \end{equation}
    where $n_0 = (0,0,0,-1)$. Using $\Diamond \in M_{\R}$, in the construction outlined above in \Cref{app:polytopes} results in a compact hypersurface in the toric variety $\mathbb{P}_{\Sigma_n(\Diamond)}$ which in general is not smooth. A fan refinement
    \begin{equation}
     \widetilde{\Sigma}_n(\Diamond) \rightarrow \Sigma_n(\Diamond)\, .
    \end{equation}
    for which all rays introduced have generators which are lattice points on $\Diamond$ gives rise to a crepant partial desingularisation. The associated 
    MPCP (again for a regular triangulation) then defines a smooth hypersurface which we denote by $Z_{\Diamond, \Diamond^\circ}$. The defining equation of 
    $Z_{\Diamond, \Diamond^\circ}$ is 
    \begin{equation}
    F(z) = \sum_{m \in \left(\Diamond \cup (0,0,0,1)\right)} \alpha_m z_0^{\langle m, n_0 \rangle} \prod_{n \in \Diamond^\circ} z_n^{\langle m, n \rangle + 1} = 0\, .
    \end{equation}
    Here $z_i$ are the homogeneous coordinates associated with the ray generators $n_i \in \Diamond^\circ$, note that $n_0$ is always a ray generator of $\Sigma_n(\Diamond)$.  
    
    The hypersurface $Z_{\Diamond, \Diamond^\circ}$ admits a K3 fibration with base $\mathbb{P}^1$ such that 
    \begin{equation}
     c_1(Z_{\Diamond, \Diamond^\circ}) = [S_0]
    \end{equation}
    where $[S_0]$ is the cohomology class dual to the divisor class of a generic K3 fibre, i.e. $Z_{\Diamond, \Diamond^\circ}$ is a building block and $X_{\Diamond, \Diamond^\circ} = Z_{\Diamond, \Diamond^\circ}\setminus S_0$ is an asymptotically cylindrical Calabi-Yau threefold. 
    
    The topological data of $Z_{\Diamond, \Diamond^\circ}$ required for our purposes can be described by combinators \cite{Braun:2016igl}. Denoting $k$-dimensional faces of $\Diamond$ by $\Theta^{[k]}$, the Hodge numbers of $Z_{\Diamond, \Diamond^\circ}$ are $h^{i,0}(Z_{\Diamond, \Diamond^\circ}) = 0$ for all $i>0$ and 
    \be
    \begin{split}
     h^{1,1}(Z_{\Diamond, \Diamond^\circ}) &= -4 + \sum_{\Theta^{[3]}} 1 + \sum_{\Theta^{[2]}} \ell^*(\sigma_n(\Theta^{[2]}))
     + \sum_{\Theta^{[1]}} \ell^*(\Theta^{[1]}) \ell^*(\sigma_n(\Theta^{[1]}))\\
     h^{2,1}(Z_{\Diamond, \Diamond^\circ}) &= \ell(\Diamond) - \ell(\Delta_F) + \sum_{\Theta^{[2]}} \ell^*(\Theta^{[2]})\ell^*(\sigma_n(\Theta^{[2]}))
     - \sum_{\Theta^{[3]}} \ell^*(\Theta^{[3]})
    \end{split}
\ee
    where $\ell^*(\sigma_n(\Theta^{[k]}))$ counts lattice points on $\Diamond^\circ$ in the relative interior of the normal cone to $\Theta^{[k]}$, and $\ell$ and $\ell^*$ of polytopes/faces are defined as before.    
    
    The ranks of the lattices $N$ and $K$ defined for building blocks in \eqref{eq:NK_buildingblocks} are 
    \begin{equation}
     \begin{aligned}
      |N(Z_{\Diamond, \Diamond^\circ}) | &= \ell^1 (\Delta_F) - 3 + \sum_{ve \,\,\Theta_F^{\circ [1]}} \ell^*(\Theta_F^{\circ [1]}) \ell^*(\Theta_F^{ [1]})\\
      |K(Z_{\Diamond, \Diamond^\circ}) | &= h^{1,1}(Z_{\Diamond, \Diamond^\circ}) - |N| -1  
     \end{aligned}
    \end{equation}
    where $\ell^1()$ counts points on the one-skeleton, and $ve \,\,\Theta_F^{[1]}$ denotes only those one-dimensional faces of $\Delta$ which are bounding a face of $\Diamond^\circ$ that is `vertical', i.e. parallel to $n_0$. 
    
    As for reflexive polyhedra, one may interchange the roles played by $\Diamond$ and $\Diamond^\circ$ which results in another building block 
    $Z_{\Diamond^\circ, \Diamond}$ that satisfies \cite{Braun:2017ryx}
    \be
    \begin{split}
     h^{2,1}(Z_{\Diamond, \Diamond^\circ}) &= |K(Z_{\Diamond^\circ, \Diamond}) | \\
     h^{2,1}(Z_{\Diamond^\circ, \Diamond}) &= |K(Z_{\Diamond, \Diamond^\circ}) | \, .
     \end{split}
     \ee
     Furthermore, the lattice $N(Z_{\Diamond, \Diamond^\circ})$ and $N(Z_{\Diamond^\circ, \Diamond})$ admit a primitive embedding
    \begin{equation}
     N(Z_{\Diamond, \Diamond^\circ}) \oplus N(Z_{\Diamond^\circ, \Diamond}) \oplus U \hookrightarrow \Gamma^{3,19}
    \end{equation}
    for $\Gamma^{3,19} = H^2(S_0,\mathbb{Z})$ the unique even self-dual lattice of signature $(3,19)$. This implies that the K3 fibres of 
    $Z_{\Diamond, \Diamond^\circ}$ and $Z_{\Diamond^\circ, \Diamond}$ are from algebraic mirror families \cite{Aspinwall:1994rg}. 
    The above relations play a crucial role in the construction of mirror $G_2$ manifolds of TCS type.

    Any two projecting tops $\Diamond_1^\circ$ and $\Diamond_2^\circ$ for which $\Delta^\circ_{1F}= \Delta^\circ_{2F}$ 
    can be joined to create a reflexive polytope $\Delta^\circ_{12}$ \cite{Candelas:2012uu}. A large fraction of the polytopes in the Kreuzer-Skarke list 
    are of this type, and as their Hodge numbers can be understood from this decomposition as well, a number of patterns in the plot of Hodge numbers can be explained by this. Conversely, 
    given a reflexive polytope $\Delta^\circ$ that can be decomposed into two projecting tops, the Calabi-Yau threefold $X_{\Delta,\Delta^\circ}$ admits a stable degeneration limit in which it becomes reducible into the two building blocks $Z_{\Diamond, \Diamond^\circ}$ and $Z_{\Diamond^\circ, \Diamond}$ \cite{Braun:2017ryx}. This limit can be understood as stretching the base $\mathbb{P}^1$ of the K3 fibration on $X_{\Delta,\Delta^\circ}$, separating the singular K3 fibres to its two ends. Cutting along the stretched base along the middle then decomposes $X_{\Delta,\Delta^\circ}$ into the asymptotically cylindrical threefolds $X_{\Diamond_1, \Diamond^\circ_1}$ and $X_{\Diamond_2, \Diamond^\circ_2}$.

\providecommand{\href}[2]{#2}\begingroup\raggedright\endgroup


\begin{thebibliography}{10}

\bibitem{Dixon:1988}
L.~Dixon, \emph{{Proceedings, Summer Workshop in High-energy Physics and
  Cosmology: Superstrings, Unified Theories and Cosmology}: {Trieste, Italy,
  June 29-August 7, 1987}},  (Singapore, Singapore), World Scientific, 1988.

\bibitem{lerche1989chiral}
W.~Lerche, C.~Vafa and N.P.~Warner, \emph{Chiral rings in n= 2 superconformal
  theories}, {\emph{Nuclear Physics B} {\bfseries 324} (1989) 427}.

\bibitem{Candelas:1989hd}
P.~Candelas, M.~Lynker and R.~Schimmrigk, \emph{{Calabi-Yau Manifolds in
  Weighted P(4)}},
  \href{https://doi.org/10.1016/0550-3213(90)90185-G}{\emph{Nucl. Phys. B}
  {\bfseries 341} (1990) 383}.

\bibitem{Candelas:1990rm}
P.~Candelas, X.C.~De~La~Ossa, P.S.~Green and L.~Parkes, \emph{{A Pair of
  Calabi-Yau manifolds as an exactly soluble superconformal theory}},
  \href{https://doi.org/10.1016/0550-3213(91)90292-6}{\emph{Nucl. Phys. B}
  {\bfseries 359} (1991) 21}.

\bibitem{Kontsevich:1994dn}
M.~Kontsevich, \emph{{Homological Algebra of Mirror Symmetry}},
  \href{https://arxiv.org/abs/alg-geom/9411018}{{\ttfamily alg-geom/9411018}}.

\bibitem{Batyrev94dualpolyhedra}
V.V.~Batyrev, \emph{{Dual polyhedra and mirror symmetry for Calabi-Yau
  hypersurfaces in toric varieties}}, {\emph{J. Alg. Geom} (1994) 493}
  [\href{https://arxiv.org/abs/alg-geom/9310003}{{\ttfamily
  alg-geom/9310003}}].

\bibitem{Hosono:1993qy}
S.~Hosono, A.~Klemm, S.~Theisen and S.-T.~Yau, \emph{{Mirror symmetry, mirror
  map and applications to Calabi-Yau hypersurfaces}},
  \href{https://doi.org/10.1007/BF02100589}{\emph{Commun. Math. Phys.}
  {\bfseries 167} (1995) 301}
  [\href{https://arxiv.org/abs/hep-th/9308122}{{\ttfamily hep-th/9308122}}].

\bibitem{greene1990duality}
B.R.~Greene and M.R.~Plesser, \emph{Duality in calabi-yau moduli space},
  {\emph{Nuclear Physics B} {\bfseries 338} (1990) 15}.

\bibitem{Aspinwall:1990xe}
P.S.~Aspinwall, C.A.~Lutken and G.G.~Ross, \emph{{Construction and Couplings of
  Mirror Manifolds}},
  \href{https://doi.org/10.1016/0370-2693(90)91659-Y}{\emph{Phys. Lett. B}
  {\bfseries 241} (1990) 373}.

\bibitem{hori2000mirror}
K.~Hori and C.~Vafa, \emph{{Mirror symmetry}},
  \href{https://arxiv.org/abs/hep-th/0002222}{{\ttfamily hep-th/0002222}}.

\bibitem{Gepner:1987vz}
D.~Gepner, \emph{{Exactly Solvable String Compactifications on Manifolds of
  SU(N) Holonomy}},
  \href{https://doi.org/10.1016/0370-2693(87)90938-5}{\emph{Phys. Lett. B}
  {\bfseries 199} (1987) 380}.

\bibitem{gepner1989space}
D.~Gepner, \emph{Space-time supersymmetry in compactified string theory and
  superconformal models},  in \emph{Current Physics--Sources and Comments},
  vol.~4, pp.~381--402, Elsevier (1989).

\bibitem{Vafa:1988uu}
C.~Vafa and N.P.~Warner, \emph{{Catastrophes and the Classification of
  Conformal Theories}},
  \href{https://doi.org/10.1016/0370-2693(89)90473-5}{\emph{Phys. Lett. B}
  {\bfseries 218} (1989) 51}.

\bibitem{Greene:1988ut}
B.R.~Greene, C.~Vafa and N.P.~Warner, \emph{{Calabi-Yau Manifolds and
  Renormalization Group Flows}},
  \href{https://doi.org/10.1016/0550-3213(89)90471-9}{\emph{Nucl. Phys. B}
  {\bfseries 324} (1989) 371}.

\bibitem{Cecotti:1989jc}
S.~Cecotti, L.~Girardello and A.~Pasquinucci, \emph{{Nonperturbative Aspects
  and Exact Results for the $N=2$ Landau-ginzburg Models}},
  \href{https://doi.org/10.1016/0550-3213(89)90226-5}{\emph{Nucl. Phys. B}
  {\bfseries 328} (1989) 701}.

\bibitem{Cecotti:1990kz}
S.~Cecotti, \emph{{N=2 Landau-Ginzburg versus Calabi-Yau sigma models:
  Nonperturbative aspects}},
  \href{https://doi.org/10.1142/S0217751X91000939}{\emph{Int. J. Mod. Phys. A}
  {\bfseries 6} (1991) 1749}.

\bibitem{Witten:1993yc}
E.~Witten, \emph{{Phases of N=2 theories in two-dimensions}},
  \href{https://doi.org/10.1016/0550-3213(93)90033-L}{\emph{Nucl. Phys. B}
  {\bfseries 403} (1993) 159}
  [\href{https://arxiv.org/abs/hep-th/9301042}{{\ttfamily hep-th/9301042}}].

\bibitem{Strominger:1996it}
A.~Strominger, S.-T.~Yau and E.~Zaslow, \emph{{Mirror symmetry is T duality}},
  \href{https://doi.org/10.1016/0550-3213(96)00434-8}{\emph{Nucl. Phys. B}
  {\bfseries 479} (1996) 243}
  [\href{https://arxiv.org/abs/hep-th/9606040}{{\ttfamily hep-th/9606040}}].

\bibitem{vafa1995orbifolds}
C.~Vafa and E.~Witten, \emph{{On orbifolds with discrete torsion}},
  \href{https://doi.org/10.1016/0393-0440(94)00048-9}{\emph{J. Geom. Phys.}
  {\bfseries 15} (1995) 189}
  [\href{https://arxiv.org/abs/hep-th/9409188}{{\ttfamily hep-th/9409188}}].

\bibitem{Shatashvili:1994zw}
S.L.~Shatashvili and C.~Vafa, \emph{{Superstrings and manifold of exceptional
  holonomy}}, \href{https://doi.org/10.1007/BF01671569}{\emph{Selecta Math.}
  {\bfseries 1} (1995) 347}
  [\href{https://arxiv.org/abs/hep-th/9407025}{{\ttfamily hep-th/9407025}}].

\bibitem{joyce1996compactI}
D.D.~Joyce, \emph{Compact riemannian 7-manifolds with holonomy $ g\_2 $. i},
  {\emph{Journal of differential geometry} {\bfseries 43} (1996) 291}.

\bibitem{joyce1996compactII}
D.D.~Joyce, \emph{Compact riemannian 7-manifolds with holonomy g2. ii},
  {\emph{J. Diff. Geom} {\bfseries 43} (1996) 329}.

\bibitem{Acharya:1997rh}
B.S.~Acharya, \emph{{On mirror symmetry for manifolds of exceptional
  holonomy}}, \href{https://doi.org/10.1016/S0550-3213(98)00140-0}{\emph{Nucl.
  Phys. B} {\bfseries 524} (1998) 269}
  [\href{https://arxiv.org/abs/hep-th/9707186}{{\ttfamily hep-th/9707186}}].

\bibitem{gaberdiel2004generalised}
M.R.~Gaberdiel and P.~Kaste, \emph{{Generalized discrete torsion and mirror
  symmetry for g(2) manifolds}},
  \href{https://doi.org/10.1088/1126-6708/2004/08/001}{\emph{JHEP} {\bfseries
  08} (2004) 001} [\href{https://arxiv.org/abs/hep-th/0401125}{{\ttfamily
  hep-th/0401125}}].

\bibitem{eguchi2002string}
T.~Eguchi and Y.~Sugawara, \emph{{String theory on G(2) manifolds based on
  Gepner construction}},
  \href{https://doi.org/10.1016/S0550-3213(02)00187-6}{\emph{Nucl. Phys. B}
  {\bfseries 630} (2002) 132}
  [\href{https://arxiv.org/abs/hep-th/0111012}{{\ttfamily hep-th/0111012}}].

\bibitem{blumenhagen2002superconformal}
R.~Blumenhagen and V.~Braun, \emph{{Superconformal field theories for compact
  G(2) manifolds}},
  \href{https://doi.org/10.1088/1126-6708/2001/12/006}{\emph{JHEP} {\bfseries
  12} (2001) 006} [\href{https://arxiv.org/abs/hep-th/0110232}{{\ttfamily
  hep-th/0110232}}].

\bibitem{roiban2002rational}
R.~Roiban and J.~Walcher, \emph{{Rational conformal field theories with G(2)
  holonomy}}, \href{https://doi.org/10.1088/1126-6708/2001/12/008}{\emph{JHEP}
  {\bfseries 12} (2001) 008}
  [\href{https://arxiv.org/abs/hep-th/0110302}{{\ttfamily hep-th/0110302}}].

\bibitem{Roiban:2002iv}
R.~Roiban, C.~Romelsberger and J.~Walcher, \emph{{Discrete torsion in singular
  G(2) manifolds and real LG}},
  \href{https://doi.org/10.4310/ATMP.2002.v6.n2.a2}{\emph{Adv. Theor. Math.
  Phys.} {\bfseries 6} (2003) 207}
  [\href{https://arxiv.org/abs/hep-th/0203272}{{\ttfamily hep-th/0203272}}].

\bibitem{Harvey:1999as}
J.A.~Harvey and G.W.~Moore, \emph{{Superpotentials and membrane instantons}},
  \href{https://arxiv.org/abs/hep-th/9907026}{{\ttfamily hep-th/9907026}}.

\bibitem{Partouche:2000uq}
H.~Partouche and B.~Pioline, \emph{{Rolling among G(2) vacua}},
  \href{https://doi.org/10.1088/1126-6708/2001/03/005}{\emph{JHEP} {\bfseries
  03} (2001) 005} [\href{https://arxiv.org/abs/hep-th/0011130}{{\ttfamily
  hep-th/0011130}}].

\bibitem{Braun:2017ryx}
A.P.~Braun and M.~Del~Zotto, \emph{{Mirror Symmetry for $G_2$-Manifolds:
  Twisted Connected Sums and Dual Tops}},
  \href{https://doi.org/10.1007/JHEP05(2017)080}{\emph{JHEP} {\bfseries 05}
  (2017) 080} [\href{https://arxiv.org/abs/1701.05202}{{\ttfamily
  1701.05202}}].

\bibitem{Braun:2017csz}
A.P.~Braun and M.~Del~Zotto, \emph{{Towards Generalized Mirror Symmetry for
  Twisted Connected Sum $G_2$ Manifolds}},
  \href{https://doi.org/10.1007/JHEP03(2018)082}{\emph{JHEP} {\bfseries 03}
  (2018) 082} [\href{https://arxiv.org/abs/1712.06571}{{\ttfamily
  1712.06571}}].

\bibitem{Aganagic:2001ug}
M.~Aganagic and C.~Vafa, \emph{{G(2) manifolds, mirror symmetry and geometric
  engineering}},  \href{https://arxiv.org/abs/hep-th/0110171}{{\ttfamily
  hep-th/0110171}}.

\bibitem{Braun:2019wnj}
A.P.~Braun, \emph{{M-Theory and Orientifolds}},
  \href{https://doi.org/10.1007/JHEP09(2020)065}{\emph{JHEP} {\bfseries 09}
  (2020) 065} [\href{https://arxiv.org/abs/1912.06072}{{\ttfamily
  1912.06072}}].

\bibitem{joyce2000compact}
D.~Joyce, \emph{Compact Manifolds with Special Holonomy}, Oxford mathematical
  monographs, Oxford University Press (2000).

\bibitem{Braun:2019lnn}
A.P.~Braun, S.~Majumder and A.~Otto, \emph{{On Mirror Maps for Manifolds of
  Exceptional Holonomy}},
  \href{https://doi.org/10.1007/JHEP10(2019)204}{\emph{JHEP} {\bfseries 10}
  (2019) 204} [\href{https://arxiv.org/abs/1905.01474}{{\ttfamily
  1905.01474}}].

\bibitem{Grimm:2004ua}
T.W.~Grimm and J.~Louis, \emph{{The Effective action of type IIA Calabi-Yau
  orientifolds}},
  \href{https://doi.org/10.1016/j.nuclphysb.2005.04.007}{\emph{Nucl. Phys. B}
  {\bfseries 718} (2005) 153}
  [\href{https://arxiv.org/abs/hep-th/0412277}{{\ttfamily hep-th/0412277}}].

\bibitem{Grigorian:2009nx}
S.~Grigorian, \emph{{Betti numbers of a class of barely G(2) manifolds}},
  \href{https://doi.org/10.1007/s00220-010-1152-2}{\emph{Commun. Math. Phys.}
  {\bfseries 301} (2011) 215}
  [\href{https://arxiv.org/abs/0909.4681}{{\ttfamily 0909.4681}}].

\bibitem{2017arXiv170709325J}
D.~{Joyce} and S.~{Karigiannis}, \emph{{A new construction of compact
  torsion-free $G_2$-manifolds by gluing families of Eguchi-Hanson spaces}},
  \href{https://doi.org/10.48550/arXiv.1707.09325}{\emph{arXiv e-prints} (2017)
  arXiv:1707.09325} [\href{https://arxiv.org/abs/1707.09325}{{\ttfamily
  1707.09325}}].

\bibitem{MR2024648}
A.~Kovalev, \emph{Twisted connected sums and special {R}iemannian holonomy},
  \href{https://doi.org/10.1515/crll.2003.097}{\emph{J. Reine Angew. Math.}
  {\bfseries 565} (2003) 125}.

\bibitem{MR3109862}
A.~Corti, M.~Haskins, J.~Nordstr{\"o}m and T.~Pacini, \emph{Asymptotically
  cylindrical {C}alabi-{Y}au 3-folds from weak {F}ano 3-folds},
  \href{https://doi.org/10.2140/gt.2013.17.1955}{\emph{Geom. Topol.} {\bfseries
  17} (2013) 1955}.

\bibitem{Corti:2012kd}
A.~Corti, M.~Haskins, J.~Nordstr{\"o}m and T.~Pacini,
  \emph{{$\mathrm{G}_{2}$-manifolds and associative submanifolds via semi-Fano
  $3$-folds}}, \href{https://doi.org/10.1215/00127094-3120743}{\emph{Duke Math.
  J.} {\bfseries 164} (2015) 1971}
  [\href{https://arxiv.org/abs/1207.4470}{{\ttfamily 1207.4470}}].

\bibitem{Halverson:2014tya}
J.~Halverson and D.R.~Morrison, \emph{{The landscape of M-theory
  compactifications on seven-manifolds with G$_{2}$ holonomy}},
  \href{https://doi.org/10.1007/JHEP04(2015)047}{\emph{JHEP} {\bfseries 04}
  (2015) 047} [\href{https://arxiv.org/abs/1412.4123}{{\ttfamily 1412.4123}}].

\bibitem{Halverson:2015vta}
J.~Halverson and D.R.~Morrison, \emph{{On gauge enhancement and singular limits
  in G$_{2}$ compactifications of M-theory}},
  \href{https://doi.org/10.1007/JHEP04(2016)100}{\emph{JHEP} {\bfseries 04}
  (2016) 100} [\href{https://arxiv.org/abs/1507.05965}{{\ttfamily
  1507.05965}}].

\bibitem{Braun:2016igl}
A.P.~Braun, \emph{{Tops as building blocks for G$_{2}$ manifolds}},
  \href{https://doi.org/10.1007/JHEP10(2017)083}{\emph{JHEP} {\bfseries 10}
  (2017) 083} [\href{https://arxiv.org/abs/1602.03521}{{\ttfamily
  1602.03521}}].

\bibitem{Guio:2017zfn}
T.C.d.C.~Guio, H.~Jockers, A.~Klemm and H.-Y.~Yeh, \emph{{Effective action from
  M-theory on twisted connected sum $G_2$-manifolds}},
  \href{https://arxiv.org/abs/1702.05435}{{\ttfamily 1702.05435}}.

\bibitem{Braun:2017uku}
A.P.~Braun and S.~Schäfer-Nameki, \emph{{Compact, Singular $G_2$-Holonomy
  Manifolds and M/Heterotic/F-Theory Duality}},
  \href{https://doi.org/10.1007/JHEP04(2018)126}{\emph{JHEP} {\bfseries 04}
  (2018) 126} [\href{https://arxiv.org/abs/1708.07215}{{\ttfamily
  1708.07215}}].

\bibitem{Aspinwall:1994rg}
P.S.~Aspinwall and D.R.~Morrison, \emph{{String theory on K3 surfaces}},
  {\emph{AMS/IP Stud. Adv. Math.} {\bfseries 1} (1996) 703}
  [\href{https://arxiv.org/abs/hep-th/9404151}{{\ttfamily hep-th/9404151}}].

\bibitem{gross1999special}
M.~Gross, \emph{Special lagrangian fibrations ii: Geometry},  1999.

\bibitem{Morrison:2012np}
D.R.~Morrison and W.~Taylor, \emph{{Classifying bases for 6D F-theory models}},
  \href{https://doi.org/10.2478/s11534-012-0065-4}{\emph{Central Eur. J. Phys.}
  {\bfseries 10} (2012) 1072}
  [\href{https://arxiv.org/abs/1201.1943}{{\ttfamily 1201.1943}}].

\bibitem{nikulin1976finite}
V.V.~Nikulin, \emph{Finite groups of automorphisms of k\"ahlerian surfaces of
  type k3}, {\emph{Uspekhi Matematicheskikh Nauk} {\bfseries 31} (1976) 223}.

\bibitem{0025-5726-14-1-A06}
V.V.~Nikulin, \emph{Integral symmetric bilinear forms and some of their
  applications}, {\emph{Mathematics of the USSR-Izvestiya} {\bfseries 14}
  (1980) 103}.

\bibitem{2004math......6536A}
V.~{Alexeev} and V.V.~{Nikulin}, \emph{{Classification of log del Pezzo
  surfaces of index $\le 2$}}, {\emph{ArXiv Mathematics e-prints} (2004) }
  [\href{https://arxiv.org/abs/math/0406536}{{\ttfamily math/0406536}}].

\bibitem{cox1999mirror}
D.~Cox and S.~Katz, \emph{Mirror Symmetry and Algebraic Geometry}, Mathematical
  surveys and monographs, American Mathematical Society (1999).

\bibitem{hori2003mirror}
K.~Hori and R.~Pandharipande, \emph{Mirror symmetry}, vol.~1, American
  Mathematical Soc. (2003).

\bibitem{Aspinwall:2009isa}
P.S.~Aspinwall, T.~Bridgeland, A.~Craw, M.R.~Douglas, A.~Kapustin, G.W.~Moore
  et~al., \emph{{Dirichlet branes and mirror symmetry}}, vol.~4 of \emph{Clay
  Mathematics Monographs}, AMS, Providence, RI (2009).

\bibitem{blumenhagen2009introduction}
R.~Blumenhagen and E.~Plauschinn, \emph{Introduction to conformal field theory:
  with applications to string theory}, vol.~779, Springer Science \& Business
  Media (2009).

\bibitem{Howe:1991ic}
P.S.~Howe and G.~Papadopoulos, \emph{{Holonomy groups and W symmetries}},
  \href{https://doi.org/10.1007/BF02097022}{\emph{Commun. Math. Phys.}
  {\bfseries 151} (1993) 467}
  [\href{https://arxiv.org/abs/hep-th/9202036}{{\ttfamily hep-th/9202036}}].

\bibitem{witten1982constraints}
E.~Witten, \emph{Constraints on supersymmetry breaking}, {\emph{Nuclear Physics
  B} {\bfseries 202} (1982) 253}.

\bibitem{odake1989extension}
S.~Odake, \emph{Extension of n= 2 superconformal algebra and calabi-yau
  compactification}, {\emph{Modern Physics Letters A} {\bfseries 4} (1989)
  557}.

\bibitem{fiset2018superconformal}
M.-A.~Fiset, \emph{{Superconformal algebras for twisted connected sums and
  G$_{2}$ mirror symmetry}},
  \href{https://doi.org/10.1007/JHEP12(2018)011}{\emph{JHEP} {\bfseries 12}
  (2018) 011} [\href{https://arxiv.org/abs/1809.06376}{{\ttfamily
  1809.06376}}].

\bibitem{figueroa1997extended}
J.M.~Figueroa-O'Farrill, \emph{{A Note on the extended superconformal algebras
  associated with manifolds of exceptional holonomy}},
  \href{https://doi.org/10.1016/S0370-2693(96)01506-7}{\emph{Phys. Lett. B}
  {\bfseries 392} (1997) 77}
  [\href{https://arxiv.org/abs/hep-th/9609113}{{\ttfamily hep-th/9609113}}].

\bibitem{vafa1989string}
C.~Vafa, \emph{String vacua and orbifoldized lg models}, {\emph{Modern Physics
  Letters A} {\bfseries 4} (1989) 1169}.

\bibitem{greene1997string}
B.R.~Greene, \emph{{String theory on Calabi-Yau manifolds}},  in
  \emph{{Theoretical Advanced Study Institute in Elementary Particle Physics
  (TASI 96): Fields, Strings, and Duality}}, pp.~543--726, 6, 1996
  [\href{https://arxiv.org/abs/hep-th/9702155}{{\ttfamily hep-th/9702155}}].

\bibitem{GriffithsI}
P.A.~Griffiths, \emph{On the periods of certain rational integrals: I},
  {\emph{Annals of Mathematics} {\bfseries 90} (1969) 460}.

\bibitem{gepner1987modular}
D.~Gepner and Z.~Qiu, \emph{Modular invariant partition functions for
  parafermionic field theories}, {\emph{Nuclear Physics B} {\bfseries 285}
  (1987) 423}.

\bibitem{Braun:2018fdp}
A.P.~Braun, M.~Del~Zotto, J.~Halverson, M.~Larfors, D.R.~Morrison and
  S.~Sch\"afer-Nameki, \emph{{Infinitely many M2-instanton corrections to
  M-theory on G$_{2}$-manifolds}},
  \href{https://doi.org/10.1007/JHEP09(2018)077}{\emph{JHEP} {\bfseries 09}
  (2018) 077} [\href{https://arxiv.org/abs/1803.02343}{{\ttfamily
  1803.02343}}].

\bibitem{Acharya:2018nbo}
B.S.~Acharya, A.P.~Braun, E.E.~Svanes and R.~Valandro, \emph{{Counting
  associatives in compact $G_{2}$ orbifolds}},
  \href{https://doi.org/10.1007/JHEP03(2019)138}{\emph{JHEP} {\bfseries 03}
  (2019) 138} [\href{https://arxiv.org/abs/1812.04008}{{\ttfamily
  1812.04008}}].

\bibitem{Joyce:2016fij}
D.~Joyce, \emph{{Conjectures on counting associative 3-folds in
  $G_2$-manifolds}},  \href{https://arxiv.org/abs/1610.09836}{{\ttfamily
  1610.09836}}.

\bibitem{Joyce:1999nk}
D.~Joyce, \emph{{A new construction of compact 8-manifolds with holonomy ${\rm
  Spin}(7)$}}, {\emph{J. Diff. Geom.} {\bfseries 53} (1999) 89}
  [\href{https://arxiv.org/abs/math/9910002}{{\ttfamily math/9910002}}].

\bibitem{Braun:2018joh}
A.P.~Braun and S.~Sch\"afer-Nameki, \emph{{Spin(7)-manifolds as generalized
  connected sums and 3d $\mathcal{N}=1$ theories}},
  \href{https://doi.org/10.1007/JHEP06(2018)103}{\emph{JHEP} {\bfseries 06}
  (2018) 103} [\href{https://arxiv.org/abs/1803.10755}{{\ttfamily
  1803.10755}}].

\bibitem{Blumenhagen:2001qx}
R.~Blumenhagen and V.~Braun, \emph{{Superconformal field theories for compact
  manifolds with spin(7) holonomy}},
  \href{https://doi.org/10.1088/1126-6708/2001/12/013}{\emph{JHEP} {\bfseries
  12} (2001) 013} [\href{https://arxiv.org/abs/hep-th/0111048}{{\ttfamily
  hep-th/0111048}}].

\bibitem{cox2011toric}
D.A.~Cox, J.B.~Little and H.K.~Schenck, \emph{Toric varieties}, vol.~124,
  American Mathematical Soc. (2011).

\bibitem{fulton}
W.~Fulton, \emph{Introduction to Toric Varieties. (AM-131)}, Princeton
  University Press (1993).

\bibitem{de2010triangulations}
J.~De~Loera, J.~Rambau and F.~Santos, \emph{Triangulations: Structures for
  Algorithms and Applications}, Algorithms and Computation in Mathematics,
  Springer Berlin Heidelberg (2010).

\bibitem{Klemm:1995tj}
A.~Klemm, W.~Lerche and P.~Mayr, \emph{{K3 Fibrations and heterotic type II
  string duality}},
  \href{https://doi.org/10.1016/0370-2693(95)00937-G}{\emph{Phys. Lett. B}
  {\bfseries 357} (1995) 313}
  [\href{https://arxiv.org/abs/hep-th/9506112}{{\ttfamily hep-th/9506112}}].

\bibitem{Avram:1996pj}
A.C.~Avram, M.~Kreuzer, M.~Mandelberg and H.~Skarke, \emph{{Searching for K3
  fibrations}},
  \href{https://doi.org/10.1016/S0550-3213(97)00214-9}{\emph{Nucl. Phys. B}
  {\bfseries 494} (1997) 567}
  [\href{https://arxiv.org/abs/hep-th/9610154}{{\ttfamily hep-th/9610154}}].

\bibitem{Candelas:1996su}
P.~Candelas and A.~Font, \emph{{Duality between the webs of heterotic and type
  II vacua}},
  \href{https://doi.org/10.1016/S0550-3213(96)00410-5}{\emph{Nucl.Phys.}
  {\bfseries B511} (1998) 295}
  [\href{https://arxiv.org/abs/hep-th/9603170}{{\ttfamily hep-th/9603170}}].

\bibitem{Candelas:2012uu}
P.~Candelas, A.~Constantin and H.~Skarke, \emph{{An Abundance of K3 Fibrations
  from Polyhedra with Interchangeable Parts}},
  \href{https://doi.org/10.1007/s00220-013-1802-2}{\emph{Commun. Math. Phys.}
  {\bfseries 324} (2013) 937}
  [\href{https://arxiv.org/abs/1207.4792}{{\ttfamily 1207.4792}}].

\end{thebibliography}
\end{document}